\documentclass[a4paper,11pt]{article}
\usepackage{jheppub} 
\usepackage{lineno}
\usepackage{wrapfig}
\usepackage{cancel}
\usepackage{graphicx}
\usepackage{dcolumn}
\usepackage{bigints}
\usepackage{bm}
\usepackage[usenames,dvipsnames]{color}
\usepackage{soul}
\usepackage{subcaption}
\usepackage{enumitem}
\usepackage{xspace}
\usepackage{booktabs}
\usepackage{lipsum}
\usepackage[per-mode=power]{siunitx} 
\usepackage[ISO]{diffcoeff}
\usepackage[capitalise]{cleveref}
\usepackage{ragged2e}
\usepackage{mathrsfs}
\usepackage{cleveref}
\usepackage{hyperref}
\usepackage[normalem]{ulem}
\usepackage{amsmath}
\usepackage{wrapfig}
\usepackage{tcolorbox}
\usepackage{esint}

\DeclareSIUnit\year{yr} 
\DeclareCaptionJustification{justified}{\justifying}
\newcommand{\bx}{{\bm x}}
\newcommand{\by}{{\bm y}}
\newcommand{\bk}{{\bm k}}

\newcommand{\bp}{{\bm p}}
\newcommand{\bq}{{\bm q}}

\newcommand{\beq}{\begin{equation}\begin{aligned}}
\newcommand{\eeq}{\end{aligned}\end{equation}}

\newcommand{\m}{m} 

\definecolor{rp}{cmyk}{0.2, 1, 0.6, 0}
\definecolor{rp}{cmyk}{0.2, 1, 0.6, 0}
\definecolor{green2}{cmyk}{0.27, 0, 1, 0.52}

\usepackage{color}
\hypersetup{
    colorlinks=true,       
    linkcolor=green2,          
    citecolor=green2,        
    filecolor=magenta,      
    urlcolor=green2           
}

\usepackage{physics}
\usepackage{tensor}

\title{Free Streaming in Warm Wave Dark Matter} 
\author[a]{Siyang Ling}
\emailAdd{siyang.ling@rice.edu}
\author[a]{, Mustafa A. Amin}
\emailAdd{mustafa.a.amin@rice.edu}

\affiliation[a]{Department of Physics and Astronomy, Rice University,
Houston, TX, 77005, U.S.A.}

\abstract{We provide a framework for numerically computing the effects of free-streaming
in scalar fields produced after inflation. First, we provide a
detailed prescription for setting up initial conditions in the field. This prescription allows us to specify the power
spectra of the fields  (peaked on subhorizon length scales and without a homogeneous field mode), and importantly, also  correctly reproduces the
behaviour of density perturbations on large length scales consistent with 
superhorizon adiabatic perturbations. We then evolve
the fields using a spatially inhomogeneous Klein-Gordon equation, including the effects of expansion and radiation-sourced metric perturbations. We show how gravity enhances, and how free streaming erases the initially
adiabatic density perturbations of the field, revealing more of the underlying, non-evolving, white-noise isocurvature density contrast. Furthermore, we explore the effect of
non-gravitational self-interactions of the field, including oscillon formation, on the suppression dynamics. As part of this paper, we make our code, {\sf{Cosmic-Fields-Lite}} ({\sf{CFL}}), publicly available. 
For observationally accessible signatures, our work is particularly relevant for structure formation in light/ultralight dark matter fields.}

\begin{document}

\maketitle
\flushbottom

\newpage

\section{Introduction}
\label{sec:intro}
Dark matter (DM) makes up $\sim 85\%$ of the non-relativistic matter content in our cosmos \cite{Planck:2018vyg}. However, the  identity of dark matter particles/fields is unknown. Apart from the fact that they must interact gravitationally, we do not know their mass, spin, and other potential interactions~\cite{Cirelli:2024ssz}. Astrophysical observations allow for a broad range of masses for the dark matter ``particles": $10^{-19}{\rm eV}\lesssim m\lesssim {\rm few} \times M_{\odot}$~\cite{ParticleDataGroup:2020ssz,Dalal:2022rmp,Amin:2022nlh}. 

For  $m\lesssim 100\,\rm eV$, dark matter is necessarily bosonic \cite{Tremaine,DiPaolo:2017geq} because of phase space occupation number considerations (see, however, \cite{PhysRevD.103.055014}). For the $m\ll$ eV regime in particular, the occupation numbers of the bosonic field are sufficiently high that it is possible to describe dark matter as a classical field with wave-dynamics (rather than as discrete point particles governed by classical mechanics) \cite{Hui:2021tkt}. Examples of such dark matter fields include the QCD axion, other scalar fields, dark photon or vector dark matter, etc. For recent reviews, see for example \cite{Ferreira:2020fam,Hui:2021tkt,OHare:2024nmr} for scalars, and  \cite{Antypas:2022asj} for ultralight vectors (and scalars).  The production mechanisms for wave-like fields is typically non-thermal and can be either inflationary or post-inflationary. A canonical example is the QCD axion, with Peccei-Quinn (PQ) symmetry breaking happening before or after inflation. If the PQ symmetry is broken before inflation, typically a field with a dominant homogeneous mode is produced, whereas post-inflationary PQ symmetry breaking leads to a field without a homogeneous mode and with topological defects \cite{Dine:1982ah,Sikivie:1982qv,Hogan:1988mp,Gorghetto:2020qws,OHare:2021zrq,Buschmann:2021sdq}.

\begin{figure}[t]
  \centering
  \includegraphics[width=\textwidth]{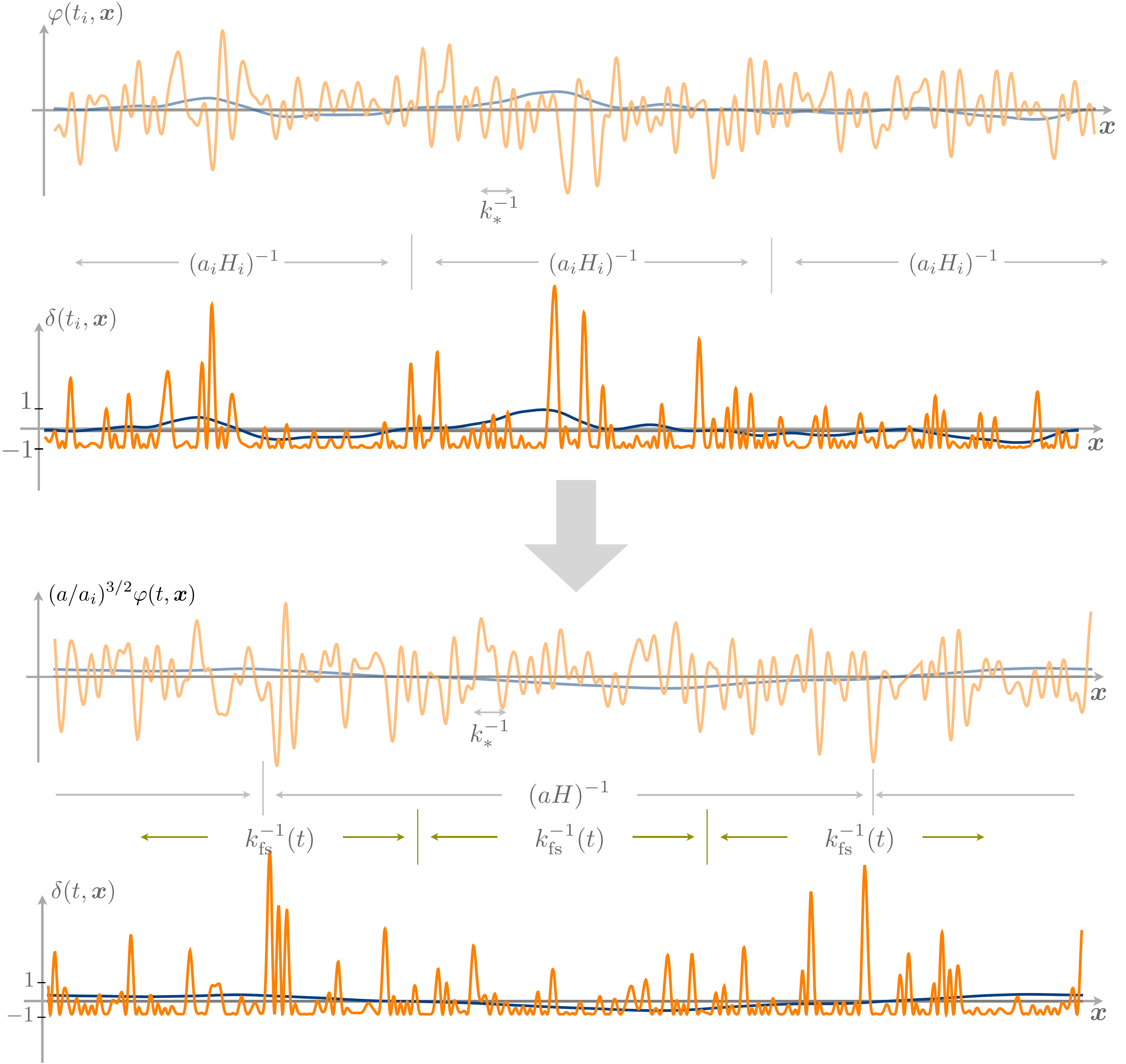}
  \caption{\small{A schematic of our initial conditions, and how free-streaming removes pre-existing correlations in the density perturbations. {\it Top two panels:} The field (light orange) at the initial time, $t_i$, is a  Gaussian random field with most of its power on subhorizon scales $k_*^{-1}\ll (a_iH_i)^{-1}$, with a spatially-dependent variance: $\langle\varphi^2(\bx)\rangle\propto 1+\mathcal{O}[\Psi(t,\bx)]$ ($\Psi$, in blue, is the gravitational potential). This spatial dependence is due to long-wavelength adiabatic perturbations on superhorizon scales. The fractional density perturbations $\delta$ (dark orange) at the initial time are dominated by (i) subhorizon scale features, (ii) a Poisson like distribution of these density features on length-scales larger than their separation, (iii) with even larger-scale density perturbations related to the spatial dependence of the field variance (equivalently the gravitational potentials). {\it Bottom Panels:} At later times, the field  looks statistically similar (apart from red-shifting). However, in detail, free-streaming removes existing correlations in the density fluctuations on length scales smaller than $k_{\rm fs}^{-1}(t)<(aH)^{-1}$. That is, the Poisson distribution of density perturbations now extends to larger length scales. }}
  \label{fig:cartoon}
\end{figure}

More generally, if the fields are produced after inflation, causality considerations typically lead to fields with a subhorizon correlation length at the time of production and significant spatial variations around the correlation length. Such fields lack a dominant spatially homogeneous component. 
The lack of a homogeneous mode and the presence of significant small-scale variation in the field leads to two effects: (1) the free-streaming of the fields, which leads to a suppression of the adiabatic density perturbations in the field, and (2) enhanced isocurvature density perturbations on small scales, which do not evolve during radiation domination. See our Fig.~\ref{fig:cartoon} for a heuristic overview of these effects. Note that such a field can still have an almost spatially homogeneous density with approximately adiabatic density perturbations on sufficiently large length scales (as required by observations).

In this paper, using 3+1 dimensional lattice simulations, we numerically explore the two aforementioned effects during the radiation dominated era for wave-like dark matter. We provide a detailed framework to generate initial conditions for such fields, which lack a zero mode, but are nevertheless consistent with adiabatic density perturbations on sufficiently large length scales. Our code, {\sf Cosmic-Fields-Lite (CFL)}, for generating initial conditions, as well as carrying out the time-evolution of the fields in the presence of expansion and metric perturbations sourced by radiation, can be downloaded at \href{https://github.com/hypermania/Cosmic-Fields-Lite}{https://github.com/hypermania/Cosmic-Fields-Lite}. We also provide a starting point for simulations in the late
universe, where self-gravity of the fields needs to be included.

For readers familiar with free-streaming and Poisson noise fluctuations of classical point particles, the following connection from field to particle description might be useful. Our field configuration (in Fig.~\ref{fig:cartoon}) can be viewed as a collection of quasi-particles with {\it comoving} size $\sim k_*^{-1}$, moving with typical comoving momenta $k_*$, or equivalently with physical velocities $v_*\sim k_*/am$, which is simply the group velocity of the waves. The free-streaming length is the characteristic length over which these quasi-particles move (in random directions) within some time $t$. Below this length, pre-existing large-scale correlations in densities are erased. However, the small-scale isocurvature density fluctuations arise from the spatial Poisson distribution of these quasi-particles. This distribution cannot be changed by quasi-particle motion with random velocities, hence the white noise remains even on length scales smaller than free-streaming length. Finally, it might also be helpful to think of the field power spectrum as the momentum distribution of the particles in phase-space.

Observations of structure on large length scales $\gtrsim \rm Mpc$ (comoving) have confirmed the existence of dark matter through its gravitational effects. Observational probes of smaller scale structure now hold the potential for revealing its identity and its production mechanism.  Existing and upcoming measurements of the small-scale matter power spectrum including Ly$\alpha$,  galaxy satellite populations, gravitational lensing, stellar streams, 21 cm intensity mapping \cite{Mondino:2020rkn,Drlica-Wagner:2022lbd,Chung:2023syw,Irsic:2023equ,Delos:2023dwq,Nadler:2024ims,Xiao:2024qay,Ji:2024ott,deKruijf:2024voc}, etc. provide a strong motivation for our work. We hope that future calculations based on our numerical framework will be able to provide robust predictions for features in the small-scale power spectrum, encompassing free-streaming suppression and isocurvature enhancement. With observations, these features can be used to constrain or confirm the properties of dark matter and its production mechanism. 

To put our work in context, we provide a brief (and incomplete) overview of the existing literature related to free-streaming dynamics of fields in the early universe. 

Our paper follows and builds on the analytic work of \cite{Amin:2022nlh}, where the expected effect on the matter power spectrum due to free-streaming suppression and isocurvature enhancement is explored. The lack of observation of these features for length scales larger than $\sim \rm Mpc$ was then used to constrain the mass of dark matter particles to $m\gtrsim 10^{-19}\,\rm eV$. In \cite{Liu:2024pjg}, the authors focused on an axion-like scenario, and explored the relative importance of the two effects for constraining the mass (also see  \cite{Irsic:2019iff} for the isocurvature part). Here, we perform detailed numerical simulations of the field dynamics to explore both effects, and do not rely on analytic estimates. Our numerical results confirm the expected lack of evolution of isocurvature density perturbations, and the free-streaming suppression of the adiabatic perturbations. 

There is a large body of literature on the evolution of ultralight dark matter fields in the early universe. However, contrary to the scenario considered in this work, the majority of the literature deals with the case where the fields have a dominant homogeneous mode~\cite{Hwang:2009js,Marsh:2010wq,Nelson:2011sf,Arias:2012az,Noh:2017sdj,Cembranos:2015oya}. In this case, the fields can be linearized with respect to the homogeneous mode, and the equations for the linearized density perturbations can be solved mode by mode~\cite{Hu:1998kj,Poulin:2018dzj}. Typically, these equations are solved via software packages such as AxionCAMB or CLASS~\cite{Blas:2011rf,2022ascl.soft03026G}. This approach had been exploited to determine the cosmological consequences in the case of single and multiple scalar fields~\cite{Hlozek:2014lca,Marsh:2015xka}. Typically, free-streaming effects are negligible, and the most remarkable effect on the density power spectrum comes from Jeans suppression~\cite{Hu:2000ke,Irsic:2017yje,PhysRevD.101.123026,PhysRevLett.126.071302,Passaglia:2022bcr} (at $k_J\sim a\sqrt{mH}$). The implications of the Jeans suppression in the late, matter dominated universe (starting with an almost homogeneous field) has been explored using Schr\"{o}dinger Poisson simulations \cite{Schive:2014dra,Mocz:2019pyf,Chan:2021bja,May:2021wwp,May:2022gus}. In contrast, the lack of a zero mode in our scenario means that the fields can be ``warm", and free-streaming will provide a stronger bound than Jeans suppression.

Another widely used approach for understanding the time evolution of dark matter fields is numerical integration of the field equations on a lattice in the early universe. These simulations have been performed for fields with and without a homogeneous mode, and with self-interactions~\cite{Kolb:1993hw, Amin:2019ums,Lozanov:2017hjm,Buschmann:2021sdq,Garcia:2022vwm,Agrawal:2018vin}. However, typically such simulations do not include adiabatic density perturbations, so the free-streaming effects of the light fields are not apparent. To see substantial effects from free-streaming, one must include density perturbations beyond isocurvature perturbations in the fields' initial conditions.

There is a long history of studies related to free-streaming phenomenology in the context of particle-like dark matter~\cite{Narayanan:2000tp,Hansen:2001zv,Lewis:2002nc,Green:2003un,Viel:2005qj,Lesgourgues:2006nd,Viel:2007mv,Boyarsky:2008xj,Erickcek:2011us,Lancaster:2017ksf,Irsic:2017ixq,wdm,Erickcek:2021fsu,Ballesteros:2020adh,Sarkar:2021pqh,Garcia:2023qab,Irsic:2023equ}. Notable examples include the suppression of the density spectrum due to free-streaming of massive neutrinos and warm dark matter (thermal and non-thermal).
In these studies, the matter content is typically coarse grained over a length scale much larger than the deBroglie wavelength of the particles and particle separations, and the linearized density perturbations evolve according to the fluid equations (or more generally, Boltzmann equations).  For our case, we also consider the density perturbations on the deBroglie scale, which necessitates a treatment that includes the wave dynamics. As we mentioned earlier, for length scales large compared to the deBroglie scale, there is a useful map between the particle and wave dynamical picture using the dynamics of deBroglie scale quasi-particles.

The rest of the paper is organized as follows. We introduce our model, and key equations in Sec.~\ref{sec:model}. In Sec.~\ref{sec:IC} we provide a detailed discussion of the field initial conditions, including a step-by-step numerical algorithm for generating such initial conditions. The evolution of the field and its implication are discussed in Sec.~\ref{sec:evol}. We discuss the analytical expectations of field evolution in Sec.~\ref{sec:evolution_of_isocurvature} and Sec.~\ref{sec:FreeStreamingPhysics}, and the numerical results from Sec.~\ref{sec:W/GravPertEvol} to Sec.~\ref{sec:monodromy_potential_evolution}. Our simulations include cases with and without gravitational perturbations, as well as with and without non-gravitational self-interactions. We discuss immediate future directions in Sec.~\ref{sec:future}, and summarize our key results in Sec.~\ref{sec:summary}. In a series of appendices (\ref{sec:EOS}-\ref{sec:smoothing}), we flesh out details of our analytic and numerical calculations, and provide some heuristic insights to make our results easily reproducible and understandable.

Before moving on to the main section, we specify our notation and conventions below.
\subsection*{Notation and Conventions}
\label{sec:NotCon}
We set $\hbar=c=1$ and use $-+++$ metric signature. We will use the following conventions for Fourier transform and spectrum for any function $f(\bx)$ (which might itself be a random variable),
\begin{align}
  \label{eq:fourier_conventions}
    & f_\bk \equiv \int_\bx e^{- i \bk \cdot \bx} f(\bx),\quad
    f(\bx) \equiv \int_\bk e^{i \bk \cdot \bx} f_\bk\,,\\
    &P_f(k) \equiv (2\pi)^3\int_{\bk'} \frac{\abs{f_{\bk'}}^2}{\mathcal{V}} \frac{\delta_D(\abs{\bk'}-k)}{4\pi k^2},\quad
    \Delta^2_f(k) \equiv \frac{k^3}{2\pi^2} P_f(k)\,.
\end{align}
where $\mathcal{V}=(2\pi)^3\delta_D(\bm{0})$ is the formal spatial volume factor, and we defined $\int_{\bx}=\int d^3\bx$ and $\int_\bk=\int d^3\bk/(2\pi)^3$ to reduce clutter.
Also, $\overline{f}$ will denote spatial average of $f$:
\beq
\overline{f}= \frac{1}{\mathcal{V}}\int_{\bx} f(\bx)\,, 
\eeq
We have suppressed the time-dependence of the function $f$ and related quantities. These definitions are valid for each realization of $f$.\footnote{While we define Fourier transforms and power spectra in the infinite spatial volume limit, when necessary, we will restrict the field to a finite volume $\mathcal{V} = L^3$, so that $\bk$ become discrete, the integral over $\bk$ becomes a sum, and Dirac Delta functions in Fourier space become Kronecker Delta functions. Specifically, we have $f_\bk\rightarrow \sqrt{L^3}f^L_{\bk}$, $\int_\bk\rightarrow \sum_\bk/L^3$ and $(2\pi)^3\delta_D(\bk-\bk')\rightarrow L^3\delta_{\bk,\bk'}$.}

We will also refer to ensemble averaged versions of the above quantities. Whenever we use expectation value $\expval{\hdots}$ or underline $\underline{\hdots}$, we mean averaging over an ensemble, instead of space.  For example, $\expval{f(\bx)}$ or $\underline{f(\bx)}$ would mean the value of $f$ at $\bx$ averaged over the ensemble, instead of the spatial average $\overline{\varphi}$. We also introduce symbols for ensemble averaged power spectra: $\underline{P}_f(k)=\expval{ P_f(k) }$ and $\underline{\Delta}_f(k)=\expval{\Delta_f(k)}$. For our purposes, it is important to distinguish between ensemble averaged spectrum and realization spectrum. For example, we typically specify an ensemble of homogeneous Gaussian random field with an ensemble averaged spectrum $\underline{P}_f$; if we draw realizations from that ensemble, each realization spectrum $P_f$ can be different from $\underline{P}_f$, but taking the average of $P_f$ over many realizations will lead to a result converging to $\underline{P}_f$. 

Note that any given realization can have spatial inhomogeneities even if the ensemble averaged quantities are spatially homogeneous. By statistical inhomogeneity, we mean that the ensemble averaged quantities are also spatially inhomogeneous, $\expval{f(\bx)} \neq \expval{f({\bm y})}$ for $\bx \neq {\bm y}$.  Whenever we use the word ``statistical", we refer to a property of an underlying field ensemble, instead of a single field realization.

Finally, we note that the ensembles under consideration are ensembles for field $\varphi$ and $\dot{\varphi}$; these ensembles are such that in an ensemble-averaged sense, they reproduce the correct density perturbations on long length scales. The long length scale density perturbations are themselves part of a spatially homogeneous and isotropic ensemble. Unless stated otherwise, the ensemble averages mean averages  over the $\varphi$ and $\dot{\varphi}$ ensemble.

\section{Model}
\label{sec:model}

We are interested in studying the evolution of a light scalar field $\varphi$ (dark matter) during radiation domination. In this section, we introduce our model for the scalar field and the equations that we will evolve numerically.
\subsection{Geometry}
We use Ma \& Bertschinger's convention for Newtonian gauge metric \cite{Ma:1995ey}:
\begin{align}
  \label{eq:metric}
  & ds^2 = -e^{2\Psi} dt^2 + a^2 e^{-2\Phi} d\bx^2\,,
\end{align}
where $t$ is cosmic time and $|\Phi|,|\Psi|\ll 1$. In radiation domination, the scale factor $a(t)$ and Hubble rate $H=\dot{a}/a$ are
\begin{align}
  \label{eq:radiation_domination}
  a(t) = a_i (t / t_i)^{1/2},\quad H(t) = H_i(t / t_i)^{-1},\quad H_i = 1 / (2 t_i)\,.
\end{align}
In what follows, $\dot{f} =\partial_tf$ and $t_i$ represents an initial time during radiation domination.

We assume that the field $\varphi$ is a subdominant component of the total energy content deep in the radiation era (but $\varphi$ could be all of dark matter), so its contribution to the gravitational potentials $\Phi$ and $\Psi$ are negligible. Consistent with this assumption, the Fourier components of the gravitational potentials during this era are taken to be:
\begin{align}
  \label{eq:gravitational_potentials_in_radiation_domination}
  \Psi_\bk = \Phi_\bk = 2 \mathcal{R}_\bk \frac{\sin(k \eta / \sqrt{3}) - (k \eta / \sqrt{3}) \cos(k \eta / \sqrt{3})}{(k \eta / \sqrt{3})^3} \qq{where} \eta = \frac{(2 H_i t)^{1/2}}{a_i H_i}\,,
\end{align}
where $k$ is the comoving wavenumber, and $\mathcal{R}$ is the comoving curvature perturbation. We ignore anisotropic stress, and set $\Phi=\Psi$. This completely specifies the geometry of spacetime for this work.
\subsection{Field Evolution}
We take the scalar field action to be
\begin{align}
  \label{eq:action}
  & S_\varphi = \int d^4\bx \sqrt{-g} \left[-\frac12 \nabla_\mu \varphi \nabla^\mu \varphi - V(\varphi) \right]\,,
\end{align}
and include gravitational effects via the metric and covariant derivatives as usual. The scalar field potential we use in most of  the paper is that of a free field
\begin{align}
  \label{eq:varphi_potential}
  V(\varphi) = \frac12 m^2 \varphi^2\,.
\end{align}
Later in the paper, we also consider $V(\varphi)=m^2M^2\left[\sqrt{1+(\varphi/M)^2}-1\right]$ when we briefly discuss the effect of self-interactions on free-streaming. The equations below and our numerical schemes are valid for a general $V(\varphi)$.

The equation of motion is 
\begin{align}
  \label{eq:varphi_eom}
  0 &= \nabla^\nu \nabla_\nu \varphi - V'(\varphi)\,, \nonumber \\
  0 &= \ddot{\varphi} + (3 H - 4 \dot{\Psi}) \dot{\varphi} - e^{4\Psi} \frac{\nabla^2\varphi}{a^2} + e^{2\Psi} V'(\varphi)\,,
\end{align}
The stress-energy tensor and its components are
\begin{align}
  \label{eq:varphi_stress_energy}
  T\indices{_\mu_\nu} &= \frac{-2}{\sqrt{-g}} \fdv{S_\varphi}{g\indices{^\mu^\nu}}\,, \nonumber \\
  \rho &= -T\indices{^0_0} = e^{-2\Psi} \frac12 \dot{\varphi}^2 + e^{2\Psi} \frac{1}{2 a^2} (\nabla \varphi)^2 + V(\varphi)\,,\nonumber \\
  p &= \frac13 T\indices{^i_i} = e^{-2\Psi} \frac12 \dot{\varphi}^2 - e^{2\Psi} \frac{1}{6 a^2} (\nabla \varphi)^2 - V(\varphi)\,, \nonumber \\
  Q_i &= \frac{1}{a} T\indices{^0_i} = - \frac{e^{-2\Psi}}{a} \dot{\varphi} \partial_i \varphi\,,
\end{align}
where $\rho$, $p$ and $\bm{Q}$ are the energy density, pressure and momentum density respectively. The continuity equation is 
\begin{align}
  \label{eq:fluid_equations}
  & \nabla_\mu T\indices{^\mu_\nu} = 0\,, \nonumber \\
  & \dot{\rho} + 3 (H - \dot{\Psi}) (\rho + p) + \frac{\nabla}{a} \cdot {\bm Q} = (3 - e^{4\Psi}) {\bm Q} \cdot \frac{\nabla}{a} \Psi\,.
\end{align}
We will always work in the regime where $|\Psi|\ll 1$, so the above expressions are consistent with the ones where the exponentials are expanded to linear order in $\Psi$. To reduce clutter, here the $\Psi$'s are kept within the exponents.

\section{Initial Conditions}
\label{sec:IC}

We consider field realizations generated by local processes after inflation, which lead to isocurvature density perturbations on small subhorizon scales. Nevertheless, these field realizations must also be consistent with the adiabatic density perturbations correlated to curvature on larger superhorizon scales. We describe a novel way to generate such field realizations as initial conditions for lattice simulations of the field in a spatially perturbed cosmological background. The key feature we aim to capture is the adiabatic density perturbations related to our field, which is crucial for simulating the free-streaming suppression of the density perturbations. Section \ref{sec:statistical_properties} contains the expected statistical properties of the field, including field and energy overdensity power spectra, and spatially dependent variances of the field.
Section \ref{sec:generating_fields} contains the numerical procedure for generating field realizations and detailed explanation for its validity. 

\subsection{Statistical properties of the scalar field}
\label{sec:statistical_properties}
In this section we discuss the statistical properties of the field at the initial time $t_i$, including field and energy overdensity power spectra, and spatially dependent variances of the field. 
\subsubsection{No homogeneous mode and field power spectra}
\label{sec:field_spectra}
Our focus in this paper is on scalar fields without a zero mode, that is $\varphi(\bx)$ such that  $\overline{\varphi}=0$ on the largest scale of interest. Such fields arise naturally when they are produced post-inflation via local production mechanisms. Example production mechanisms for $\varphi$ include parametric resonance from a misaligned parent scalar (for a review, see \cite{Lozanov:2019jxc}), as well as a PQ phase transition after inflation for an axion-like field (see \cite{OHare:2024nmr} for a review).

\begin{figure}[t]
  \centering
    \includegraphics[width=\textwidth]{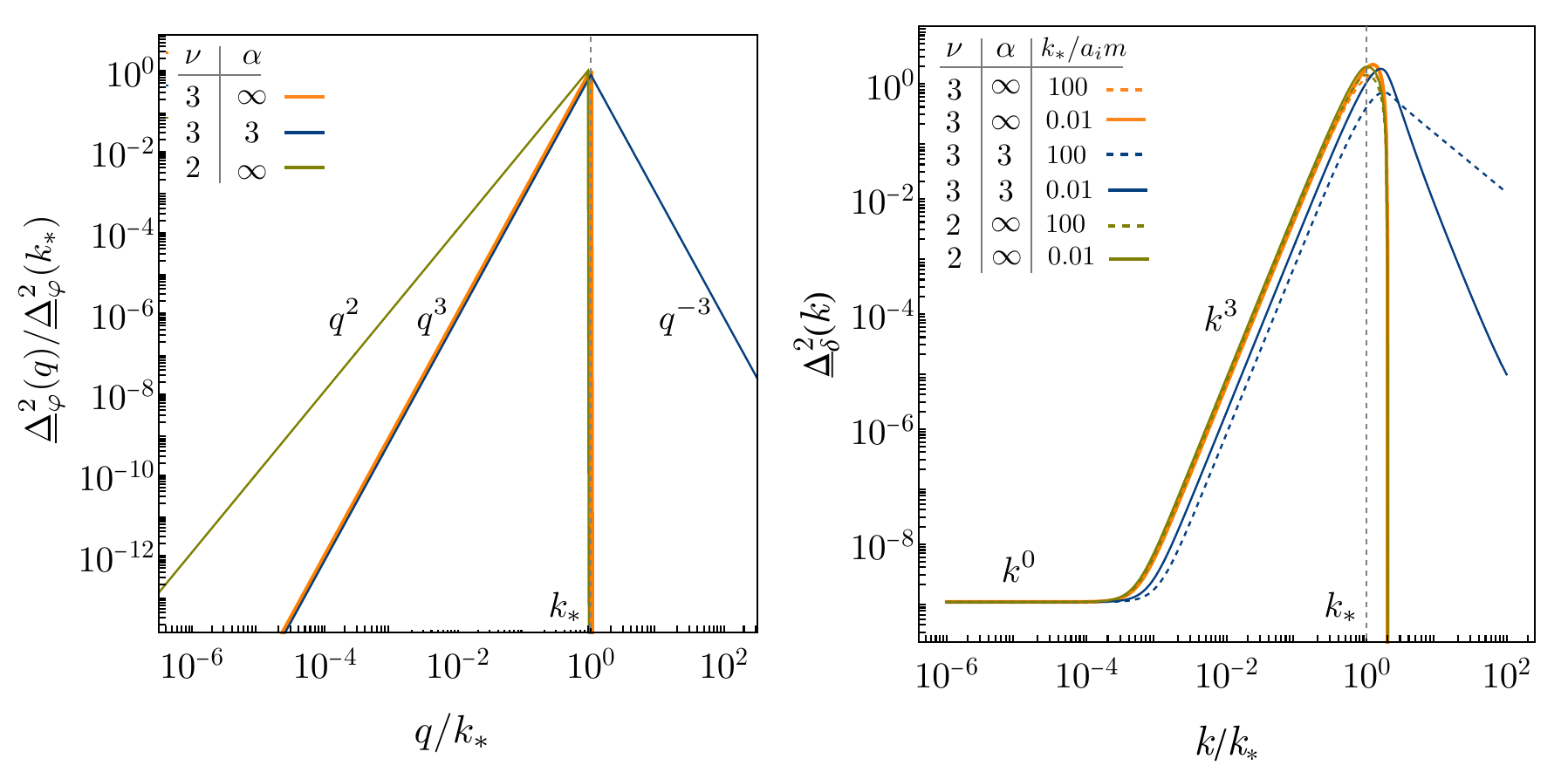}
  \caption{Initial conditions for the field spectrum $\underline{\Delta}^2_{\varphi}(q)$ and density spectrum $\underline{\Delta}^2_{\delta}(k)$ for some sample values of $\alpha$, $\nu$ and $k_*/a_i m$ in Eq.~\eqref{eq:P_varphi_candidates}. The $\underline{\Delta}^2_{\varphi}(q)$ is peaked at $k_\ast$ by choice. The $\underline{\Delta}^2_{\delta}(k)$, spectra, evaluated from the field variations,  exhibit different shapes depending on the high-$k$ tail and $k_\ast / a_i m$. One can see that small length scale power in $\underline{\Delta}^2_{\varphi}(k)$ leads to small length scale power in $\underline{\Delta}^2_{\delta}(q)$. On large length scales with $k \ll k_\ast$, all spectra exhibit $k^3$ power law, as expected for quantities with finite correlation length. Furthermore,  $\underline{\Delta}^2_\delta(k\ll k_*)\propto k^3$  even if the field spectra did not have $k^3$ behavior for $k\ll k_*$. The scale-invariant $\Delta^2_\delta\propto k^0$ low-$k$ tail discussed in Sec.~\ref{sec:density_spectra} is also added here with amplitude $\Delta_{\mathcal{R}}^2\sim 10^{-9}$. For $\alpha=\infty$, if the field is non-relativistic, then the density power spectrum is higher than the ultra-relativistic one around $k_\ast$.}
  \label{fig:sample_field_spectrum}
\end{figure}

Beyond (the lack of) a homogeneous mode, we wish to specify the statistical properties of the inhomogeneous field. We assume that it is a Gaussian random field. The specific form of the field power spectrum is model-dependent. A simple parametrization for the field spectrum that captures qualitative features of a number of scenarios (with some inflationary and most post-inflationary production mechanisms) was given by \cite{Amin:2022nlh}:
\begin{align}
    \label{eq:P_varphi_candidates}
     \underline{\Delta}^2_\varphi(q) = \underline{\Delta}^2_\varphi(k_\ast) \left[ \left(\frac{q}{k_\ast}\right)^\nu \Theta(k_\ast - q) + \left(\frac{q}{k_\ast}\right)^{-\alpha} \Theta(q - k_\ast)\right]\,,
\end{align}
where $\nu,\alpha>0$, and $\Theta(x)$ is the Heaviside step function. In this work, we take $\nu=3$, $\alpha=\infty$ and $k_*\gg a_i H_i$. From the expectation of equipartition of energy between positive and negative frequency modes, we use $\underline{\Delta}^2_{\dot{\varphi}}(q) = \omega_q^2  \underline{\Delta}^2_\varphi(q) $ where $\omega_q=\sqrt{q^2/a^2+m^2}$.   The left panel of Fig.~\ref{fig:sample_field_spectrum} shows some example field spectra based on Eq.~\eqref{eq:P_varphi_candidates}.

To justify our choice $\nu=3$, consider a scalar field $\varphi$ that is produced in different spatial locations (separated by $r>r_*\sim 1/k_*$) by independent physical processes. That is, $\expval{\varphi(\bx)\varphi(\bx')} = \expval{\varphi(\bx)} \expval{\varphi(\bx')} = 0$ for $\abs{\bx - \bx'} \gg r_\ast$. Such a field will exhibit a $q^3$ power law for the low $q$ part of the spectrum, that is $\nu=3$ for $q\ll k_*$. This fact can be checked by directly evaluating the expectation for $\Delta^2_\varphi(q)$ with \eqref{eq:fourier_conventions}:
\begin{align}
    \underline{\Delta}^2_\varphi(q) = \frac{q^3}{2\pi^2}  \int  \left[ \frac{1}{\mathcal{V}} \int \int e^{i {\bm r} \cdot \bq'} \expval{\varphi(\bx) \varphi(\bx + {\bm r})} d^3{\bm r} d^3\bx \right]  \frac{\delta_D(\abs{\bq'}-q)}{4 \pi q^2} d^3\bq'.
\end{align}
By assumption, the two-point correlation function in the integrand becomes $0$ when $\abs{\bm r} > r_\ast$, so the integral tends to a constant for $q \ll r_\ast^{-1}$, and $\underline{\Delta}^2_\varphi(q) \propto q^3$.  Note that we have not assumed statistical isotropy or homogeneity in the above argument; in fact, as we will show in the next subsections, these assumptions are false in our case of interest. Relaxing the assumption that the correlation strictly vanishes beyond $r_*$ will lead to a different power law $\nu$.\footnote{We note that even if we change $\nu=3$ to $\nu=2$ (more generally, to any $\nu>3/2$), the $k^3$ behaviour of the density perturbations, $\delta$, discussed in the next section would persist -- that is, the $k^3$ in the density spectrum is not determined by $q^3$ in the field spectrum.}

The small-scale physics of the production mechanism determines the shape of the spectrum at around $k_\ast \equiv r_\ast^{-1}$ and above, and the specific shape of $\underline{\Delta}^2_\varphi(q) $ is model dependent.  For instance, if $\varphi$ is produced via parametric resonance from a field $X$ with mass $m_X$, the peak $k_\ast / a$ would be determined by the particle mass $m_X$ \cite{Amin:2014eta, Lozanov:2019jxc}.  Typically $k_*$ is subhorizon, or becomes subhorizon soon after initial production. For $q\gg k_*$, $\underline{\Delta}^2_\varphi(q)$ must decrease, since the field has finite energy density, hence assuming $\alpha >0$ is sensible. For production mechanisms that are fast compared to the Hubble time at the time of production, we expect $\alpha$ to be large (sharp cutoff) \cite{Ballesteros:2020adh,Garcia:2022vwm}. The normalization $\underline{\Delta}^2_\varphi(k_*)$ can be fixed by requiring that the total density of $\varphi$: $\expval{\overline{\rho}} \approx (1/2)\int d\ln q \left[ \underline{\Delta}_{\dot{\varphi}}^2(q)+\omega_q^2 \underline{\Delta}_\varphi^2(q) \right]$ matches the necessary dark matter density at $t_i$.\footnote{It's also worth mentioning that since $\overline{\varphi^2} =  \sigma_\varphi^2 + \overline{\varphi}^2 = \sigma_\varphi^2 = \int d\ln{q} \Delta^2_\varphi(q)$ where $\sigma_\varphi^2 = \overline{(\varphi - \overline{\varphi})^2}$, we expect field amplitudes to be of order $\sqrt{\Delta^2_\varphi(k_\mathrm{peak})}$.}
\subsubsection{Density power spectra}
\label{sec:density_spectra}
In this subsection, we discuss features of the spectrum $\Delta^2_\delta(k)$, where $\delta = (\rho - \overline{\rho}) / \overline{\rho}$ is the density contrast in the energy density of  $\varphi$. 

On sufficiently large length scales (superhorizon at $t_i$, and eventually probed by CMB and large-scale structure observations),  we assume that $\delta$ is consistent with adiabatic initial conditions
\begin{align}
  \label{eq:adiabatic_ic}
  \delta_\bk = -\frac32 (1 + w) \Psi_\bk\,,\qquad
  \dot{\delta}_\bk=3(1+w)\dot{\Psi}_\bk+\frac{\dot{w}}{1+w}\delta_\bk\,,
\end{align}
where $w\equiv \overline{p}/\overline{\rho}$ is the equation of state of the field in a given realization, and $\overline{\rho}$ and $\overline{p}$ are averages over a scale much larger than the $k^{-1}$ (see Appendix \ref{sec:EOS}). From observations we know that $\Delta^2_\mathcal{R}$ (or equivalently $\Delta^2_\Psi$, since $\Psi=(2/3)\mathcal{R}$), $\Delta^2_\delta$ and $\Delta^2_{\dot{\delta}}$ should be approximately scale-invariant on these scales.

On sufficiently small length scales inside the horizon at $t_i$, where the gravitational potentials due to radiation are negligible, (isocurvature) density fluctuations are predominantly determined by small-scale field fluctuations. In other words, the density spectrum $\Delta_\delta^2(k)$ is dependent on the field spectra $\Delta_\varphi^2(q)$ and $\Delta_{\dot{\varphi}}^2(q)$.  Since $\rho(\bx)$ is a function of $\varphi(\bx)$ and $\dot{\varphi}(\bx)$, if a field has typical wavenumber $k_\ast$, then its density fluctuations would have a typical length scale of around $1 / k_\ast$.
For illustration, let us explicitly compute the $\underline{\Delta}_\delta^2(k)$ spectrum for an ensemble, assuming $\Psi = 0$.
Given a field realization, we have
\begin{align}
    \abs{\delta_{\bk}}^2&= \frac{1}{\overline{\rho}^{2}}\abs{ \int_\bx e^{-i \bk \cdot \bx} \left[ \frac12 \dot{\varphi}^2(\bx) + \frac{1}{2a^2} (\nabla\varphi(\bx))^2 + \frac12 m^2 \varphi^2(\bx) \right] }^2\,,  \nonumber \\
    &=  \frac{1}{\overline{\rho}^{2}}\abs{ \int_{\bm q} \left( \frac12 \dot{\varphi}_{\bm q} \dot{\varphi}_{\bk - {\bm q}} + \frac{m^2 + {\bm q} \cdot ({\bm q} - \bk) / a^2}{2} \varphi_{\bm q} \varphi_{\bk - {\bm q}} \right) }^2 \;.
\end{align}
Suppose $\varphi$ is a homogeneous Gaussian random field specified by $\expval{\varphi_\bq^\ast \varphi_{\bq'}} = (2\pi)^3 \delta_D(\bq - \bq') \underline{P}_\varphi(q)$, $\expval{\dot{\varphi}_\bq^\ast \dot{\varphi}_{\bq'}} = (2\pi)^3 \delta_D(\bq - \bq') \underline{P}_{\dot{\varphi}}(q)$ and  $\expval{\dot{\varphi}_\bq^\ast \varphi_{\bq'}} = 0$ at $t_i$. Taking the ensemble average of $\abs{\delta_\bk}^2$ and putting it in Eq.~\eqref{eq:fourier_conventions} gives:
\begin{align}
\label{eq:predicted_isocurvature_density_perturbation}
  \underline{P}_\delta(k) 
  &= \frac{1}{2{\overline{\rho}}^2} \int_\bq \left[ \underline{P}_{\dot\varphi}(q) \underline{P}_{\dot{\varphi}}(\abs{\bk - {\bm q}}) + \left( m^2 + \frac{{\bm q} \cdot ({\bm q} - \bk)}{a^2} \right)^2 \underline{P}_{\varphi}(q) \underline{P}_{\varphi}(\abs{\bk - {\bm q}}) \right] \;.
\end{align}
Note that $\overline{\rho}=\langle \overline{\rho} \rangle$ for fields in an infinite volume.\footnote{ 
  The ensemble variance of $\overline{\rho}$ scales as $1 / \mathcal{V}$ if the two point correlation of $\rho$ drops sufficiently quickly with distance.
  More specifically, the ensemble variance of $\overline{\rho}$ is $\expval{(\frac{1}{\mathcal{V}} \int_\bx \rho(\bx) - \expval{\overline{\rho}})^2} = \frac{1}{\mathcal{V}^2} \int_{\bx, {\bm y}} \expval{(\rho(\bx)-\expval{\overline{\rho}})(\rho(\bm y)-\expval{\overline{\rho}})} = \frac{1}{\mathcal{V}^2} \int_{\bx} \left[ \int_{\bm r} \expval{(\rho(\bx)-\expval{\overline{\rho}})(\rho(\bx + {\bm r})-\expval{\overline{\rho}}) } \right]$, so as long as the integral $\int_{\bm r} \expval{(\rho(\bx)-\expval{\overline{\rho}})(\rho(\bx + {\bm r})-\expval{\overline{\rho}})}$ is finite, the variance is $1/\mathcal{V}$ suppressed.
  This implies that, in the limit $\mathcal{V}\to\infty$, we have $\overline{\rho} = \expval{\overline{\rho}}$ for each field realization.
  In light of this, we will use $\overline{\rho}$ and $\expval{\overline{\rho}}$ ($\overline{\varphi^2}$ and $\expval{\overline{\varphi^2}}$, etc) interchangeably.
}
For $q^3\underline{P}_\varphi$ and $q^3\underline{P}_{\dot\varphi}$ peaked at $k_*$, one can check that $\underline{\Delta}^2_\delta(k_*)=k_*^3\underline{P}_\delta(k_*)/2\pi^2\sim 1$ and $\underline{\Delta}_\delta^2(k)\sim (k/k_*)^3$ for $k\ll k_*$. The $(k/k_\ast)^3$ dependence can be seen by noting that the integral over $\bq$ becomes a constant for $k\ll k_*$ (independent of $k$), and hence $k^3\underline{P}_\delta(k)\propto k^3$. As mentioned earlier, this $k^3$ power law does not rely in detail on the power laws used in the field spectrum. On the other hand, for $k \gg k_\ast$, $\underline{\Delta}_\delta^2(k)$ decreases in a way that does depend on the shape of the field power spectra. Some sample spectra are shown in the right panel of Fig.~\ref{fig:sample_field_spectrum}. Note the differences between non-relativistic ($k_*\ll a_i m$) and relativistic ($k_*\gg a_i m$) fields.\footnote{Here, the isocurvature density spectrum is derived assuming no phase correlation between field modes. While there could be phase correlations in realistic scenarios, they generally lead to an enhanced density spectrum and stronger constraints. We now provide a rough argument for this claim: note that the energy density is $\rho_{\bm k} \sim m^2 \int_{\bm q} \varphi_{\bm q}^\ast \varphi_{{\bm k} - {\bm q}}$. If there is no phase correlation, then ensemble averages should not depend on the relative phase between $\varphi_{\bm k}$ and $\varphi_{{\bm k} - {\bm q}}$, so for arbitrary phase $\theta$ we have $\expval{\varphi_{\bm q}^\ast \varphi_{{\bm k} - {\bm q}}} = e^{i\theta} \expval{\varphi_{\bm q}^\ast \varphi_{{\bm k} - {\bm q}}}$, which means $\expval{\varphi_{\bm q}^\ast \varphi_{{\bm k} - {\bm q}}} = 0$ and $\expval{\abs{\rho_{\bm k}}^2}$ is minimized.}

Combining our above discussions, we expect $\underline{\Delta}_\delta^2(k)$ to be approximately scale invariant for very small $k$ (due to the adiabatic initial conditions), for somewhat larger $k$ it should connect with the $k^3$ spectrum (resulting from small length scale isocurvature perturbations). The scale where they connect can be obtained by solving $(k/k_*)^3\sim (1+w)^2\Delta^2_{\mathcal{R}}(k)$, which yields $k\equiv k_{\rm dev}\sim (1+w)^{2/3}10^{-3}k_*$ (we used $\Delta_\mathcal{R}^2(k)\sim 10^{-9}$~\cite{Planck:2018jri}). In summary, we have
\begin{equation}
\label{eq:approxDeltain}
\underline{\Delta}_\delta^2(k)\sim 
\begin{cases}
(1+w)^2\Delta^2_\mathcal{R}(k)& \textrm{when}\qquad k<k_{\rm dev}\,,\\
(k/k_*)^3& \textrm{when}\qquad k_{\rm dev}<k<k_*\,.
\end{cases}
\end{equation}
These arguments provide a reasonable rough shape for $\underline{\Delta}_\delta^2(k)$ only if the $k_{\rm dev}$ above is superhorizon at $t_i$, since the adiabatic initial conditions we used for the low-$k$ tail were derived from the superhorizon limit. A realistic spectrum will have deviations from this simple shape that depend on the details of the actual production mechanism (especially if the productions takes longer than $H_i^{-1}$, or if the scale $k_*$ is not too far from the Horizon), the impact of the gravitational dynamics around the horizon scale, etc.

\subsubsection{Inhomogeneities of field variances}
\label{sec:variance_inhomogeneities}
In this section, we argue that $\varphi$ and $\dot{\varphi}$ must be statistically inhomogeneous, with spatially changing variances given by Eq.~\eqref{eq:generic_variance_inhomogeneity} and \eqref{eq:variance_inhomogeneity_solution}.

Suppose the gravitational potential $\Psi$ is given.
The adiabatic initial conditions in Eq.~\eqref{eq:adiabatic_ic} then tell us what the large-scale density fluctuations should be.
However, the adiabatic initial conditions make no claim on what the small-scale density fluctuations are, and the value of $\delta(\bx)$ cannot be determined by Eq.~\eqref{eq:adiabatic_ic} alone.
It is useful to consider an ensemble of fields, such that the field has spectrum given by $\underline{\Delta}_\varphi^2$, and also respects the adiabatic initial conditions.\footnote{We stress that large-scale quantities such as $\Psi$ and $\delta$ are fixed and are not treated as random variables here, and ensemble averaging is only over $\varphi$ and $\dot{\varphi}$, not over $\Psi$. If we also take $\Psi$ as random, then the expectation here should be understood as conditional expectations with $\Psi$ fixed. }
Given Eq.~\eqref{eq:adiabatic_ic}, this ensemble must satisfy
\begin{align}
\label{eq:adEns}
  \expval{\delta(\bx)} = -\frac32 (1 + w) \Psi(\bx),\quad
  \expval{\dot{\delta}(\bx)} = 3(1+w)\dot{\Psi}(\bx) + \frac{\dot{w}}{1+w} \expval{\delta(\bx)} \;.
\end{align}
To see why this requires the ensemble averages of field variances to be statistically inhomogeneous, note that taking the ensemble average of $ \rho(\bx)\equiv\overline{\rho}(1+\delta(\bx))$  (using the expression for $\rho$  in Eq.~\eqref{eq:varphi_stress_energy}) gives
\begin{align}
  \label{eq:rho_perturbation_expectations}
  \expval{\delta(\bx)} &= \frac{1}{2\overline{\rho}} \left[ \left( \expval{\dot{\varphi}^2(\bx)} - \overline{\dot{\varphi}^2} \right) + \frac{\expval{(\nabla \varphi(\bx))^2} - \overline{(\nabla\varphi)^2}}{a^2}  + m^2 \left( \expval{\varphi^2(\bx)} - \overline{\varphi^2} \right) \right] \nonumber \\
                     &\quad + \frac{1}{\overline{\rho}} \Psi(\bx) \left[- \overline{\dot{\varphi}^2} + \frac{\overline{(\nabla \varphi)^2}}{a^2} \right] + \order{\Psi^2}
\end{align}
where we have assumed $\expval{\dot{\varphi}^2(\bx)} / \overline{\dot{\varphi}^2} - 1$ is at most $\order{\Psi}$, etc.
One can check that the above expression cannot be consistent with Eq.~\eqref{eq:adEns} if $\expval{\varphi^2(\bx)} = \overline{\varphi^2}$, $\expval{\dot{\varphi}^2(\bx)} = \overline{\dot{\varphi}^2}$ and $\expval{(\nabla\varphi(\bx))^2} = \overline{(\nabla\varphi)^2}$ for all $\bx$; for instance, in the ultra-relativistic limit where $\overline{\dot{\varphi}^2} = \overline{(\nabla \varphi)^2} / a^2$, we would have $\expval{\delta(\bx)} = 0 \neq -\frac32 (1 + w) \Psi(\bx)$.
This argument tells us that the expectation values $\expval{\varphi^2(\bx)}$ and $\expval{\dot{\varphi}^2(\bx)}$ must vary with $\bx$, that is, $\varphi$ and $\dot{\varphi}$ must be statistically inhomogeneous. 

We now give a systematic derivation of the statistical variance inhomogeneities, such that the large-scale density and pressure fluctuations are consistent with Eq.~\eqref{eq:adEns}.
We parameterize the spatial perturbation of the field variances by
\begin{align}
  \label{eq:generic_variance_inhomogeneity}
  \expval{\varphi^2(\bx)} = \overline{\varphi^2} ( 1 + f_1(\bx) ),\quad
  \expval{\dot{\varphi}^2(\bx)} = \overline{\dot{\varphi}^2} ( 1 + f_2(\bx) ) \;,
\end{align}
where $f_i(\bx)=\order{\Psi}$. The idea is to first express $\langle\dot{\delta}\rangle$ and $\langle{\delta}\rangle$ appearing in \eqref{eq:adEns} in terms of $f_1$ and $f_2$, and solve for $f_1$ and $f_2$ from the two equations \eqref{eq:adEns}. This is done by replacing the field variances $\langle \varphi^2(\bx)\rangle$ and $\expval{\dot{\varphi}^2(\bx)}$ appearing in the definitions \eqref{eq:varphi_stress_energy} of $\expval{\rho}$, $\expval{p}$ with $f_1$ and $f_2$, and by using the expectation value of the continuity equation \eqref{eq:fluid_equations}. 

We work through these straighforward (but somewhat tedious) steps below. Using Eq.~\eqref{eq:rho_perturbation_expectations}, we find that density perturbations are given by
\begin{align}
  \expval{\delta} &= \frac{\overline{\dot{\varphi}^2}}{2\overline{\rho}} f_2 + \left(1 - \frac{\overline{\dot{\varphi}^2}}{2\overline{\rho}} - \frac{m^2 \overline{\varphi^2}}{2\overline{\rho}} \right) f_1  +  \frac{m^2 \overline{\varphi^2}}{2\overline{\rho}} f_1
                    + \Psi \left[ 2\left(1 - \frac{\overline{\dot{\varphi}^2}}{2\overline{\rho}} - \frac{m^2 \overline{\varphi^2}}{2\overline{\rho}} \right) - \frac{\overline{\dot{\varphi}^2}}{\overline{\rho}} \right] + \order{\Psi^2} 
\end{align}
where we have used $\overline{(\nabla\varphi)^2} / (a^2 \overline{\rho}) = 2 (1 - \overline{\dot{\varphi}^2}/(2\overline{\rho}) - m^2 \overline{\varphi^2} / (2\overline{\rho}) )$.
Similarly, pressure perturbations (using Eq.~\eqref{eq:varphi_stress_energy}) are given by
\begin{align}
  \label{eq:p_perturbation_expectations}
  \frac{\expval{p} - \overline{p}}{\overline{\rho}} &= \frac{\overline{\dot{\varphi}^2}}{2\overline{\rho}} f_2 - \frac13 \left(1 - \frac{\overline{\dot{\varphi}^2}}{2\overline{\rho}} - \frac{m^2 \overline{\varphi^2}}{2\overline{\rho}} \right) f_1  -  \frac{m^2 \overline{\varphi^2}}{2\overline{\rho}} f_1
                    - \Psi \left[ \frac{\overline{\dot{\varphi}^2}}{\overline{\rho}} + \frac23 \left(1 - \frac{\overline{\dot{\varphi}^2}}{2\overline{\rho}} - \frac{m^2 \overline{\varphi^2}}{2\overline{\rho}} \right) \right]\nonumber\\
                    &\quad + \order{\Psi^2} \;.
\end{align}
The above equations are valid for general values of $\overline{\dot{\varphi}^2} / (2\overline{\rho})$ and $m^2 \overline{\varphi^2} / (2\overline{\rho})$.
In our case of interest, the fact that $P_{\dot{\varphi}}(q) = \omega_k^2 P_\varphi(q)$ and $\overline{p} = w \overline{\rho}$ implies $\overline{\dot{\varphi}^2} = {\overline{(\nabla\varphi)^2}}/{a^2} + m^2 \overline{\varphi^2}, 
  {\overline{\dot{\varphi}^2}} /{(2\overline{\rho})} = {1}/{2}$ and $
  {m^2 \overline{\varphi^2}}/{(2\overline{\rho})} = (1-3w)/{2}$. The perturbations then simplify to
\begin{align}
  \label{eq:rho_p_perturbations_in_w}
  \expval{\delta} = \frac{f_1 + f_2}{2} - (1-3w) \Psi,\quad
  \frac{\expval{p} - \overline{p}}{\overline{\rho}} = \frac{f_1(2w-1) + f_2 }{2} - (1+w) \Psi \;.
\end{align}

The $\expval{\delta}$ equation in Eq.~\eqref{eq:adEns} can be directly matched with the $\expval{\delta}$ in Eq.~\eqref{eq:rho_p_perturbations_in_w}. A bit more work is needed for the left hand side of the $\expval{\dot{\delta}}$ equation, which can be obtained from 
the continuity equation~\eqref{eq:fluid_equations}:
\begin{align}
  \expval{\dot{\delta}} 
                        &= 3\dot{\Psi} (1+w) - 3 H \frac{\expval{p} - \overline{p}}{\overline{\rho}} + 3 H w \expval{\delta} - \frac{1}{\overline{\rho}} \expval{\frac{\nabla}{a} \cdot {\bm Q}} + \order{\Psi^2} \;,
\end{align}
where we have used $\dot{\overline{\rho}} = - 3 H (\overline{\rho} + \overline{p})$.
Combining with Eq.~\eqref{eq:rho_p_perturbations_in_w}, the adiabatic initial conditions \eqref{eq:adEns} can now be used to solve for $f_1$ and $f_2$:
\begin{align}
  \label{eq:variance_inhomogeneity_solution}
  f_1 &= \frac{3w^2-8w-3}{2(1-w)} \Psi - \frac{\dot{w}}{2 H (1-w)} \Psi + \frac{1}{3 H (1-w)} \frac{1}{\overline{\rho}} \expval{\frac{\nabla}{a} \cdot {\bm Q}}\,, \nonumber \\
  f_2 &= \frac{15w^2-8w+1}{2(1-w)} \Psi + \frac{\dot{w}}{2 H (1-w)} \Psi - \frac{1}{3 H (1-w)} \frac{1}{\overline{\rho}} \expval{\frac{\nabla}{a} \cdot {\bm Q}} \;.
\end{align}
We have thus derived the general form of the inhomogeneity of variances $\expval{\varphi^2}$ and $\expval{\dot{\varphi}^2}$.

We will assume that the $\expval{\nabla \cdot {\bm Q}}$ term in Eq.~\eqref{eq:variance_inhomogeneity_solution} is zero for the rest of this work.
This term is divergence of momentum; upon ensemble averaging, the term signifies the large-scale spreading or converging of energy.
In the superhorizon case that we consider, $\expval{\nabla \cdot {\bm Q}} = - e^{-2\Psi} \expval{\nabla \cdot (\dot{\varphi} \nabla\varphi)} / a$ should be negligible compared to others, since it involves two spatial gradients and is $k^2$ suppressed.
Statistically, this term reflects correlations between the random fields $\varphi$ and $\dot{\varphi}$.
If $\varphi$ and $\dot{\varphi}$ are statistically independent, then $\expval{\bm Q} = - e^{-2\Psi} \expval{\dot{\varphi}} \expval{\nabla\varphi} / a = 0$, so $\expval{\nabla \cdot {\bm Q}} \neq 0$ necessitates statistical correlations between $\varphi$ and $\dot{\varphi}$.
In our case with $\expval{\nabla \cdot {\bm Q}}$ negligible, it suffices to treat $\varphi$ and $\dot{\varphi}$ as independent random variables initially.

Finally, we evaluate \eqref{eq:variance_inhomogeneity_solution} for some specific cases, including the one we use in our work.
Using the $\nu\alpha$ parametrization of the initial field spectrum (with $\alpha>2$) (see \eqref{eq:P_varphi_candidates} and Appendix \ref{sec:EOS}), we can obtain analytic expressions for $w$ and $\dot{w}$: 
\begin{align}
  w = \frac{1}{3}\frac{(k_*/a m)^2} {(k_*/a m)^2+(1-2\alpha^{-1})(1+{2}{\nu}^{-1})},\quad
  \dot{w} = \frac{-2}{3} \frac{(k_*/a m)^2 (1-2\alpha^{-1})(1+{2}{\nu}^{-1}) H} {\left((k_*/a m)^2+(1-2\alpha^{-1})(1+{2}{\nu}^{-1})\right)^2} \;.
\end{align}
Note that $w$ changes with time due to redshifting, and $w$ is only roughly constant in the ultra-relativisitic or non-relativistic limit.
Eq.~\eqref{eq:variance_inhomogeneity_solution} now yields
\begin{align}
\label{eq:genf1f2nualpha}
  f_1 &= \frac{-\left(\alpha  \left(\left(4 (k_\ast/am)^2+3\right) \nu +6\right)-6 (\nu +2)\right)^2}{2 \left(\alpha \left((k_\ast/am)^2 \nu +\nu +2\right)-2 (\nu +2)\right) \left(\alpha  \left(\left(2 (k_\ast/am)^2+3\right) \nu +6\right)-6 (\nu +2)\right)} \Psi \nonumber \\
  f_2 &= \frac{- (\alpha -2) (\nu +2) \left(\alpha  \left(\left(4 (k_\ast/am)^2-3\right) \nu -6\right)+6 (\nu +2)\right)}{2  \left(\alpha  \left((k_\ast/am)^2 \nu +\nu +2\right)-2 (\nu +2)\right) \left(\alpha  \left(\left(2 (k_\ast/am)^2+3\right) \nu +6\right)-6 (\nu +2)\right)} \Psi \;.
\end{align}
We find that $f_1 \to -4\Psi$, $f_2 \to 0$ in the ultra-relativistic limit ($k_\ast / am \to \infty$), and $f_1 \to -(3/2)\Psi$, $f_2 \to (1/2)\Psi$ in the non-relativistic limit ($k_\ast / am \to 0$).
Between the two limits, we typically have $|f_2| < |f_1|$.
In the simple case where the field power spectrum with $\nu=3\,,\alpha=\infty$, we have:
\begin{align}
  \label{eq:f1f2_our_case}
  f_1 = \frac{- 3 \left(4 (k_\ast/am)^2+5\right)^2}{12 (k_\ast/am)^4+50 (k_\ast/am)^2+50} \Psi,\quad
  f_2 = \frac{25-20 (k_\ast/am)^2}{12 (k_\ast/am)^4+50 (k_\ast/am)^2+50} \Psi \;.
\end{align}
For the parameters we use in our simulation, we have $k_*/am=1$, $w=1/8$, $f_1(\bx)=-2.17\Psi(\bx)$ and $f_2(\bx)=0.04\Psi(\bx)$.

\subsection{Generating field realizations}
\label{sec:generating_fields}
In this subsection, we present a concrete procedure for generating field initial conditions $(\varphi(\bx), \dot{\varphi}(\bx))$ that enjoy the statistical properties described in previous sections.
In Sec.~\ref{sec:generating_fields_density}, we define the procedure and discuss the statistical features of the generated field.
In Sec.~\ref{sec:numerical_implementation}, we provide a numerical implementation of the procedure.

\subsubsection{Generating field with given density perturbations}
\label{sec:generating_fields_density}

We first present the procedure for generating an inhomogeneous Gaussian random field $\varphi$.
Suppose we are given power spectrum $\underline{P}_\varphi(q)$ with characteristic wavenumber $k_\ast$, and a function $f(\bx)$ such that $\overline{f}=0$, $\abs{f} \ll 1$, and $\abs{\nabla f} / f \ll k_\ast$.
We want to generate a Gaussian random field $\varphi$ with spectrum $\underline{P}_\varphi(q)$, and that $\expval{\varphi^2(\bx)} \approx \overline{\varphi^2} e^{f(\bx)}$.

To achieve this, we introduce a new Gaussian field $\hat{\varphi}$ and define $\varphi$ in Fourier space via:
\begin{align}
  \label{eq:inhomogeneous_grf_definition}
  \varphi_{\bq} = \sqrt{\underline{P}_\varphi(q)} {\hat{\varphi}}_{\bq}\,, \qq{where}
  \expval{\hat{\varphi}(\bx) \hat{\varphi}(\bx')} = e^{f(\bx)} \delta_D(\bx-\bx') \;.
\end{align}
Here, each $\hat{\varphi}(\bx)$ is an independent Gaussian random variable with variance $e^{f(\bx)}$; in other words, $\hat{\varphi}$ is a white noise field with spatially varying variance. We get $\hat{\varphi}_\bq$ by taking a Fourier transform of $\hat{\varphi}(\bx)$, and we set $\varphi_\bk=\sqrt{\underline{P}_\varphi(q)}\hat{\varphi}_\bq$. We claim that the $\varphi_\bq$ defined above (and its inverse Fourier transform $\varphi(\bx)$) have the desired statistical properties.

To understand this procedure better, we introduce an equivalent definition of $\varphi$ in terms of a spatial convolution:
\begin{align}
  \label{eq:inhomogeneous_grf_convolution_definition}
  \varphi(\bx) = \int_{\bm y} K(\bx-{\bm y}) \hat{\varphi}({\bm y}) \qq{where}
  K(\bx) = K(\abs{\bx}) = \int_{\bm q} e^{i {\bm q} \cdot \bx} \sqrt{\underline{P}_\varphi(q)} \;.
\end{align}
For typical choices of $\underline{P}_\varphi(q)$, the convolution kernel $K(\bx)$ is peaked at $\bx=0$ and drops rapidly for $\abs{\bx} > k_\ast^{-1}$.\footnote{
For example, if $\underline{P}_\varphi(k) = \mathcal{A} \Theta(k_\ast - k)$, then the kernel is given by:
\begin{align}
    K(r) = \frac{\sqrt{\mathcal{A}}}{2 \pi^2} \frac{\sin(k_\ast r) - k_\ast r \cos(k_\ast r)}{r^3} \;.
\end{align}
This function is concentrated at $r=0$ and rapidly decreasing for $r > k_\ast^{-1}$, as expected.
Formally, this computation is similar to evaluating the RMS overdensity within a spherical top hat, except that now the role of $\bk$ and $\bx$ are interchanged, and $\underline{P}_\varphi(k)$ replaces the role of the spatial spherical top hat. For more general expressions related to different field power spectra, see Appendix \ref{sec:kernel}.
}
This means the integrand $K(\bx-{\bm y}) \hat{\varphi}({\bm y})$ is mainly supported around $\bx$, and the value of $\varphi(\bx)$ will inherit the variance of the underlying white noise field $\hat{\varphi}(\bx)$.
Moreover, the spectral power in the kernel $K(\bx)$ is given by the spectrum $\sqrt{\underline{P}_\varphi(q)}$, so convolving the kernel with any function $\hat{\varphi}$ will pick out the corresponding Fourier modes in $\hat{\varphi}$.
This explains why the generated field $\varphi$ should have spectrum $\underline{P}_\varphi(q)$.

We now confirm that $\varphi$ has the desired properties by direct computation.
The spatial 2-point function is given by
\begin{align}
  \expval{\varphi(\bx)\varphi({\bm y})} 
  &= \int_{\bm z} K(\bx-{\bm z}) K({\bm z}-{\bm y}) e^{f({\bm z})}\,, \nonumber \\
    &\approx
    \begin{cases}
      e^{f(\bx)} \int_{\bm z} K(\bx-{\bm z})^2 & \abs{\bx - {\bm y}} \ll k_\ast^{-1} \\
      0 & \abs{\bx - {\bm y}} \gg k_\ast^{-1}
    \end{cases}
      \;.
\end{align}
In the $\abs{\bx - {\bm y}} \ll k_\ast^{-1}$ case, the integrand is supported only around ${\bm z} \approx \bx$, where $e^{f({\bm z})} \approx e^{f(\bx)}$ since $\abs{\nabla f} / f \ll k_\ast$.
In particular, taking $\bx = {\bm y}$ tells us that $\expval{\varphi^2(\bx)} \propto e^{f(\bx)}$.
In the $\abs{\bx - {\bm y}} \gg k_\ast^{-1}$ case, we recover the fact that the 2-point function should fall beyond the correlation length $k_\ast^{-1}$.
Also, the Fourier space 2-point function is
\beq
\label{eq:phiqphiq'withf}
    \expval{\varphi_\bq^\ast \varphi_{\bq'}} &= \underline{P}_\varphi(q) (2\pi)^3 \delta_D(\bq-\bq') + \sqrt{\underline{P}_\varphi(q)\underline{P}_\varphi(q')} \left(f_{\bq'-\bq} + \order{f^2}\right)\,, \\
    \expval{\abs{\varphi_\bq}^2} &= \underline{P}_\varphi(q)\mathcal{V} + \order{f^2} \;.
\eeq
We have thus checked that $\varphi$ has the required spectrum and variance inhomogeneities.\footnote{The connection between the first and second line for $\bq=\bq'$ is best understood by thinking of the field being confined to a large but finite volume ($L^3$) with periodic boundary conditions. Using $\varphi_\bq\rightarrow\sqrt{L^3}\varphi_\bq^L$ (and other changes mentioned in footnote 1), we have $\expval{ \varphi_\bq^L (\varphi_{\bq'}^L)^\ast } = \underline{P}_\varphi(q)\delta_{\bq,\bq'}+\sqrt{\underline{P}_\varphi(q)\underline{P}_\varphi(q')}f_{\bq-\bq'}^L /\sqrt{L^3}$, with $\expval{ |\varphi_\bq^L|^2 } =\underline{P}_\varphi(q). $
We used the fact that $f$ has a zero average, so that $f_{\bq-\bq=0}^L=0$. }

The above generic procedure can now be applied to generate a $(\varphi,\dot{\varphi})$ pair that satisfies the adiabatic initial conditions.
One can first generate $\varphi$ with spectrum $\underline{P}_\varphi$ and variance inhomogeneity $f_1$, and then generate $\dot{\varphi}$ with spectrum $\underline{P}_{\dot{\varphi}} = \omega_q^2 \underline{P}_{\varphi}$ and variance inhomogeneity $f_2$, where $f_1$ and $f_2$ are derived in Sec.~\ref{sec:variance_inhomogeneities}.
The generated fields will lead to the large-scale perturbations $\expval{\delta}$ and $\expval{\dot\delta}$, as required by adiabatic initial conditions.
As discussed in Sec.~\ref{sec:variance_inhomogeneities}, that $\varphi$ and $\dot{\varphi}$ are generated independently is not an issue.

We now show that the $(\varphi,\dot{\varphi})$ pair generated by the above procedure recovers all features of density spectrum $\underline{P}_\delta(k)$ discussed in Sec.~\ref{sec:density_spectra}.
More specifically, we show that $\underline{P}_\delta(k)$ contains (large-scale) adiabatic fluctuations and (small-scale) isocurvature perturbations, in direct parallel to the discussion in Sec.~\ref{sec:density_spectra}.
To begin, note from Eq.~\eqref{eq:inhomogeneous_grf_convolution_definition} that $\varphi(\bx)$ is an integral of Gaussian random variables $\hat{\varphi}(\bx)$, so $\varphi(\bx)$ guaranteed to be Gaussian.
This Gaussianity allows us to use Isserlis' theorem to evaluate 4-point functions of $\varphi$ and $\dot{\varphi}$:
$ \expval{\varphi^2(\bx) \varphi^2({\bm y})} = \expval{\varphi^2(\bx)} \expval{\varphi^2({\bm y})} + 2 \expval{\varphi(\bx)\varphi({\bm y})}^2,
  \expval{\dot{\varphi}^2(\bx) \dot{\varphi}^2({\bm y})} = \expval{\dot{\varphi}^2(\bx)} \expval{\dot{\varphi}^2({\bm y})} + 2 \expval{\dot{\varphi}(\bx)\dot{\varphi}({\bm y})}^2,
  \expval{\dot{\varphi}^2(\bx) \varphi^2({\bm y})} = \expval{\dot{\varphi}^2(\bx)} \expval{\varphi^2({\bm y})}.$
  All terms in the density 2-point function $\expval{\rho(\bx)\rho(\bm y)}$ can be expanded into 4-point functions of $\varphi$ and $\dot{\varphi}$, which can be evaluated in this manner.
We will distinguish terms of the form $\expval{\varphi^2(\bx)}\expval{\varphi^2(\bm y)}$ as ``long'' length scale terms, and terms of the form $\expval{\varphi(\bx)\varphi(\bm y)}^2$ as ``short'' length scale terms.
The ``long'' length scale terms are responsible for perturbations from adiabatic initial condition, and the ``short'' length scale terms are responsible for isocurvature perturbations.\footnote{Also see a related discussion in \cite{Liu:2024pjg} on the different meanings of time averaged quantities $\expval{ \abs{\vb{E}}^2(\bx) \abs{\vb{E}}^2(\bx') }$ and $\expval{ \abs{\vb{E}}^2(\bx)} \expval{ \abs{\vb{E}}^2(\bx') }$, where $\vb{E}$ is an electric field.}

We now explicitly evaluate the 2-point correlation $\expval{\rho(\bx) \rho(\bm y)}$:
\begin{align}
  \expval{\rho(\bx) \rho({\bm y})} &= \xi_{\rho\rho}^{\mathrm{(long)}}(\bx, {\bm y}) + \xi_{\rho\rho}^{\mathrm{(short)}}(\bx, {\bm y}) \qq{where} \nonumber \\
  \xi_{\rho\rho}^{\mathrm{(long)}}(\bx, {\bm y}) &= \expval{\rho(\bx)} \expval{\rho({\bm y})} = \overline{\rho}^2 \left(1 + \expval{\delta(\bx)}\right) \left(1 + \expval{\delta({\bm y})}\right) \nonumber \\
  \xi_{\rho\rho}^{\mathrm{(short)}}(\bx, {\bm y}) &= e^{-2\Psi(\bx)-2\Psi({\bm y})} \frac12 \expval{\dot{\varphi}(\bx)\dot{\varphi}({\bm y})}^2 \nonumber \\
                    &\quad + e^{2\Psi(\bx)+2\Psi({\bm y})} \frac{1}{2 a^4} \expval{\partial_i\varphi(\bx) \partial_j\varphi({\bm y})} \expval{\partial_i\varphi(\bx) \partial_j\varphi({\bm y})} \nonumber \\
                    &\quad + e^{2\Psi(\bx)} \frac{1}{2 a^2} m^2 \expval{\partial_i\varphi(\bx) \varphi({\bm y})} \expval{\partial_i\varphi(\bx) \varphi({\bm y})} \nonumber \\
                    &\quad + e^{2\Psi({\bm y})} \frac{1}{2 a^2} m^2 \expval{\varphi(\bx) \partial_i\varphi({\bm y})} \expval{\varphi(\bx) \partial_i\varphi({\bm y})} \nonumber \\
                    &\quad + \frac12 m^4 \expval{\varphi(\bx)\varphi({\bm y})}^2 \;.
\end{align}
The ``long'' length scale term $\xi_{\rho\rho}^{\mathrm{(long)}}$ is simply the product of $\expval{\rho}$ at two spatial locations.
Since $\expval{\delta}$ is fixed by the adiabatic initial condition \eqref{eq:adEns}, $\xi_{\rho\rho}^{\mathrm{(long)}}/ \overline{\rho}^2$ is determined without reference to the field spectra $\underline{P}_\varphi$ and $\underline{P}_{\dot{\varphi}}$.
In other words, we can vary our choice of the field spectra while keeping $\expval{\delta}$ fixed, and $\xi_{\rho\rho}^{\mathrm{(long)}} / \overline{\rho}^2$ would remain unchanged.
The $\xi_{\rho\rho}^{\mathrm{(short)}}$ contains information beyond such large-scale fluctuations, including the small-scale isocurvature perturbations.

To find the density spectrum, we Fourier transform the correlation functions and expand up to first order in $\Psi$:
\begin{align}
\label{eq:d_longshort}
  \expval{\rho_\bk^\ast \rho_{\bk'}} &= \xi_{\rho\rho}^{\mathrm{(long)}}(\bk, \bk') + \xi_{\rho\rho}^{\mathrm{(short)}}(\bk, \bk') \qq{where} \nonumber \\
  \xi_{\rho\rho}^{\mathrm{(long)}}(\bk, \bk') &= \overline{\rho}^2 \expval{\delta_\bk^\ast} \expval{\delta_{\bk'}} \nonumber \\
  \xi_{\rho\rho}^{\mathrm{(short)}}(\bk, \bk') &= \frac12 \int_{{\bm q}} \Big[ (2\pi)^3 \delta_D(\bk-\bk') \underline{P}_{\dot\varphi}(q) \underline{P}_{\dot{\varphi}}(\abs{{\bm q} + \bk'}) \nonumber \\
                                     &\qquad -2 \Psi_{\bk'-\bk} \underline{P}_{\dot\varphi}(q) \underline{P}_{\dot{\varphi}}(\abs{{\bm q} + \bk'})
                                       -2 \Psi_{\bk'-\bk} \underline{P}_{\dot\varphi}(q) \underline{P}_{\dot{\varphi}}(\abs{{\bm q} + \bk}) \nonumber \\
                                     &\qquad + 2 (f_2)_{\bk'-\bk} \sqrt{\underline{P}_{\dot\varphi}(\abs{{\bm q}+\bk}) \underline{P}_{\dot\varphi}(\abs{{\bm q}+\bk'})}  \underline{P}_{\dot{\varphi}}(q)
                                       \Big] + \textrm{($\underline{P}_\varphi$ terms)}
\end{align}
where $\underline{P}_\varphi$ terms include $f_1$ in them entering via $\expval{{\varphi}(\bx){\varphi}(\by)}$, similar to how the terms written above include $f_2$ from $\expval{\dot{\varphi}(\bx)\dot{\varphi}(\by)}$.  See App.~\ref{sec:full_expr_for_2_point_function} for full details. Moreover, for $\bk = \bk'$, the terms involving $\Psi$ and $f_i$ are zero, since $\Psi_{\bk-\bk'=0}=\overline{\Psi}=0$ and $\overline{f}_i\propto \overline{\Psi}=0$ by choice.
In summary:
\begin{align}
  \label{eq:d_longshort_fourier}
  \underline{P}_\delta(k) &= (2\pi)^3 \int_{\bk'} \frac{\expval{|\delta_{\bk'}|^2}}{\mathcal{V}} \frac{\delta_D(\abs{\bk'}-k)}{4 \pi k^2} 
  = (2\pi)^3 \int_{\bk'} \frac{\expval{|\rho_{\bk'}|^2}}{\mathcal{V} \overline{\rho}^2} \frac{\delta_D(\abs{\bk'}-k)}{4 \pi k^2} \nonumber \\
                          &= \frac{1}{2\overline{\rho}^2} \int_{{\bm q}} \left[ \underline{P}_{\dot\varphi}(q) \underline{P}_{\dot{\varphi}}(\abs{{\bm q} + \bk}) + \left(m^2 + \frac{{\bm q} \cdot (\bk + {\bm q})}{a^2}\right)^2 \underline{P}_{\varphi}(q) \underline{P}_{\varphi}(\abs{{\bm q} + \bk}) \right] \left(1 + \order{\Psi^2}\right) \nonumber \\
                          &\quad + (2\pi)^3 \int_{\bk'} \frac{|\expval{\delta_{\bk'}}|^2}{\mathcal{V}} \frac{\delta_D(\abs{\bk'}-k)}{4 \pi k^2} \;.
\end{align}
Note that the third line above is in fact $P_{\expval{\delta}}(k)$, and that $P_{\expval{\delta}}(k) \neq \underline{P}_{\delta}(k)$.
The density spectrum is now clearly separated into ``long'' (third line) and ``short'' (second line) contributions.
The ``long'' contribution $P_{\expval{\delta}}(k)$ is solely determined by adiabatic initial conditions, as previously discussed.
The ``short'' contribution is solely determined by the field spectra $\underline{P}_{\varphi}$ and $\underline{P}_{\dot\varphi}$, as is manifest from the above integral expression.
Note that the ``short'' contribution is the same as Eq.~\eqref{eq:predicted_isocurvature_density_perturbation}, which was derived for homogeneous Gaussian random fields.
We see that this procedure of adding inhomogeneity in variances leads to additional power $P_{\expval{\delta}}(k)$ in the density spectrum, and nothing more.

\subsubsection{Numerical implementation}
\label{sec:numerical_implementation}

Suppose our 3D lattice has $N^3$ grid points and volume $L^3$.
To generate an inhomogeneous Gaussian random field $\varphi$ on the lattice, we essentially discretize the formulas in \eqref{eq:inhomogeneous_grf_definition}.
More specifically, we do the following:
\begin{itemize}
\item[0.] Before we can apply the procedure, we must specify $f(\bx)$.
  The perturbation $f(\bx)$ can be any function of choice as long as $\overline{f} = 0$.
  To obtain the adiabatic initial conditions described in Sec.~\ref{sec:variance_inhomogeneities}, we first generate the gravitational potential $\Psi$ as a homogeneous Gaussian random field, and use $f(\bx) \propto \Psi(\bx)$.\footnote{We omit the well-known procedures for generating homogeneous Gaussian random fields. See, for example, Garrett Goon's tutorial at \url{https://garrettgoon.com/gaussian-fields/}.}
\item[1.] At each lattice point $\bx$, generate a value for $\hat{\varphi}(\bx)$ from Gaussian distribution $\hat{\varphi}(\bx) \sim \mathcal{N}(0, e^{f(\bx)})$.  The random variables $\hat{\varphi}(\bx)$ and $\hat{\varphi}(\bm{y})$ are independent as long as $\bx \neq \bm{y}$.
\item[2.] Take the discrete Fourier transform on $\hat{\varphi}(\bx)$ to get $\hat{\varphi}_\bq$. The discrete Fourier transform is given by $ \hat{\varphi}_{\bq_{a,b,c}} = {N^{-3/2}} \sum_{r,s,t=0}^{N-1} e^{-2\pi i (a,b,c)\cdot(r,s,t) / N} \hat{\varphi}(\bx_{r,s,t})$, 
  where $\bx_{r,s,t}$ and $\bq_{a,b,c}$ are lattice sites in position space and momentum space.\footnote{In the continuous ($N\to\infty$) limit, we can recover $\varphi_\bq$ via $\varphi_\bq = \sqrt{(\Delta{x})^3} \varphi_{\bq_{a,b,c}}$, where $\bq = \frac{2\pi}{L}(a,b,c)$.}
\item[3.] Compute $\varphi_\bq = \sqrt{\underline{P}_\varphi(q)} \hat{\varphi}_\bq$, where $\underline{P}_\varphi(q)$ is the spectrum that we want for $\varphi$.
  Note that $q = \frac{2\pi}{L} \sqrt{a^2 + b^2 + c^2}$, where $(a,b,c)$ is the index of $\bq$ on the reciprocal lattice.
\item[4.] Take the inverse discrete Fourier transform on $\varphi_\bq$ to get $\varphi(\bx)$: \\  $\varphi(\bx_{r,s,t}) = {N^{-3/2}} \sum_{a,b,c=0}^{N-1} e^{2\pi i (a,b,c)\cdot(r,s,t) / N} \varphi_{\bq_{a,b,c}} $
\end{itemize}
The procedure for $\dot{\varphi}$ is similar.

\section{Evolution}
\label{sec:evol}
In Sec.~\ref{sec:analytics}, we first discuss the main features we expect to see in our simulations based on analytic understanding of density spectrum evolution. We then discuss the results of our numerical simulations in Sec.~\ref{sec:simres} and compare them with the analytic expectations.

\subsection{Analytical results}
\label{sec:analytics}
Our discussion in this section  summarizes and builds upon the results in Ref.~\cite{Amin:2022nlh}. A discussion of the evolution of isocurvature density perturbations as well as free-streaming for free fields can also be found in \cite{Liu:2024pjg}. 

Before diving into the technical details, for a heuristic understanding of the main results: (i) a lack of evolution of the isocurvature white-noise spectrum, and (ii) the free-streaming suppression of the adiabatic spectrum, it might be useful to think in terms of the quasi-particle picture discussed in the introduction. The random velocities of the quasi-particles leads to their re-arrangement; they cannot change the statistics of the initial Poisson distribution of their locations. Hence the white noise density spectrum does not evolve. However, existing correlations (adiabatic density perturbations) can be erased by these random motions of the quasi-particles. This leads to the free-streaming suppression of the adiabatic density spectrum.

\subsubsection{Evolution of isocurvature}
\label{sec:evolution_of_isocurvature}
In this section, we discuss how the isocurvature density fluctuations evolves over time. As a matter of semantics, when we refer to isocurvature on subhorizon scales, we simply mean perturbations that are not necessarily correlated with the curvature perturbations. In Sec.~\ref{sec:density_spectra} we argued that the initial isocurvature spectrum is given by \eqref{eq:predicted_isocurvature_density_perturbation}. We now show that, after the field has become nonrelativistic, the spectrum converges to \eqref{eq:late_time_delta_spectrum}. Moreover, the late time isocurvature spectrum can be calculated using the initial isocurvature spectrum formula \eqref{eq:predicted_isocurvature_density_perturbation}, but with $k_\ast/am \ll 1$.

We consider a simple scenario, in which field evolution is given by the Klein Gordon equation in an FRW spacetime (see Appendix Sec.~\ref{sec:WKB} for inclusion of long-wavelength metric perturbations). The mode equation and the general solution are given by
\begin{align}
\label{eq:varphi_decoupled_solution}
  & \ddot{\varphi}_\bq+ 3 H \dot{\varphi}_\bq + \omega_q^2 \varphi_\bq = 0,\quad
  \varphi_\bq(t) = f_q(t) \varphi_\bq(t_i) + g_q(t) \dot{\varphi}_\bq(t_i) \;,
\end{align}
where $f_q(t)$ and $g_q(t)$ are defined by
\begin{align}
  & \ddot{f}_q + 3 H \dot{f}_q + \omega_q^2 f_q = 0,\quad f_q(t_i) = 1,\quad \dot{f}_q(t_i) = 0 \nonumber \\
  & \ddot{g}_q + 3 H \dot{g}_q + \omega_q^2 g_q = 0,\quad g_q(t_i) = 0,\quad \dot{g}_q(t_i) = 1 \;.
\end{align}
We are interested in the isocurvature spectrum when $H(t) \ll m$ and $k_\ast \ll a(t) m$. In general, for $\omega_q\gg H$, the above equations have WKB solutions: 
\beq
f_q(t)\approx D_q(t)\cos\int_{t_i}^t \omega_q, \quad g_q(t)\approx \frac{D_q(t)}{\omega_q(t_i)}\sin\int_{t_i}^t \omega_q,\quad\textrm{where}\quad D_q(t)=\sqrt{a_i^3\omega_q(t_i)/a^3\omega_q(t)}.
\eeq
We have assumed $\dot{D}_q\sim HD_q\ll \omega_q D_q$, and with that we also have $\dot{f}_q\approx-[\omega_q\omega_q(t_i)]g_q$ and $\dot{g}_q\approx[\omega_q/\omega_q(t_i)]f_q$ which will be useful in the following.

As in Sec.~\ref{sec:density_spectra}, we take the initial conditions of $\varphi$ from an ensemble of homogeneous Gaussian random field, specified by $\expval{\varphi_\bq^\ast(t_i) \varphi_{\bq'}(t_i)} = (2\pi)^3 \delta_D(\bq - \bq') \underline{P}_\varphi(t_i, q)$, $\expval{\dot{\varphi}_\bq^\ast(t_i) \dot{\varphi}_{\bq'}(t_i)} = (2\pi)^3 \delta_D(\bq - \bq') \underline{P}_{\dot{\varphi}}(t_i, q)$ and  $\expval{\dot{\varphi}_{\bq}(t_i) \varphi_{\bq'}(t_i)} = 0$.
These assumptions are the same as that used to derive Eq.~\eqref{eq:predicted_isocurvature_density_perturbation}. Then, at later times $t$,  
\beq
&\expval{\varphi_\bq^\ast(t) \varphi_{\bq'}(t)} = (2\pi)^3 \delta_D(\bq - \bq')\left[ \underline{P}_\varphi(t_i, q) f_q^2(t)+ \underline{P}_{\dot{\varphi}}(t_i, q)g_q^2(t)\right]
\\&\expval{\dot{\varphi}_{\bq}(t)\dot{\varphi}_{\bq'}(t)} = (2\pi)^3 \delta_D(\bq - \bq') \left[ \underline{P}_\varphi(t_i, q) \dot{f}_q^2(t)+ \underline{P}_{\dot{\varphi}}(t_i, q)\dot{g}_q^2(t)\right],\\
 &\expval{\dot{\varphi}_\bq^\ast(t) \varphi_{\bq'}(t)} = (2\pi)^3 \delta_D(\bq - \bq')   \big[    \underline{P}_\varphi(t_i, q)\dot{f}_q(t) f_q(t) + \underline{P}_{\dot{\varphi}}(t_i, q) g_q(t) \dot{g}_q(t) \big],
 \eeq
where the square brackets in the first, second and third line are $\underline{P}_\varphi(t,q), \underline{P}_{\dot{\varphi}}(t,q)$ and $\underline{P}_{\varphi\dot{\varphi}}(t,q)$ respectively. Note that Klein-Gordon time evolution of an initially statistically homogeneous Gaussian random field yields another statistically homogeneous Gaussian random field at a later time, fully specified by the 2-point correlation functions.\footnote{We sketch a proof of this point. If $\varphi(t_i,\bx)$ and $\dot{\varphi}(t_i,\bx)$ are Gaussian random fields, then $\varphi(t,\bx)$ and $\dot{\varphi}(t,\bx)$ are also Gaussian random fields; this is because $\varphi(t,\bx)$ and $\dot{\varphi}(t,\bx)$ are integrals of the initial conditions in terms of Green's function, and an integral over Gaussian random variables is Gaussian.
Moreover, if $\varphi$ and $\dot{\varphi}$ are initially statistically homogeneous, then they are also statistically homogeneous at $t$, since the Klein Gordon system has translation symmetry.
Statistical homogeneity and Gaussianity implies that the ensemble of $\varphi(t,\bx)$ and $\dot{\varphi}(t,\bx)$ are determined by their 2-point correlation functions. }

We can then compute the (ensemble averaged) density spectrum by plugging the solution Eq.~\eqref{eq:varphi_decoupled_solution} into Eq.~\eqref{eq:varphi_stress_energy} with $\Psi=0$.
The density spectrum at time $t$ is
\begin{align}
  \label{eq:delta_k_evolution}
  \underline{P}_\delta(t,k) &= \frac{1}{2 \overline{\rho}(t)^{2}} \int_{\bm q} \Big[ \Big( \dot{g}_q \dot{g}_{\abs{\bk - {\bm q}}} + (m^2 + {\bm q} \cdot ({\bm q} - \bk) / a^2) g_q g_{\abs{\bk - {\bm q}}} \Big)^2 \underline{P}_{\dot{\varphi}}(t_i, \abs{\bk - {\bm q}}) \underline{P}_{\dot{\varphi}}(t_i, q) \nonumber \\
                         &\quad + \Big( \dot{f}_q \dot{f}_{\abs{\bk - {\bm q}}} + (m^2 + {\bm q} \cdot ({\bm q} - \bk) / a^2) f_q f_{\abs{\bk - {\bm q}}} \Big)^2 \underline{P}_{\varphi}(t_i, \abs{\bk - {\bm q}}) \underline{P}_{\varphi}(t_i, q) \nonumber \\
                         &\quad + \Big( \dot{f}_q \dot{g}_{\abs{\bk - {\bm q}}} + (m^2 + {\bm q} \cdot ({\bm q} - \bk) / a^2) f_q g_{\abs{\bk - {\bm q}}} \Big)^2 \underline{P}_{\dot{\varphi}}(t_i, \abs{\bk - {\bm q}}) \underline{P}_{\varphi}(t_i, q) \nonumber \\
                         &\quad + \Big( \dot{g}_q \dot{f}_{\abs{\bk - {\bm q}}} + (m^2 + {\bm q} \cdot ({\bm q} - \bk) / a^2) g_q f_{\abs{\bk - {\bm q}}} \Big)^2 \underline{P}_{\varphi}(t_i, \abs{\bk - {\bm q}}) \underline{P}_{\dot{\varphi}}(t_i, q)
  \Big] ,
\end{align}
where $\overline{\rho}(t)$ is given by
\begin{align}
  \overline{\rho}(t) = \frac12 \int_{\bm q} \Big[ (\dot{g}_q^2 + \omega_q^2 g_q^2) \underline{P}_{\dot{\varphi}}(t_i, q) + (\dot{f}_q^2 + \omega_q^2 f_q^2) \underline{P}_{\varphi}(t_i, q)  \Big] \;.
\end{align}
Using the WKB solutions, and assuming $\underline{P}_{\dot{\varphi}}(t_i, q) = \omega_k^2(t_i) \underline{P}_{{\varphi}}(t_i, q)$, the two-point correlation functions simplify to 
$\expval{\varphi_\bq^\ast(t) \varphi_{\bq'}(t)} \approx  (2\pi)^3 \delta_D(\bq - \bq')D_q^2(t){P}_\varphi(t_i, q)={\omega_q^{-2}(t)}\expval{\dot{\varphi}_{\bq}(t)\dot{\varphi}_{\bq'}(t)},$ and $\expval{\dot{\varphi}_\bq^\ast(t) \varphi_{\bq'}(t)} \approx 0$. That is, $\underline{P}_\varphi(t,q)\approx D_q^2(t)\underline{P}_\varphi(t_i,q)$ and the properties of the field ensemble at later times is the same as at $t_i$. We will return to this shortly. Moving on to the density power spectrum,  a significant simplification arises to yield:
\beq 
 \underline{P}_\delta(t,k)
 = \frac{\expval{|\delta_\bk(t)|^2}}{\mathcal{V}}
 \approx \frac{1}{2\overline{\rho}^2(t)}\int_\bq&\left[\omega_q^2(t)\omega_{|\bq-\bk|}^2(t)+\left(m^2 + \frac{{\bm q} \cdot ({\bm q} - \bk) }{a^2}\right)^2\right] \\
&\times D^2_q(t) D^2_{|\bq-\bk|}(t)\underline{P}_\varphi(t_i,|\bk-\bq|)\underline{P}_\varphi(t_i,q)\,,
\eeq
and $\overline{\rho}(t)\approx \int_\bq D_q^2(t)\omega_q^2(t_i)\underline{P}_{\varphi}(t_i, q) $. 

In the late time limit $t\gg t_i$ when all modes contributing to the integral are non-relativistic, we have $\omega_q(t)\approx \omega_{|\bq-\bk|}(t)\approx m$ and $D^2_{|\bq-\bk|}(t)/D^2_q(t)=\omega_{|\bq-\bk|}(t_i)/\omega_q(t_i)$. Further more, $D_q(t)\rightarrow (a_i/a)^{3/2}(\omega_q(t_i)/m)^{1/2}$. Then, the late time density power spectrum
\beq
\label{eq:late_time_delta_spectrum}
\underline{P}_\delta(t,k)= \frac{\expval{ |\delta_\bk(t)|^2}}{\mathcal{V}} \approx \frac{m^4}{\left[\int_\bq \omega_q^3(t_i)\underline{P}_{\varphi}(t_i, q)\right]^2}\int_\bq \omega_{|\bq-\bk|}(t_i){\omega_q(t_i)}\underline{P}_\varphi(t_i,|\bk-\bq|)\underline{P}_\varphi(t_i,q)\,,
\eeq
which is time-independent, and depends only on the field power spectrum at $t_i$. Note that no assumptions were made about whether the field configuration is relativistic or non-relativistic initially. 

If the initial field configuration is non-relativistic to begin with, then all the $\omega_q(t_i)\approx m$, and we immediately have $\underline{P}_\delta(t,k)\approx \underline{P}_\delta(t_i,k)$. That is, the power spectrum of density perturbations does not evolve at late times. Whether we initialize the field at $t_i$, evolve and then calculate the density spectrum at $t$ or initialize the fields at $t$ with $\underline{P}_\varphi(t,q)\approx D_q^2(t)\underline{P}(t_i,q)$ and use it in \eqref{eq:predicted_isocurvature_density_perturbation}, the fractional density spectrum is the same.\footnote{Furthermore, for such nonrelativistic fields, it is particularly straightforward to evaluate the isocurvature spectrum in the limit $k\ll k_\ast$ limit using \eqref{eq:P_varphi_candidates} -- the power spectrum becomes $\underline{P}_\delta(t,k)$ becomes independent of $k$:
\beq
\underline{\Delta}_\delta^2(t,k\ll k_*)=\frac{k^3}{2\pi^2}\underline{P}_\delta(t,k\ll k_\ast)\approx \left(\frac{k}{k_{\rm wn}}\right)^{\!3}
\textrm{with}\quad k_{\rm wn}\equiv k_\ast\left[\left(1+\frac{3}{2 \alpha}\right) \left(1-\frac{3}{2\nu }\right) \left(\frac{1}{\alpha}+\frac{1}{\nu}\right)\right]^{1/3}. 
\eeq
Note that for $\nu>3/2$, and not too extreme $\alpha,\nu>0$, $k_{\rm wn}\sim k_\ast$.}

Finally, we discuss the evolution of $\underline{\Delta}_\delta^2(k)$ for some concrete examples of field spectra.
We refer to the right panel of Fig.~\ref{fig:sample_field_spectrum}.
While this plot is for the initial $\underline{\Delta}_\delta^2(k)$ spectrum, we know from the above discussion that it also reveals the evolution of the spectrum.
Suppose the field has field spectra parameterized by $\nu=3$ and $\alpha=\infty$.
Initially, when the field is ultra-relativistic ($k_\ast/am \gg 1$), its isocurvature spectrum would resemble the dashed orange curve (for $k > k_{\mathrm{dev}}$).
After the field becomes nonrelativistic ($k_\ast/am \ll 1$), its isocurvature spectrum would become the solid orange curve (for $k > k_{\mathrm{dev}}$).
We can see in Fig.~\ref{fig:sample_field_spectrum} that there tend to be a slight growth at the peak around $k_\ast$ when the field transitions from relativistic to nonrelativistic.
The case is similar for other choices of field spectra, with the $\alpha = 3$ exhibiting the most significant change in $\underline{\Delta}_\delta^2(k)$.

\subsubsection{Free streaming physics}
\label{sec:FreeStreamingPhysics}
 The prototypical scenario that we consider is as follows. At $t=t_i$, the field power spectra $\Delta^2_\varphi$ and $\Delta^2_{\dot\varphi}$ are peaked around comoving wavenumber $k_*$, which is deep inside the horizon. The field configuration is dominated by spatio-temporal variations on characteristic length scales $1/k_*$ and time scales $1/\omega_{k_*}$ where  $\omega_{k_*}=\sqrt{k_*^2/a^2+m^2}$.  One can think of this system as a collection of particles with different comoving momenta $q$ and energies $\omega_{q}=\sqrt{q^2/a^2+m^2}$, moving at physical speeds $v_q=d\omega_q/(adq)=(q/a)/\omega_q$ in random directions. Given the peaked spectra, the energy density is dominated by particles with momenta $k_*$. The  comoving distance that a particle with momentum $q$ moves is \begin{align}
\frac{1}{q_{\rm fs}(t)}&=\int_{t_i}^t \frac{dt'}{a(t')}\frac{q/a(t')}{\sqrt{q^2/a^2(t')+m^2}}
=\int_{a_i}^a \frac{d\ln b}{b H(b)}\frac{q/b}{\sqrt{q^2/b^2+m^2}}\,, \nonumber \\
&=\frac{q}{a_i^2 H_i m} \left[ \ln\left\{(a m/q)+\sqrt{(a m/q)^2+1}\right\} - \ln\{a\rightarrow a_i\}\right]\,,
\end{align}
where the final equality assumes radiation dominated expansion history.\footnote{
An approximate, yet revealing way to write the free-streaming length is as follows :
\beq
q^{-1}_{\rm fs}(a)
    \approx \frac{\sqrt{2}}{k_{\rm eq}}\frac{a_{q,\rm nr}}{a_{\rm eq}}\left[\left(1-\frac{a_i}{a_{q,\rm nr}}\right)+\ln\left(\frac{a}{a_{q,\rm nr}}\right)\right].
\eeq
where $a_{q,\rm nr}\equiv q/m$ is the scalefactor at which the particle with momentum $q/a$ and mass $m$ becomes non-relativistic. The above expression is valid for $a_{q,\rm nr}<a\lesssim a_{\rm eq}$.
The first contribution is from the time interval when the particle is relativistic, the logarithm is after the particle has become non-relativistic.} 
This is the free streaming length for a particle with momentum $q$. Since the energy density configuration is dominated by particles of momentum $k_*$, the characteristic distance for a large fraction of the particles will be
\begin{equation}
\frac{1}{k_{\rm fs}^*(t)}=\frac{1}{q_{\rm fs}(t)}\Bigg|_{q=k_*} \;.
\end{equation}
This ``free-streaming" tends to wipe out existing adiabatic density perturbations (but not the small-scale white-noise part, or the peak on small-scales). 

The impact of this free-streaming of particles on the density power spectrum during radiation domination can be captured via the free-streaming transfer function $T_{\rm fs}^2(t,k)$ such that density power spectrum takes the form \cite{Amin:2022nlh}
\begin{equation}
\label{eq:Tfs2}
\Delta^2_\delta(t,k)\approx \Delta^{2({\rm long})}_\delta(t_i,k)T_{\rm ad}^2(t,k)T_{\rm fs}^2(t,k)+\Delta^{2{(\rm short)}}_\delta(t,k)\,,
\end{equation}
where 
\begin{equation}
T_{\rm fs}^2(t,k)\approx \left[\frac{\m^2}{\overline{\rho}(t)}\int d\ln q \Delta^2_\varphi(t,q)\frac{\sin[k/q_{\rm fs}(t)]}{k/q_{\rm fs}(t)}\right]^2\,,
\end{equation}
which is expected to be valid after the fields are non-relativistic ($k_*\ll a m$). For a simple understanding of this transfer function as a Fourier transform of the field power spectrum with respect to the displacement of the particles, see Appendix \ref{sec:smoothing}.\footnote{We thank Sten Delos for the discussion which lead to this appendix. Also see the appendix of \cite{Ganjoo:2023fgg}.}

During radiation domination, $T_{\rm ad}^2(t,k)$ contains information about density perturbations getting a boost when entering the horizon \cite{Hu:1995en}, and locking into a logarithmic growth (up to the Jeans scale). The free streaming aspects are captured by $T_{\rm fs}^2(t,k)$. Note that we  expect $\Delta^{2(\rm short)}_\delta(t,k)\approx \Delta^{2(\rm short)}_\delta(t_i,k) $ (see \cite{Amin:2022nlh}, and Sec.~\ref{sec:evolution_of_isocurvature}) during radiation domination, i.e., it does not evolve significantly beyond some initial transients as the field becomes non-relativistic. 

We note that for $\nu=3,\alpha=\infty$, and for non-relativistic fields at $t_i$, we can evaluate the transfer function analytically.\footnote{The transfer function in the case $\nu=3,\alpha=\infty$ is given by \beq T_{\rm fs}(t,k)=3(k_{\rm fs}^*/k)^3\left[\sin(k/k_{\rm fs}^*)-(k/k_{\rm fs}^*)\cos(k/k_{\rm fs}^*)\right].\eeq} An excellent approximation to the free-streaming transfer function (for $k\lesssim k^*_{\rm fs}$) is provided by:
\begin{equation}
\label{eq:Tfs-kfs-approx}
T_{\rm fs}^2(t,k)\approx e^{-k^2/3k_{\rm fs}^2(t)}\,,\qquad \textrm{where}\qquad\frac{1}{k_{\rm fs}^2(t)}\equiv \frac{m^2}{\overline{\rho}(t)}\int_0^{a(t)m} d\ln q \Delta^2_\varphi(t,q)\frac{1}{q^2_{\rm fs}(t)}\,.
\end{equation}
\begin{wrapfigure}{r}{0.5\textwidth}
\includegraphics[height=0.5\textwidth]{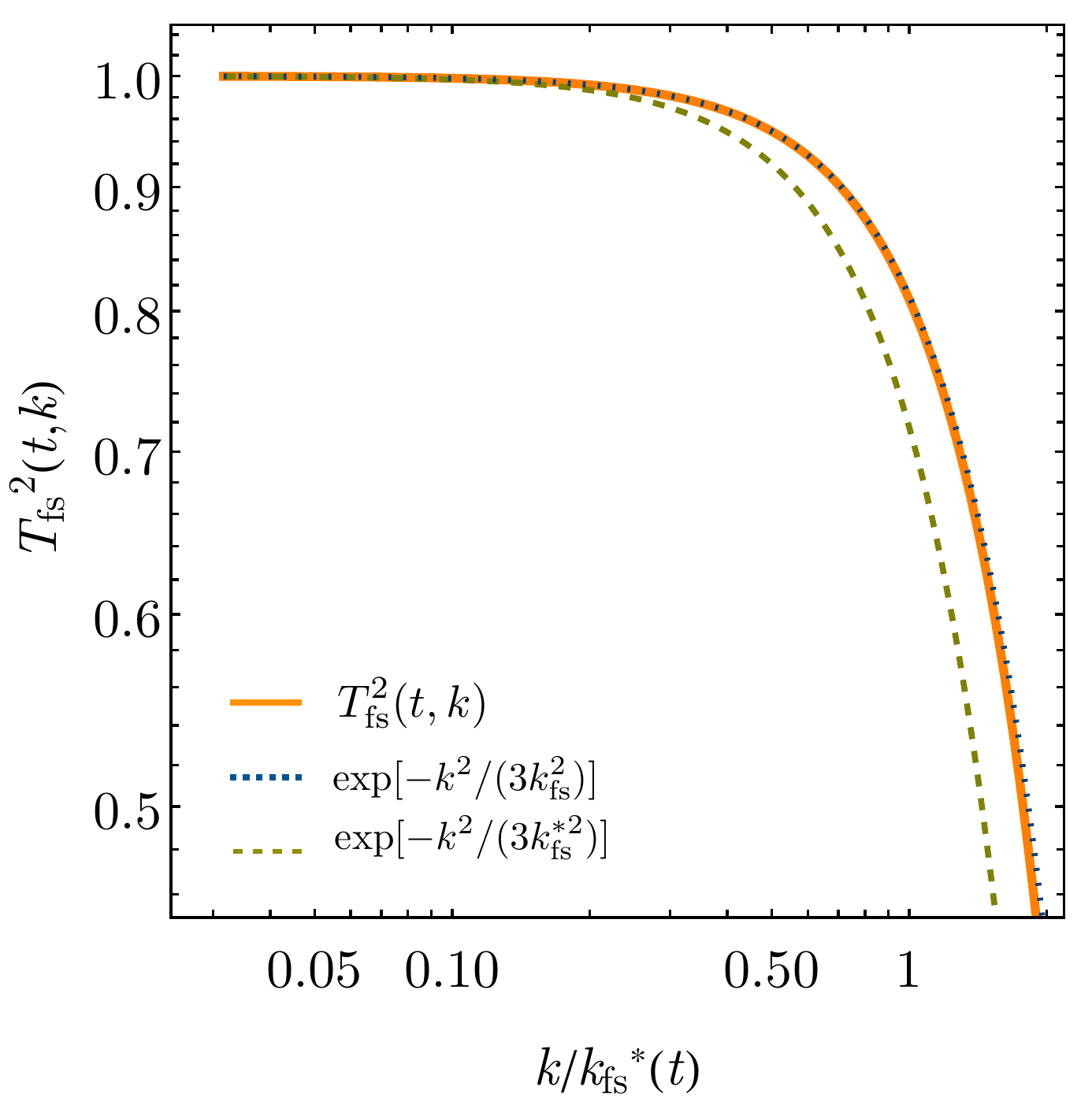}
\caption{Free-Streaming transfer function. The dotted blue curve is almost indistinguishable from the complete transfer function \eqref{eq:Tfs2} (solid orange curve).}
\label{fig:fs_transfer_function}
\end{wrapfigure}
The motivation for this form can be understood by Taylor expanding $T^2_{\rm fs}(t,k)$ around $k=0$, and defining $k_{\rm fs}$  via the coefficient of the $k^2$ term.\footnote{Expanding $T_{\rm fs}^2(t,k)$ around $k=0$, we have \begin{align}
\left[\frac{m^2}{\overline{\rho}(t)}\int d\ln q \Delta^2_\varphi(t,q)\frac{\sin [k/q_{\rm fs}(t)]}{[k/q_{\rm fs}(t)]}\right]^2
=\left[1-\frac{k^2}{3}\frac{m^2}{\overline{\rho}(t)}\int d\ln q \Delta^2_\varphi(t,q)\frac{1}{q_{\rm fs}^2(t)}+\hdots\right]
\equiv\left[1-\frac{k^2}{3k_{\rm fs}^2(t)}+\hdots\right].
\end{align}.} We caution that while this works well for the case of interest, and for $\alpha>2$, it fails when $\alpha<2$ as pointed out by \cite{Liu:2024pjg}.  In those cases it is best to use the full transfer function. A crude, but easy to use, and general approximation that works for most cases is to replace $k_{\rm fs}(t)$ by $k_{\rm fs}^*(t)$. That is, $T_{\rm fs}^2(t,k)\sim \exp\left[-k^2/3\{k^*_{\rm fs}(t)\}^2\right]$. See dashed curve in Fig.~\ref{fig:fs_transfer_function}.

In the coming sections we will verify the expected behavior encoded in the adiabatic transfer function, the free streaming transfer function, and the lack of evolution of the $k^3$ part of $\Delta^{2(\rm short)}_\delta(t,k)$ from numerical simulations.

Before moving on to the comparison, however, we wish to discuss a couple of technical points relevant for comparison with numerical simulations which will include finite box size, and resolution related effects. For the field spectrum, we plot $\left[\Delta_\varphi^2(t,q) + \Delta_{\dot\varphi}^2(t,q)/ \omega_q^2(t)\right]/2$ instead of $\Delta_\varphi^2(t,q)$.
This approach of plotting has the advantage of eliminating oscillations of the spectrum over time (particularly relevant at small wavenumbers because of the smaller number of modes available to average over in a simulation). To see this, note that:
\begin{align}
  \label{eq:field_spectrum_definition}
  &\frac{\mathcal{V}}{(2\pi)^3}\frac{2\pi^2}{q^3}\left[\Delta_\varphi^2(q) + \frac{\Delta_{\dot\varphi}^2(q)}{ \omega_q^2}\right]
  \!=  \int_{\bq'} \left( \abs{\varphi_{\bq'}}^2 + \frac{\abs{\dot{\varphi}_{\bq'}}^2}{\omega_{q'}^2} \right) \frac{\delta_D(\abs{\bq'} - q)}{4\pi q^2}
   \!=\int_{\bq'} \abs{\varphi_{\bq',+}}^2  \frac{\delta(\abs{\bq'} - q)}{4\pi q^2},\nonumber\\
  &\qq{where} \varphi_{\bq,+} \equiv \varphi_{\bq} + \frac{\dot{\varphi}_{\bq}}{i \omega_q}.
\end{align}
where we suppressed the dependence on $t$ to reduce clutter. If $\omega_q$ is slowly varying, we have $\abs{\varphi_{\bq}}^2 \sim a^{-3} \cos^2(\omega_q t+\theta)$ and $\abs{\varphi_{\bq,+}}^2 \sim a^{-3}$, which does not oscillate.

As a mild generalization, we use $m^2\Delta_\varphi^2(t,q)\rightarrow \left[\Delta^2_{\dot{\varphi}}(t,q)+\omega_q^2\Delta^2_\varphi(t,q)\right]/2$ in \eqref{eq:Tfs-kfs-approx} for calculating the effective free streaming length from simulations.

\subsection{Simulation results}
\label{sec:simres}
In the three upcoming sections, we show the results for three simulations: one for a free scalar field without gravity, one for a free scalar field in the presence of gravitational perturbations, and one for a scalar field with strong self-interactions.  The most important features of these simulations are captured by the evolution of their density spectra in Fig.~\ref{fig:delta_spectrum_without_gravity}, Fig.~\ref{fig:delta_spectrum_with_gravity} and Fig.~\ref{fig:delta_spectrum_soliton}. We also provide video versions of these figures \href{https://www.youtube.com/playlist?list=PLecJrnvnk5c7Iaqi-Wq7xvqk1Msgxn5pk}{here}.

\subsubsection{Without gravitational perturbations in the evolution}
\label{sec:W/GravPertEvol}
In this subsection, we demonstrate free streaming suppression of the density power spectrum in the absence of gravitational perturbations ($\Psi=0$) in the evolution equations for the field. We show that the large-scale inhomogeneities in the density fluctuations (resulting from spatially dependent variances in the field), gets suppressed as the free-streaming scale moves to larger and larger distance (smaller wavenumbers), revealing more and more of the white noise tail.
\begin{figure}[t]
  \centering
  \includegraphics[width=\textwidth]{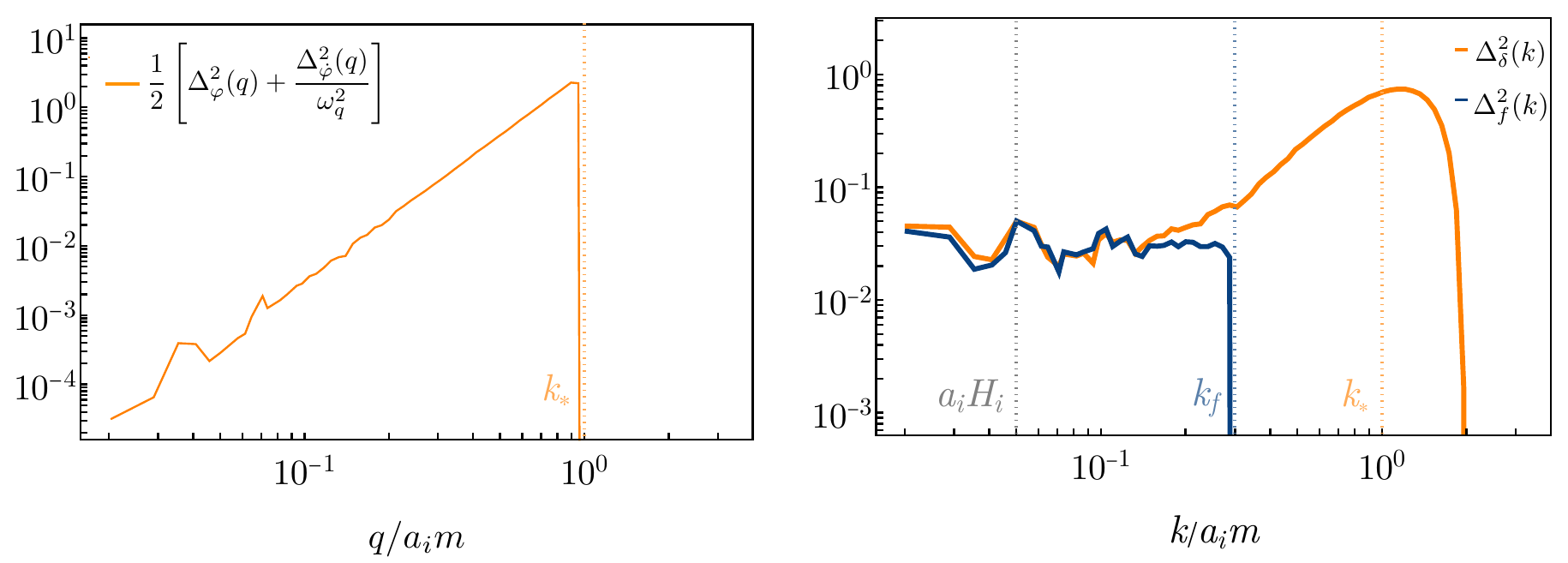}
  \caption{Initial power spectrum for the field $\varphi$ is shown on the left (in units of $m^2$), the density perturbation $\delta$, and the field variance inhomogeneity $f$ (a proxy for gravitational potential) for a given realization on the right. The field spectrum (left panel) has a $q^3$ power law below $k_\ast = a_i m$, and has a sharp cutoff at $k_\ast$. The density spectrum $\Delta_\delta^2$ (orange curve of right panel) peaks at around $k_\ast$, and is cutoff at around $2k_\ast$. Below $k_\ast$, the density spectrum exhibits a $k^3$ power law before connecting to a $k^0$ plateau below $k_f$. The field variance inhomogeneity $\Delta_f^2$ (blue curve of right panel) is approximately scale-invariant ($k^0$ power law) and has a sharp cutoff at $k_f = 0.3 a_i m$. Note that the low-$k$ plateau of $\Delta_\delta^2$ tracks the variance inhomogeneity $\Delta_f^2$, as explained in Sec.~\ref{sec:generating_fields_density}. See Sec.~\ref{sec:W/GravPertEvol} for details of $\Delta_f^2$.}
  \label{fig:spectrum_initial_1}
\end{figure}

\begin{figure}[t]
  \centering
  \includegraphics[width=0.95\textwidth]{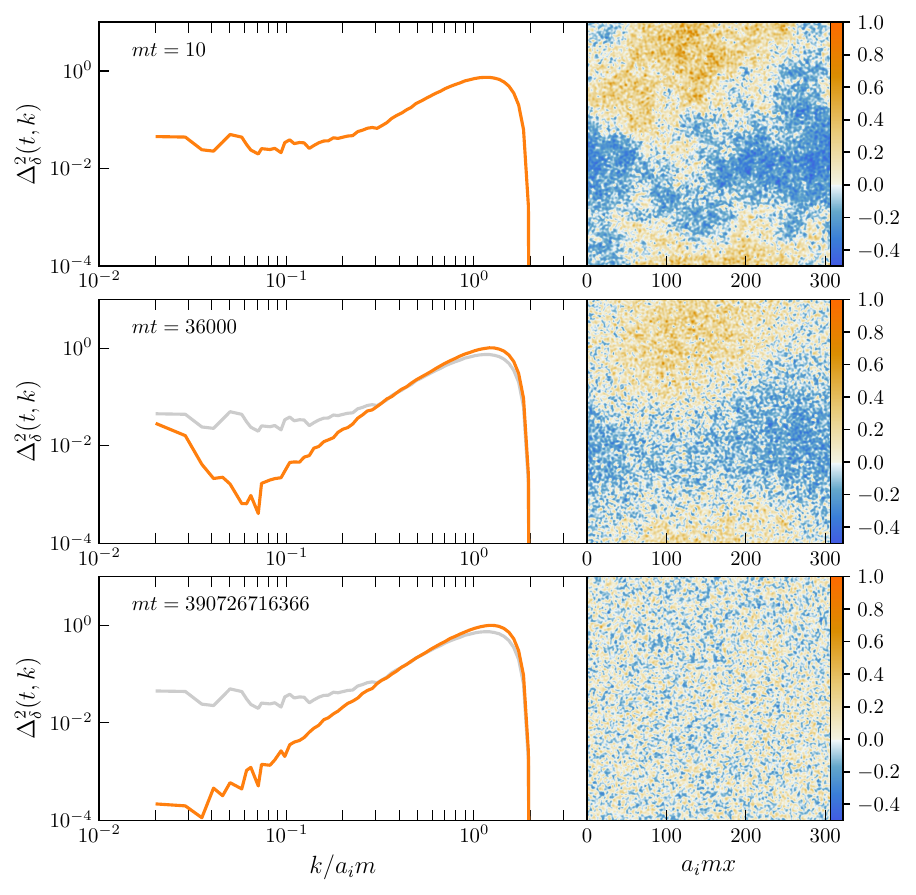}
  \caption{The left panels are the overdensity spectra $\Delta^2_\delta(k)$ over time, and the right panels are the corresponding snapshots of $\delta$.  The panels are ordered vertically by time.   The snapshots on the right panel are the $\delta$'s averaged over one axis of the lattice. One can see from the left panels that the initial $k^0$ plateau at low-$k$ (scale-invariant) of the spectra are suppressed over time as the free-streaming length increases, and only a $k^3$ white noise remains in the end.  The free-streaming scale, $k_{\mathrm{fs}}$ (not shown here, but see Fig.~\ref{fig:kfs_fit_1}) would indicate where suppression is just starting to take place, not the location of the mimimum of the orange curve. Note that only the first and second rows were from outputs of our numerical integration; the third row was produced using the WKB solution. In this entire simulation, gravitational perturbations are ignored in the evolution equations, but qualitatively included in the initial conditions. The spectra were produced using the binning scheme described in App.~\ref{sec:numerical_details}. Also note that the colorbar scaling for the snapshots on the right has different scaling for positive and negative values, chosen because of the skewness of the $\delta$ distribution.   }
  \label{fig:delta_spectrum_without_gravity}
\end{figure}

\begin{figure}[t]
  \centering
    \includegraphics[height=0.47\textwidth]{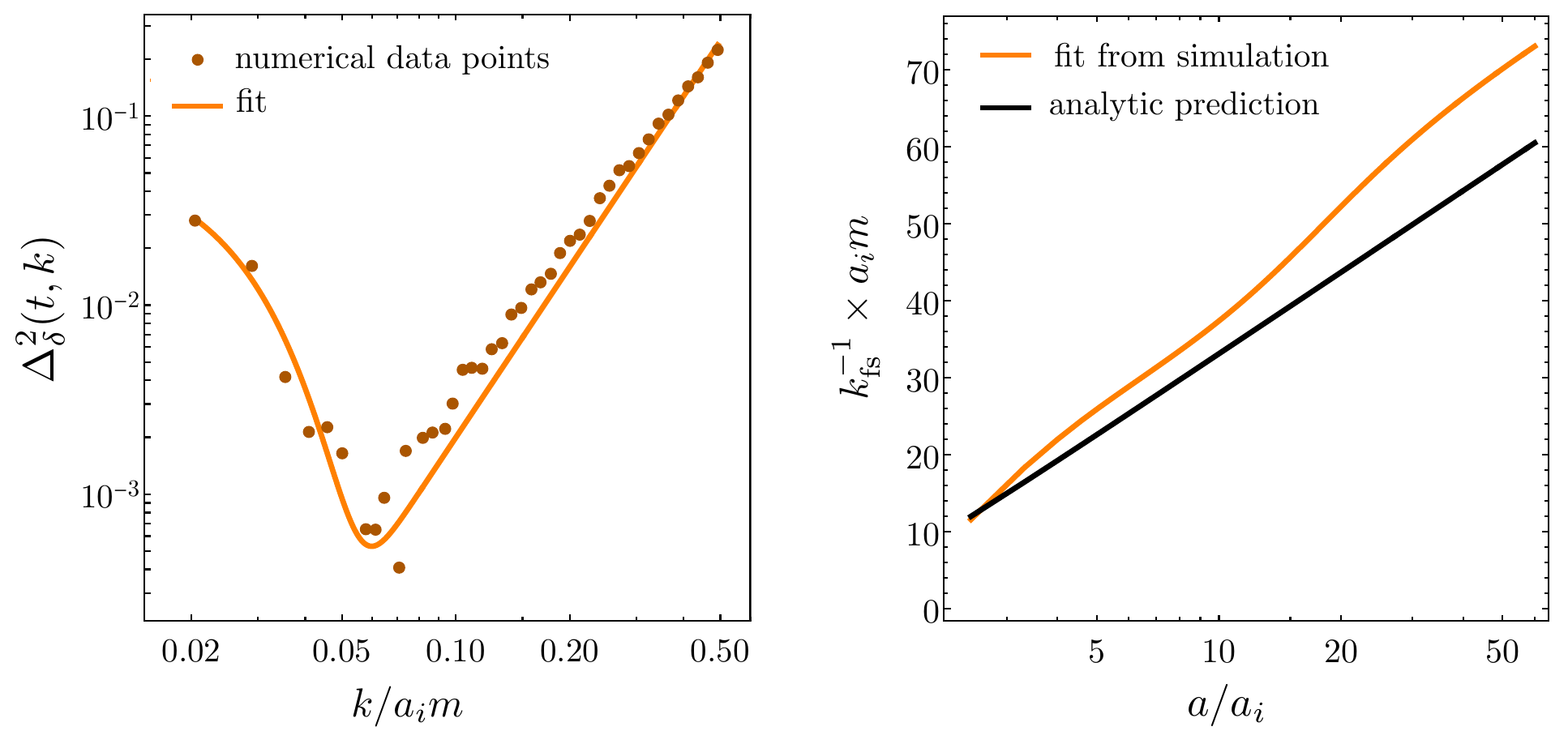}
  \caption{The left panel shows the fit of the density spectrum at $a/a_i=60$ using the ansatz \eqref{eq:FitAnsatz}. The right panel shows the free streaming length $k_\mathrm{fs}^{-1}$ versus scale factor $a$, obtained by fitting the density spectra from the simulation (orange curve) and by the analytic prediction \eqref{eq:Tfs-kfs-approx}. One can see that the free streaming length $k_\mathrm{fs}^{-1}$ is growing roughly linearly with respect to $\ln(a)$, as expected. Moreover, the orange and black curves are close to each other, with the fitted free streaming length slightly larger (less than $20\%$) than the predicted one.} 
  \label{fig:kfs_fit_1}
\end{figure}

To achieve this, we initialize $\varphi$ as a Gaussian random field, with significant large length-scale density perturbations (a {\it non}-white noise part). The field $\varphi$ is evolved numerically via a free Klein Gordon equation in a radiation dominated background.\footnote{We acknowledge that a Klein Gordon field in a radiation dominated universe, without gravitational potentials, admits mode-by-mode exact solutions in terms of special functions.  We choose to do numerical integration here in order to provide a more direct comparison to the case where gravitational perturbations are included.  The analytic solution is given by $\varphi_{\bk}(t) = C_1 e^{-imt} U(n, \frac32, 2 i m t) + C_2 e^{-imt} L_{-n}^{(1/2)}(2 i m t)$, where $n = (i k^2 + 3 a_1^2 H_1 m)/(4 a_1^2 H_1 m)$, $U$ is the Tricomi's function, and $L$ is the generalized Laguerre polynomial. For the equations in Sec.~\ref{sec:evolution_with_gravity} and Sec.~\ref{sec:monodromy_potential_evolution}, such an exact solution is not available.}
We found from the overdensity power spectra that free streaming suppression occurs as expected.
Fig.~\ref{fig:delta_spectrum_without_gravity} and \ref{fig:kfs_fit_1} contains the main results of this subsection.

\paragraph{Initialization} In order to impose large-scale density perturbations on the field $\varphi$, we generate a spatially varying field $f$ by realizing it as a homogeneous Gaussian random field with a scale invariant spectrum (we will relate this to the gravitational potential in the upcoming section) $\underline{\Delta}_f^2(k) = A \Theta(k_f - k) $, where $ k_f / (a_i m) = 0.3$, and $A$ is an amplitude chosen to satisfy $\sqrt{\overline{f^2}} = 0.3$. The spectrum $\underline{\Delta}_f^2(k)$ is chosen to be large compared to realistic expectations; see footnote 21.

The field $\varphi(\bx)$ and its time derivative $\dot{\varphi}(\bx)$ are then initialized via the procedure described in Sec.~\ref{sec:IC}, as an inhomogeneous Gaussian random field with density perturbation $f$.
More specifically, the field is generated from an ensemble specified by the following spectra:
\begin{align}
\label{eq:q^3fieldPS}
  & \underline{\Delta}_\varphi^2(q) = \underline{\Delta}_\varphi^2(k_\ast) \left(\frac{q}{k_\ast}\right)^3 \Theta(k_\ast - q),\quad \underline{\Delta}_{\dot\varphi}^2(q) = \omega_q^2 \underline{\Delta}_\varphi^2(q), \quad k_\ast / (a_i m) = 1, 
\end{align}
We decide on $\overline{\varphi^2}$ which then determines $\underline{\Delta}_\varphi^2(k_*)$.\footnote{We choose $\sqrt{\overline{\varphi^2}}$ to be unity. Its actual value is irrelevant since the equation we solve is linear in $\varphi$.}
The spatially dependent variances of $\varphi$ and $\dot{\varphi}$ are chosen to satisfy
\begin{align}
  \expval{\varphi^2(\bx)} \approx \overline{\varphi^2} e^{f(\bx)},\quad
  \expval{\dot\varphi^2(\bx)} \approx \overline{\dot\varphi^2} e^{f(\bx)} .
\end{align}
As discussed in Eq.~\eqref{eq:d_longshort_fourier} of Sec.~\ref{sec:generating_fields_density}, the density spectrum $\Delta_\delta^2(k)$ for $\varphi$ contains a small-scale component $\Delta_{\delta}^{2(\mathrm{short})}(k)$, which peaks at around $k_\ast$, and a large-scale component, which is approximately $\Delta_f^2(k)$:
\begin{align}
  &\Delta_\delta^2(k) \approx \Delta_{\delta}^{2(\mathrm{short})}(k) + \Delta_f^2(k)\,.
\end{align}
Fig.~\ref{fig:spectrum_initial_1} shows the initial power spectrum for $\varphi$, $\delta$ and $f$.

\paragraph{Evolution}
We evolve scalar field $\varphi$ satisfying  Eq.~\eqref{eq:varphi_eom} (with $\Phi = \Psi = 0$), using the fourth-order Runge-Kutta (RK4) method from time $mt=10$ to $mt = 36000$. The initial Hubble parameter was set to $H_i = 0.05 m$. We also use the WKB method to extend the numerical solution to much later times.
Figure \ref{fig:delta_spectrum_without_gravity} provide the density power spectra over time and snapshots of the  overdensity profile. The snapshots are overdensities averaged over one spatial direction, so as to suppress the large amplitude small-scale density perturbations, and highlight the smaller amplitude large length scale perturbations. Note that the shortest length scale fluctuations have comoving size $\sim k_*^{-1}$, whereas the longest length scale perturbation is $\sim 100k_*^{-1}$.

From Fig.~\ref{fig:delta_spectrum_without_gravity}, it is easy to see that the peak and white noise of the density power spectrum around $k_\ast$ is largely fixed over time, except for a slight initial growth.
These behaviors are consistent with the analytical prediction in Sec.~\ref{sec:evolution_of_isocurvature}.
The initial growth in power is apparent when one compares the first and second row of Fig.~\ref{fig:delta_spectrum_without_gravity}.
This growth happens predominantly before $mt < 100$, when the field is transitioning from relativistic to nonrelativistic, and is negligible for the rest of the evolution.
As discussed in Sec.~\ref{sec:evolution_of_isocurvature}, this growth is expected, and it reflects the transition from the dashed orange curve to solid orange curve in Fig.~\ref{fig:sample_field_spectrum}.
After the field turns nonrelativistic, the shape and amplitude of the peak become fixed, as given by \eqref{eq:late_time_delta_spectrum}.
This can be seen by comparing the second and third row of Fig.~\ref{fig:delta_spectrum_without_gravity}.

It is also evident from Fig.~\ref{fig:delta_spectrum_without_gravity} that free streaming wipes out (non-white noise) density perturbations over time.
In the second row of the figure, one can see a sharp dip in the spectrum at around $k/a_i m = 0.07$.
Initial scale-invariant density perturbations with a shorter length scale than the dip are wiped out by free streaming, hence on the right of the dip one can see a $k^3$ spectrum, as expected for random fluctuations satisfying uncorrelated white noise statistics.
On the other hand, larger length scale density perturbations remain on the left of the dip, since the free streaming length is not large enough to wipe out initial perturbations on those scales.
The third row of the figure is a snapshot taken at a much later time, at which the free streaming length is so large that almost all the initial scale-invariant density perturbations are wiped out.

We determine the free streaming scale $k_\mathrm{fs}$ by fitting the density power spectrum to the ansatz
\begin{align}
\label{eq:FitAnsatz}
  \Delta^{2({\rm fit})}_\delta(t,k) = A_\mathrm{fs} e^{- k^2 / (3 k_\mathrm{fs}^2)} + A_\mathrm{iso}( k/k_*)^3,
\end{align}
where $A_\mathrm{fs}$, $A_\mathrm{iso}$ and $k_\mathrm{fs}$ are fitting parameters.
This ansatz is inspired from Eq.~(6) of \cite{Amin:2022nlh}; also see the discussion in our Sec.~\ref{sec:FreeStreamingPhysics}. A sample fit at $a=60a_i$ is shown in the left panel of Fig.~\ref{fig:kfs_fit_1}. The fitting parameters are given by $A_\mathrm{fs} \approx 0.06$, $k_\mathrm{fs}/a_i m \approx 0.01$, $A_\mathrm{iso} \approx 2$.

In the right hand panel of Fig.~\ref{fig:kfs_fit_1}, we plot the fitted value of $k_{\rm fs}$ as a function of time, along with the analytic prediction (see Eq.~\eqref{eq:Tfs-kfs-approx}). One can see from Fig.~\ref{fig:kfs_fit_1} that the fitted $1 / k_\mathrm{fs}$ matches well with the analytic prediction, with deviations restricted to less than $30\%$.\footnote{We only perform the fit for $a/a_i \leq 60$, since the free streaming scale was approaching $k_\mathrm{IR}$ (the smallest wavenumber of the simulated box) at later times, and the fits were become unreliable due to the lack of points on the spectrum.
All fits were performed by restricting to the spectral data to the range $k \in [k_\mathrm{IR}, 0.5 a_i m]$, so that the peak of $\Delta_\delta^2(t, k)$ around $k_\ast$ do not affect the fit.}

\subsubsection{With gravitational perturbations}
\label{sec:evolution_with_gravity}

In this subsection, we discuss our simulation of a cosmological scalar field $\varphi$ during radiation domination, including the effect of gravitational perturbations.
We initialize $\varphi$ with density perturbations consistent with adiabatic initial conditions, and evolve the field in a time-dependent gravitational potential.
We have chosen the field initial conditions and the gravitational potentials in a way that ensures they are consistent with a single set of comoving curvature perturbations, as prescribed by standard cosmological perturbation theory.
In the simulation, we find both initial growth of large-scale density perturbations and subsequent free streaming suppression, as expected. We re-iterate that we do not include self-gravity of the scalar field, since it is assumed to be a subdominant component of the universe during radiation domination.
\begin{figure}[t]
  \centering
  \includegraphics[width=\textwidth]{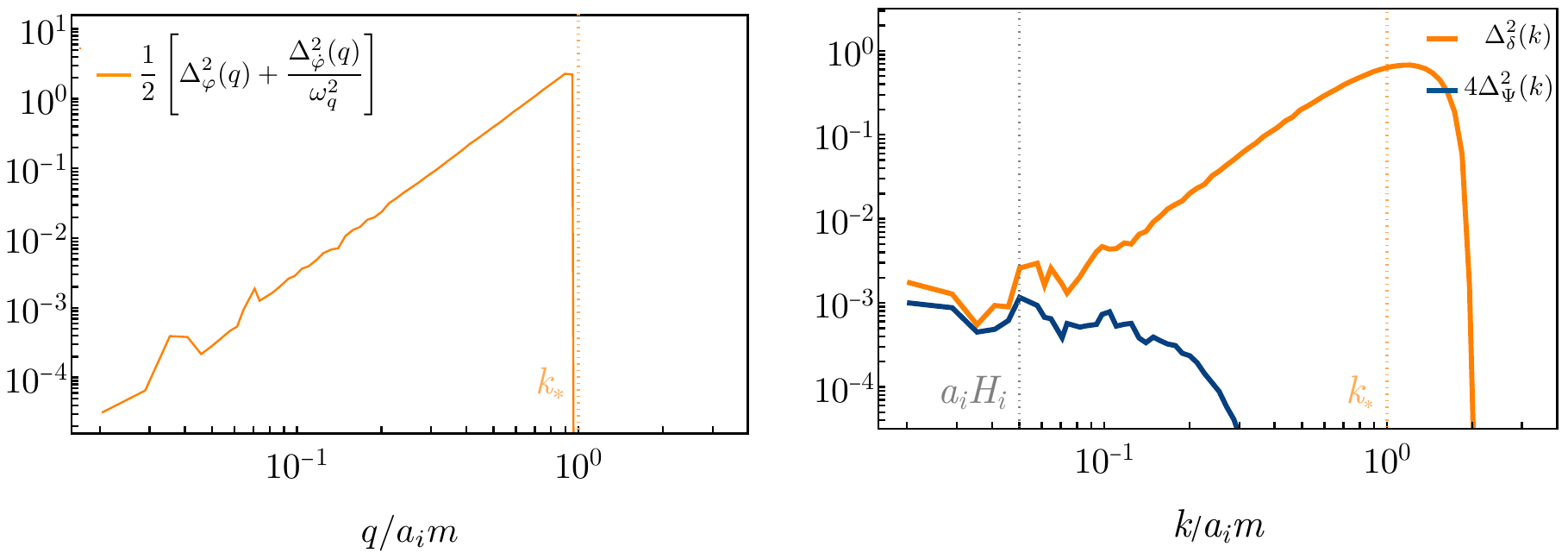}
\caption{A realization of the initial power spectrum for $\varphi$, $\delta$ and gravitational potential $\Psi$ for our simulation which will include time-dependent, radiation driven gravitational potentials, Eq.\eqref{eq:gravitational_potentials_in_radiation_domination}, in the evolution.}
  \label{fig:spectrum_initial_2}
\end{figure}
\paragraph{Initialization} The initial Hubble parameter is set to $H_i = 0.05 m$, with radiation domination $H=H_i(t/t_i)^{1/2}$ (see Eq.~\eqref{eq:radiation_domination}); a space-time dependent gravitational potential sourced by radiation is also included (see Eq.~\eqref{eq:gravitational_potentials_in_radiation_domination}).
We first generate the comoving curvature perturbations $\mathcal{R}$ by realizing it as a homogeneous Gaussian random field with a scale invariant spectrum: $\Delta_{\mathcal{R}}^2(k) = \mathcal{A}_s.$
We choose the amplitude $\mathcal{A}_s$,  such that $\sqrt{\overline{\Psi^2(t_i)}} = 0.04$, where the gravitational potential $\Psi$ is given by Eq.~\eqref{eq:gravitational_potentials_in_radiation_domination}.\footnote{Measurements from \emph{Planck} in fact require $ \mathcal{A}_s \approx 2 \times 10^{-9}$~\cite{Planck:2018jri}. We choose a much larger value for $\mathcal{A}_s$ so that we can see within our simulation the initial condition, the growth and the subsequent suppression of the large-scale density perturbations. Recall that for  $\mathcal{A}_s\sim 10^{-9}$, the wavenumber where the flat spectrum meets the $k^3$ is $k_{\rm dev}\sim 10^{-3}k_*$ (see discussion near Eq.~\eqref{eq:approxDeltain}). This requires a rather large dynamical range for the simulation, without much benefit in terms of understanding of the relevant physics.}

We choose spatially dependent variances $\expval{\varphi^2(\bx)}=\overline{\varphi^2}[1+f_1(\bx)] =\overline{\varphi^2}[1-2.17\Psi(\bx)]$ and $\expval{{\dot\varphi}^2(\bx)}=\overline{\dot{\varphi}^2}[1+f_2(\bx)]=\overline{\dot{\varphi}^2}[1+0.04\Psi(\bx)]$; the initial field configuration is generated using these variances and the procedure in Sec.~\ref{sec:generating_fields}. The variance inhomogeneities $f_1 = -2.17 \Psi$ and $f_2 = 0.04 \Psi$ are given by our choice of $k_*=a_i m$ and Eq.~\eqref{eq:f1f2_our_case}, which was derived with the assumption of adiabatic initial conditions. Similar to Sec.~\ref{sec:W/GravPertEvol}, the field power spectrum is taken to be Eq.~\eqref{eq:q^3fieldPS}, from which the density power spectrum follows using Eq.~\eqref{eq:d_longshort_fourier}. Note that the density power spectrum includes power on large length scales because of the spatially dependent field variances, as well as the gravitational potential appearing in the definition of energy density \eqref{eq:varphi_stress_energy}. 

In Fig.~\ref{fig:spectrum_initial_2}, we show the initial power spectra for the field $\varphi$, the energy overdensity $\delta$,  and the potential $\Psi$. With our choice of parameters, the $\delta$ power spectrum (right panel) includes a scale invariant piece for $k\lesssim a_i H_i$ (a consequence of spatially dependent variances of the field proportional to the gravitational potential), and a white noise component for $a_i H_i\lesssim k\lesssim k_*$. The curvature independent white noise component dominates over this curvature dependent part on subhorizon scales. The field power spectrum (left panel) is not visibly affected by the spatially dependent variances.
\begin{figure}[t]
  \centering
   \includegraphics[width=0.8\textwidth]{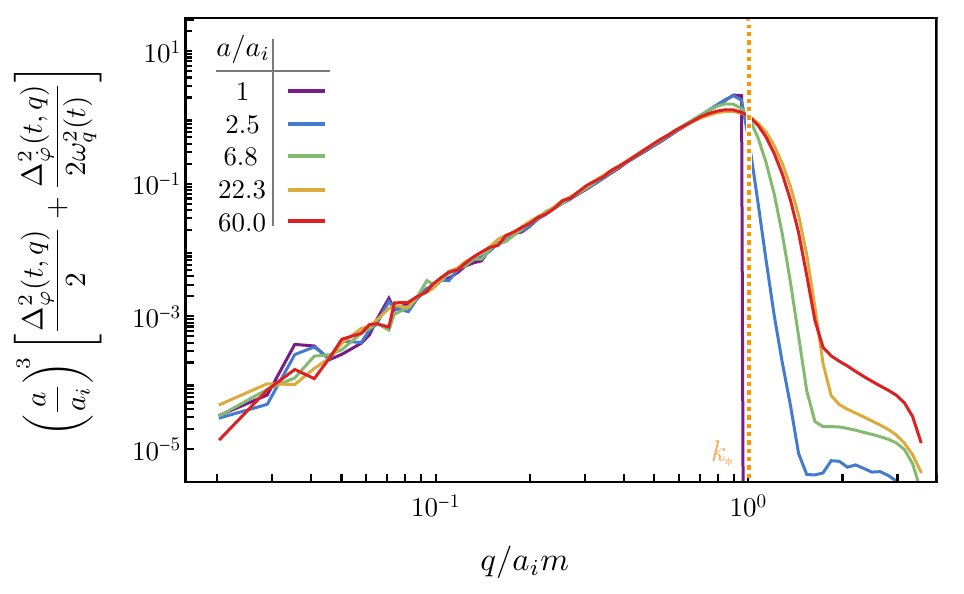}
  \caption{Evolution of the field spectrum in the case where gravitational potentials are included in the time evolution. At $k<k_*$, we essentially see redshifting of the field (not visible since we multiply the power spectrum by $(a/a_i)^3)$. However, for $k\gtrsim k_*$, the gravitational potentials do induce some additional power, likely due to gravitational infall of wavepackets. The enhancement at the highest $k$-modes (plateau) is due to an approximate treatment of the gravitational potential.  It doesn't appear in the simulation without this approximation. See appendix on details. }
  \label{fig:varphi_spectrum_evolution_2}
\end{figure}
\paragraph{Evolution}
The field $\varphi$ evolves in a radiation dominated background according to Eq.~\eqref{eq:varphi_eom}, with the gravitational potential $\Psi$ given by Eq.~\eqref{eq:gravitational_potentials_in_radiation_domination}. This choice of gravitational potential is sensible as long as the field $\varphi$ is a subdominant component of the matter composition. Like in the last subsection, we use RK4 for numerical integration, starting at $m t = 10$ and ending at $m t = 36000$.
We also use WKB solution to extend the solution up to $m t = 3.9 \times 10^{11}$. \\ \\
\noindent{\it Field Evolution}:  
Fig.~\ref{fig:varphi_spectrum_evolution_2} shows the evolution of the field spectrum, scaled by a redshift factor $(a/a_i)^3$. For $q<k_*$, the field spectrum simply redshifts without changing shape. This is to be expected for a free field evolving in an expanding universe (apart from gravitational effects). However, one can see that while the $q> k_\ast$ portion of the field spectrum has no initial support, it quickly gets populated within the time span $a / a_i \in [1, 10]$, and thereafter it remains largely unchanged. The evolution of the field spectrum differs from that in the case without gravity, wherein all modes $\varphi_{\bq}$ are decoupled, and initially absent modes stay absent throughout evolution. The presence of a spatially dependent gravitational potential allows for mode-mode coupling in the $\varphi$ field even though self-gravity of the field is ignored. To understand this intuitively, one can think of $\varphi$ wave packets as gaining kinetic energy as they fall into potential well, corresponding to an enhancement in the high-$q$ modes.
That this enhancement occurs predominantly during the early simulation period ($a / a_i \in [1, 10]$) could be due to two reasons: firstly, the gravitational potential $\Psi_\bk$ decreases in amplitude over time (on subhorizon scales); secondly, horizon entry induces a large ``kick'' on the modes, which can only happen when the horizon is smaller than the size of the simulated box.
\begin{figure}[t]
  \centering
\includegraphics[width=0.85\textwidth]{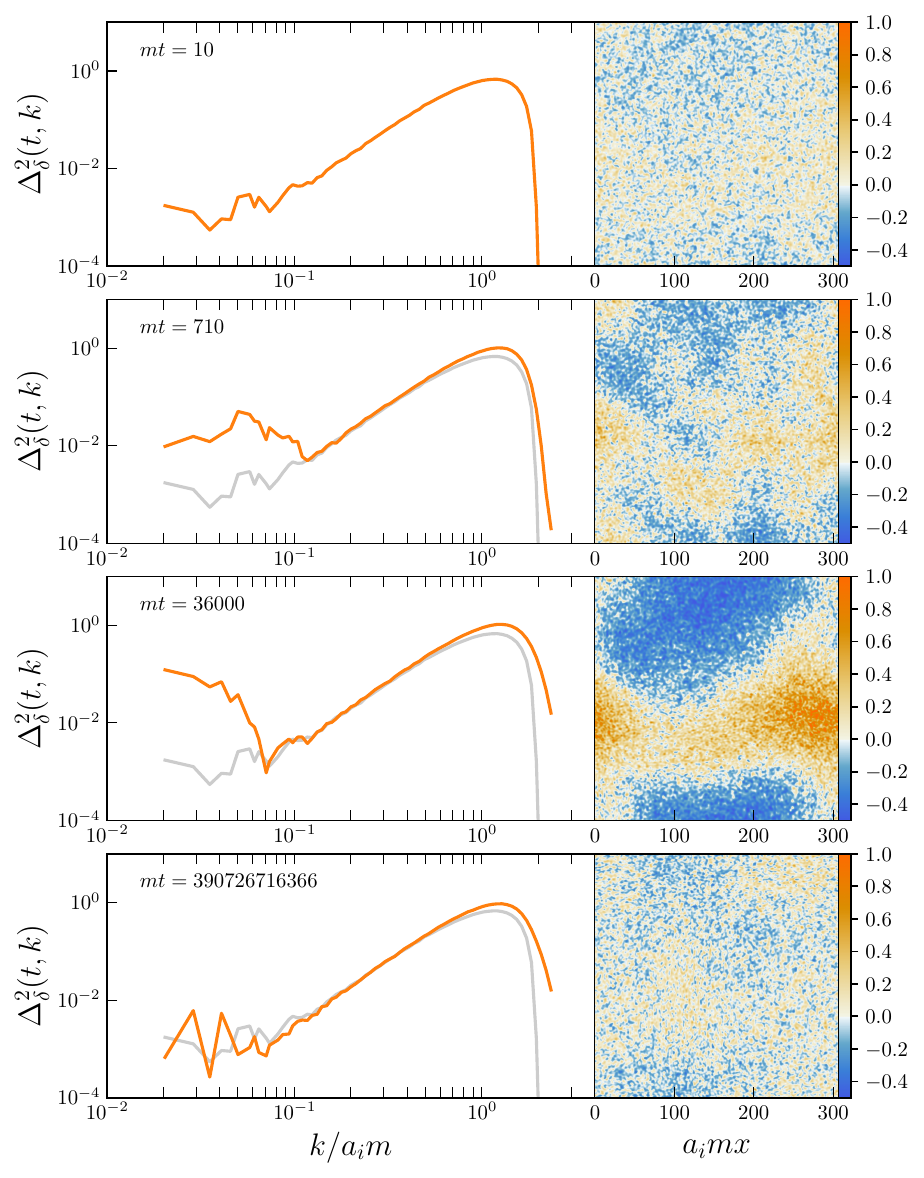}
  \caption{Left panel: Overdensity spectrum $\Delta^2_\delta(t,k)$, initial spectrum is in gray. There is an enhancement of $\delta$ at horizon entry accounting for the initial growth for $k<0.1 a_i m$. The free-streaming suppression then start to erase this growth, revealing more of the white noise spectrum.  We only show the spectrum up to $k/a_i m = 2.48$, so that a negligible numerical artifact at the high-$k$ tail does not appear in the plot. Right panel: Snapshots of the $\delta$ averaged over one axis to enhance the appearance of long wavelength growth from horizon entry, and eventual free streaming suppression. The fluctuations on the scale of $k_*^{-1}$ remain visible throughout. Note the bottom panel is a WKB extrapolation of the simulations.}
  \label{fig:delta_spectrum_with_gravity}
\end{figure}
\\ \\
\noindent{\it Density Perturbation Evolution}:
In Fig.~\ref{fig:delta_spectrum_with_gravity}, we show the evolution of the density power spectrum alongside spatial projection of the overdensity field (averaged over one axis). As discussed earlier, the averaging allows us to visually show the initial gravitational enhancement and the eventual free-streaming suppression of the small-amplitude long-wavelength density perturbations. The first three rows are based on simulations, and the fourth row is from WKB extrapolation of the simulation results.

Before turning to these effects, first note that the shape of the spectrum around $k_\ast$ changes over time. 
By the end of simulation at $mt=36000$, the high-$k$ part of $\Delta_\delta^2(t, k)$ is no longer cut-off at $k/a_i m = 2$, but exhibits modes as high as $k/a_i m = 3$ (though still heavily suppressed compared to the peak). This phenomenon can be explained by Eq.~\eqref{eq:predicted_isocurvature_density_perturbation}: as the $q \gtrsim k_\ast$ modes in the $\varphi$ and $\dot{\varphi}$ spectra are populated over time, the $\delta$ spectrum should also see its $q \gtrsim k_\ast$ modes populated.
Intuitively, the creation of smaller wavelength $\varphi$ modes naturally leads to smaller scale fluctuations in $\rho$.
At late times, the peak of the density spectrum again stabilizes, since the $\varphi$ and $\dot{\varphi}$ spectra have stabilized.

We now turn to the two main effects that we had hoped to capture: (1) a growth in the density perturbations at horizon entry, and (2) a suppression of the curvature induced density perturbations due to free-streaming. The results below confirm the key expectations of \cite{Amin:2022nlh}.
\begin{figure}[t]
  \centering
  \includegraphics[height=0.48\textwidth]{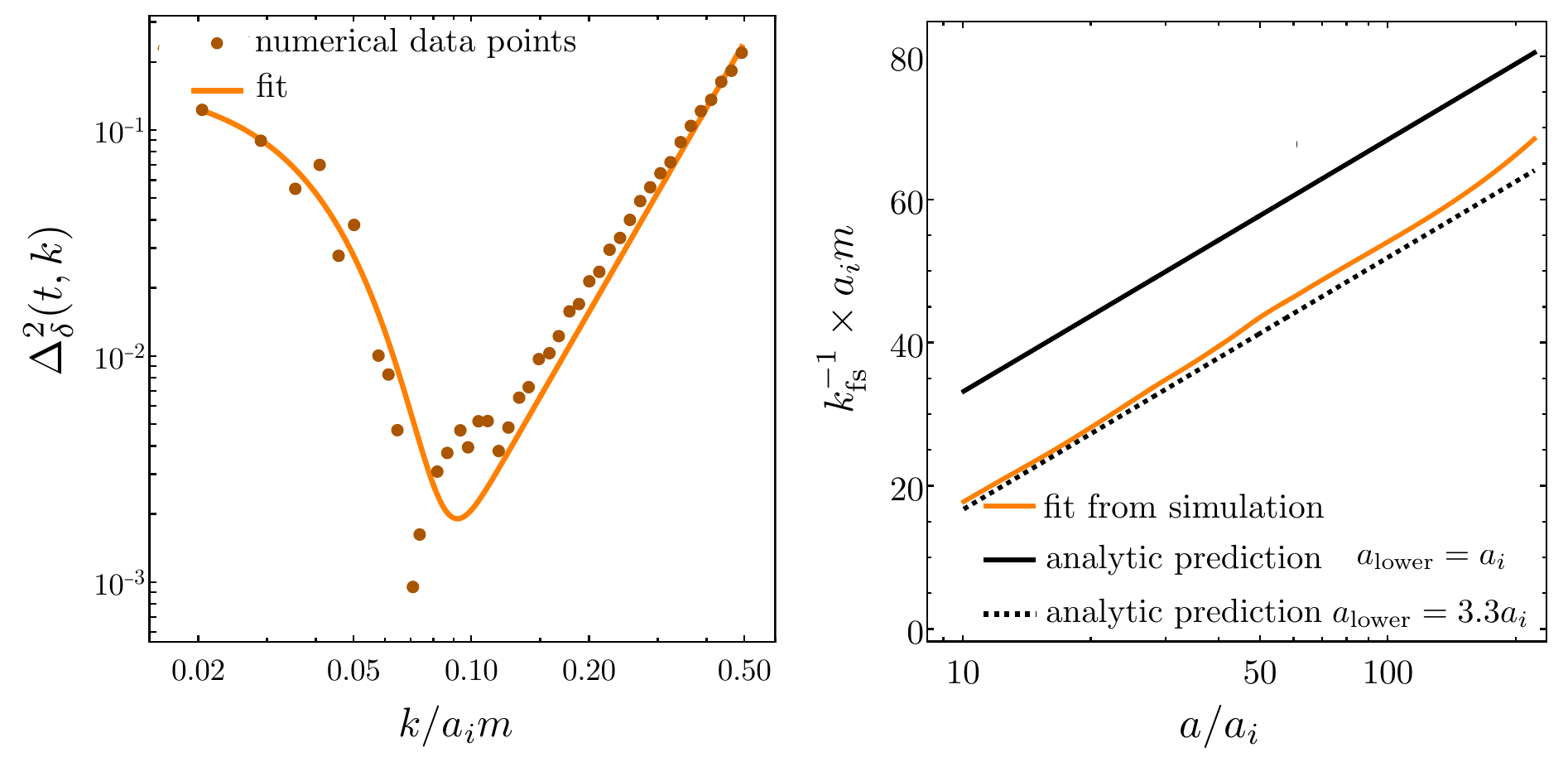}
  \caption{Same as Fig.~\ref{fig:kfs_fit_1}, but now field evolution includes time-dependent gravitational potentials in the simulation. The left panel shows the fit of the density spectrum at $a/a_i=60$ using the ansatz \eqref{eq:FitAnsatz}. The right panel shows the comoving free-streaming length $k_\mathrm{fs}^{-1}$ versus scale factor $a$, obtained by fitting the density spectra from the simulation (orange curve) and by the analytic prediction (solid and dotted black curve) \eqref{eq:Tfs-kfs-approx}. The dotted black curve gives the free-streaming length calculated after the density perturbations have grown initially; more specifically, the lower boundary of integration in \eqref{eq:Tfs-kfs-approx} is set to $am$ when $a=3.3a_i$, just after horizon crossing for all relevant modes. The solid black curve is the case where free-streaming length is calculated from the initial time $a=a_i$. }
  \label{fig:kfs_fit_2}
\end{figure}
\begin{enumerate}
\item {\bf Growth at Horizon Entry}: In Fig.~\ref{fig:delta_spectrum_with_gravity} one can see a significant growth in density perturbations for $k/a_i m < 0.1$.
In the $mt=36000$ snapshot, one can see a growth of roughly $10^2$ in the largest mode. This is to be expected based on seminal results from \cite{Hu:1995en}--linearized cosmological perturbation theory yield significant growth of (adiababtic) dark matter density perturbations at horizon entry during radiation domination. This is a well known ``textbook" result; see for example Fig.~6.6 of \cite{Baumann:2022mni}, where modes with $k > k_\mathrm{eq}$ receive a boost at horizon crossing. In our simulation, the horizon scale $k_H = aH$ becomes smaller than the IR scale of the simulated box $k_\mathrm{IR} = 2\pi / L$ at $mt=60$, so for the majority of the simulation all relevant modes are deep inside the horizon. Since our simulation is done during radiation domination, all relevant modes in our simulation satisfy $k > k_\mathrm{eq}$.
We have thus confirmed a result from linearized cosmological perturbation theory via full nonlinear field theory simulation instead of linearized equations for fluid density perturbations. 

\item {\bf Free Streaming Suppression}: Similar to the gravity-free simulation in Sec.~\ref{sec:W/GravPertEvol}, free streaming wipes out density perturbations, including those perturbations induced by gravity at horizon entry.
In Fig.~\ref{fig:delta_spectrum_with_gravity}, one can see that density perturbations within scale $k/a_im \in [0.07,0.1]$ are initially sourced by gravity, and are subsequently wiped out by free streaming. Again note that free-streaming leaves the white noise $k^3$ part unchanged in amplitude, and more of the $k^3$ tail is revealed as free-streaming eats away the curvature induced perturbations on large length scales.
\end{enumerate}

We numerically fit the overdensity power spectrum with an ansatz (see Eq.~\eqref{eq:FitAnsatz}), and fit for a free-streaming scale $k_\mathrm{fs}$. We then compare it  with theoretical expectations (see Eq.~\eqref{eq:Tfs-kfs-approx}). The left panel of Fig.~\ref{fig:kfs_fit_2} shows the fit compared to the data at a fixed time $a=60a_i$. The fitting parameters are given by $A_\mathrm{fs} \approx 0.2$, $k_\mathrm{fs}/a_i m \approx 0.02$, $A_\mathrm{iso} \approx 2$. The analytic and numerical fits for $k_{\rm fs}$ as a function of time are shown in the right panel of Fig.~\ref{fig:kfs_fit_2}. The analytic and fitted $k_{\rm fs}$ agree in their $\ln(a)$ time-dependence; however, they differ by a constant time delay, with the free streaming length from simulations being smaller (black solid curve). The difference can be eliminated by delaying $a_i$ in the analytic calculation for $k_{\rm fs}$, so as to allow for growth of structure before it gets erased (black dashed curve). A more careful calculation would evaluate free-streaming effects by taking into account time of horizon entry for density perturbations.

\subsubsection{Oscillon formation and free streaming}
\label{sec:monodromy_potential_evolution}
In this subsection, we demonstrate that free streaming can be effective in suppressing large-scale density perturbations even if the scalar field has strong self-interactions, and is forming solitons (oscillons).\footnote{Oscillons are spatially localized, exceptionally long-lived, time-periodic configurations of the field, held together by attractive self-interactions countering gradient related dispersion of the field \cite{Gleiser:1993pt,Copeland:1995fq,Kasuya:2002zs,Amin:2010jq,Zhang:2020bec,Cyncynates:2021rtf,Levkov:2023ncb}.} Unlike previous subsections, here we do not include expansion or spacetime dependent gravitational potentials in the evolution equations (though we do include density perturbations in the field initial conditions). That is, the evolution is in Minkowski space and we set $a=a_i=1$.

\begin{figure}[t]
  \centering
\includegraphics[width=\textwidth]{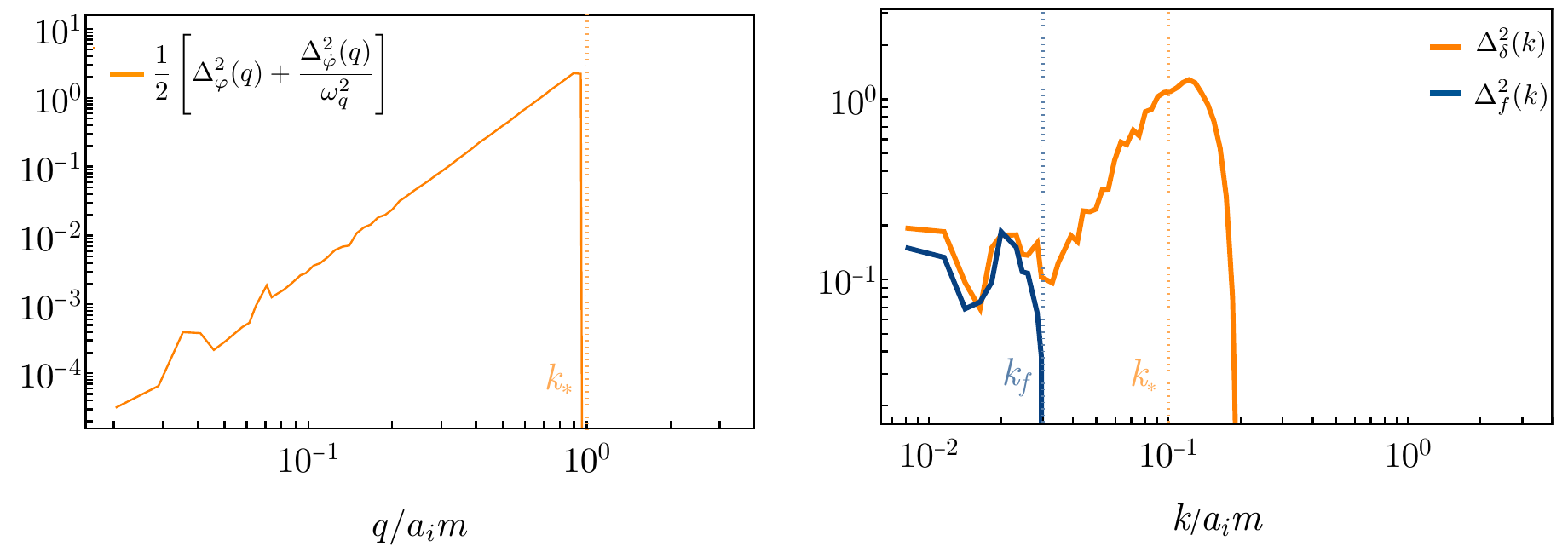}
  \caption{Initial power spectrum for the field $\varphi$, density contrast $\delta$ and $f$ (a proxy for the initial gravitational potential). The field has significant self-interactions, and will lead to the formation of oscillons.}
\label{fig:spectrum_initial_3}
\end{figure}

\paragraph{Initialization}
As in Sec.~\ref{sec:W/GravPertEvol}, we use the procedure described in Sec.~\ref{sec:generating_fields} to generate a field $\varphi$ with approximately scale-invariant large length scale initial density perturbations as well as large amplitude small-scale perturbations.
For this simulation, we let the initial field be nonrelativistic, with spectrum $\underline{P}_\varphi(q) = A \Theta(k_\ast-q)$, $k_\ast = 0.1 m$ and $\sqrt{\overline{\varphi^2}} = m$.
The perturbation in field variances is given by $\expval{\varphi^2(\bx)} = \overline{\varphi^2} e^{f(\bx)}$ and $\expval{\dot{\varphi}^2(\bx)} = \overline{\dot{\varphi}^2} e^{f(\bx)}$, with $\sqrt{\overline{f^2}} = 0.4$ and $k_f = 0.03 m$.
See Fig.~\ref{fig:spectrum_initial_3} for the initial spectra. The shape of the initial field and density spectrum here are similar to that of earlier sections (especially Sec.~\ref{sec:W/GravPertEvol}), but now the field is nonrelativistic from the beginning.  Nevertheless, the initial conditions here allow for qualitative comparison between cases with and without strong self-interaction.  \footnote{Note that these initial conditions are not identical to those in \cite{Amin:2011hj} or \cite{Lozanov:2019ylm}, where oscillon formation proceeds via parametric resonance from an approximately homogeneous field (or \cite{Gleiser:2014ipa,Shafi:2024jig}, where multiple fields are involved). Nevertheless, we expect similar features to arise in both simulations, including a peak, a white noise tail, and a scale invariant adiabatic part soon after formation of oscillons. A detailed comparison between these initial conditions and subsequent evolution is beyond the scope of this work.}

\paragraph{Evolution}
We use the monodromy potential given by $V(\varphi)=m^2M^2\left[\sqrt{1+(\varphi/M)^2}-1\right]$ to evolve a field with initial large length-scale density perturbations in a Minkowski background.
We take $M = 30 m$. The size of the box is $L=768/m$. The simulation runs from $mt=0$ to $mt=12500$. We found that oscillons form, and free streaming suppression occurs. 

\begin{figure}[t]
  \centering
    \includegraphics[width=0.8\textwidth]{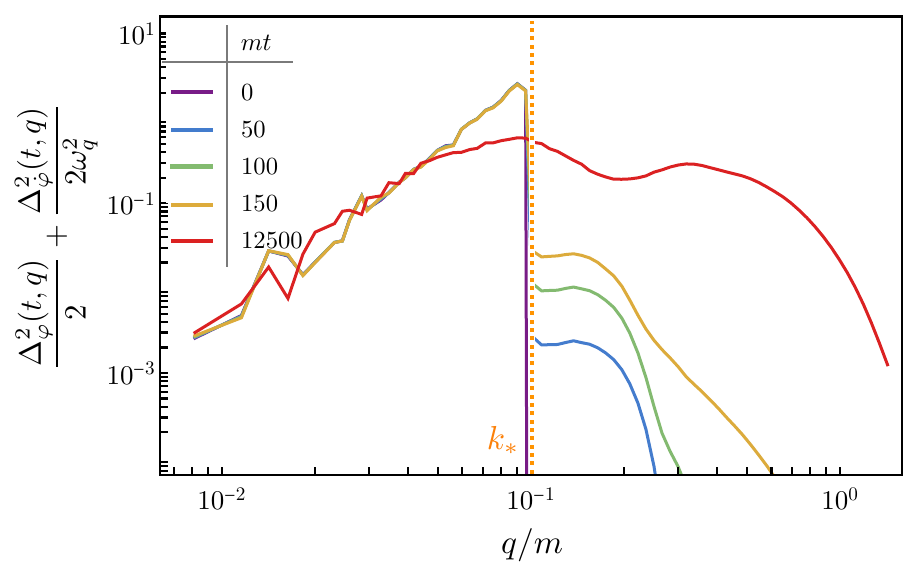}
  \caption{Evolution of the field spectrum when the field has attractive self-interactions. This simulation does not include expansion or time evolving gravitational potentials. The enhancement at the highest $k$-modes is due to formation of oscillons driven by self-interactions in the field. }
  \label{fig:varphi_spectrum_evolution_3}
\end{figure}

See Fig.~\ref{fig:varphi_spectrum_evolution_3} for evolution of the field spectrum. Although there is no gravitational interaction, the presence of terms beyond $\varphi^2$ in the potential $V(\varphi)$ ensures coupling of modes. This coupling leads to an extra growth of field power at around $q={\rm few}\times 0.1m$, which is roughly linked to the inverse scale of the oscillons. In an expanding universe, the oscillons would maintain a fixed physical size (i.e. their spatial extent does not expand with the universe), and as a result the oscillon related peak would move rightwards (in terms of comoving momenta) as the universe expands. Furthermore, the amplitude of initial conditions peak (at $k_*$) would redshift downwards with time, leaving the oscillon related peak $q_{\rm osc}/a\sim {\rm (oscillon\,\, size)}^{-1}$ to dominate the spectrum. 

\begin{figure}[t]
  \centering
  \includegraphics[width=0.86\textwidth]{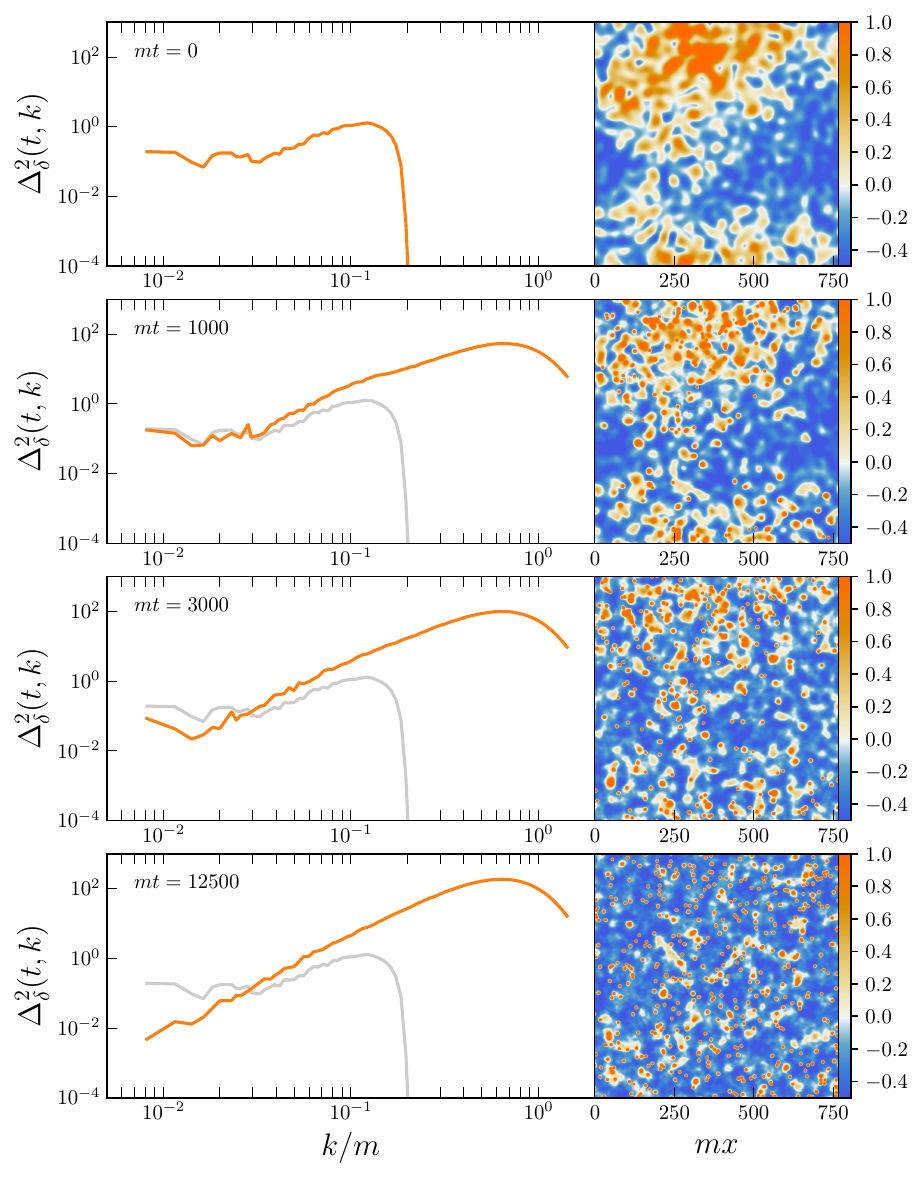}   \caption{Like in Fig.~\ref{fig:delta_spectrum_without_gravity} and Fig.~\ref{fig:delta_spectrum_with_gravity}, the left panels are the overdensity spectra $\Delta^2_\delta(k)$ over time, and the right panels are the corresponding snapshots of $\delta$.  Oscillon formation and persistence is responsible for the large peak around $k=0.7m$ in the density spectrum, and are easily seen in the $\delta$ snapshots from the circular orange objects. One can see from the left panels that the initial scale-invariant (low-$k$) part of the spectra are suppressed over time, and only a $k^3$ part remains in the end.  Unlike earlier figures in this paper, expansion and local gravitational potentials are not included in the evolution equations for these simulations, but strong non-gravitational self-interactions are included.}
  \label{fig:delta_spectrum_soliton}
\end{figure}

See Fig.~\ref{fig:delta_spectrum_soliton} for evolution of the density spectrum. Again, notice the generation of a new peak at $k\sim 0.5 m$ (initial peak is closer to $0.1 m$). This peak is due to the large over-densities $\delta \gg 1$ resulting from oscillon formation.\footnote{These are, however, subhorizon objects and not close to black hole formation \cite{Lozanov:2019ylm,Ballesteros:2024hhq}} In an expanding universe, this peak will move to the right (if plotted in terms of comoving $k$), because the oscillons maintain a fixed physical size. The low-$k$ part ($k\lesssim \textrm{few}\times 10^{-2}m$) of the density spectrum changes slowly, with free streaming suppression visible towards the end of the simulation. Once again, free-streaming reveals more of the white-noise density power spectrum over time.

The snapshots in Fig.~\ref{fig:delta_spectrum_soliton} provide further insights on the oscillons. In the snapshots, one can see oscillons (seen as quasi-circular orange regions after the first snapshot) forming and moving rapidly through the box. Moreover, in the second and third snapshots, the formed oscillons are clearly aligned with the initial high density regions, whereas in the fourth snapshot they are randomly distributed in space. This change is due to the motion of the oscillons. Using the quasi-particle picture in the introduction (with oscillons being the quasi-particles here), we intuitively expect that the random velocities of the oscillons contribute to the free streaming suppression at low $k$ also. Using the animation of the slices, we found that the solitons typically have speeds $\order{0.1} \times k_\ast / m$.
For some solitons, the velocity could be as high as $\order{k_\ast / m}$.
In an expanding universe, these velocities would of course redshift as the universe expands.

Qualitatively similar motion of oscillons/solitons was also seen in \cite{Amin:2019ums}, where the formation of solitons (starting with an almost homogeneous condensate) was driven by non-gravitational self-interactions of the field; however, the evolution also included expansion and self-gravity of the field  -- we leave a detailed investigation of soliton velocity distribution and comparison to \cite{Amin:2019ums} to later work. A comparison is also warranted in the cases discussed by \cite{Arvanitaki:2019rax,Sakharov:1994id,Sakharov:1996xg,Khlopov:1998uj} that consider the impact of self-interactions on structure formation in axions. 
Also note that free-streaming effects from the {\it decay} of oscillons was considered in \cite{Imagawa:2021sxt}, which is an additional effect (at late times) not analysed in the present work.

\section{Future Directions}
\label{sec:future}
There are a number of avenues to pursue which go beyond the present work. The three key directions we have in mind include:
\paragraph{Including self-gravity} In the present work we have ignored the self-gravity of the $\varphi$ field. While this is reasonable deep within radiation domination, as we get closer to matter radiation equality, self-gravity becomes increasingly more important. Crudely speaking, one can estimate that once $\overline{\rho}(a)\sqrt{\Delta_\delta^2(k,a)}>\overline{\rho}_\gamma(a) \sqrt{\Delta_{\delta_\gamma}^2(k,a)}$, ignoring self-gravity is no longer a reasonable assumption. Note that this is a scale-dependent statement, and will likely be satisfied first near the peak of the density spectrum. This can lead to formation of gravitationally supported solitons and miniclusters \cite{Hogan:1988mp, Kolb:1993hw,Vaquero:2018tib,Visinelli:2018wza,Eggemeier:2019khm}. Moreover, the oscillatory nature of the gravitational potentials due to radiation likely makes self-gravity important at an even earlier scale factor. Of course, for detailed understanding of the growth in the density contrast, care is needed in taking into account the Jeans scale, bulk velocities associated with the field, and effective velocity dispersion in the system.

Including self-gravity is conceptually straightforward. One simply has to solve a Poisson equation to calculate the effect of the scalar field and radiation components on $\Psi$. A lattice code such as the one used in \cite{Smith:2023fob} can be used here so as to include perturbations in both radiation and in the field itself. The added computational cost include additional Fourier transforms and energy density computations for the RHS of the Poisson equation. 

Computational cost can be reduced by employing the non-relativistic approximation. In the above approach we still solve the full Klein-Gordon equation in a perturbed spacetime, so the evolution keeps track of oscillations on timescales of $2\pi/m$. By the time self-gravity is relevant, the field is typically highly non-relativistic, hence keeping track of these fast oscillations is not completely necessary. Instead, we can adopt the non-relativistic Schrodinger-Poisson approach, where we keep track of the slowly varying dynamics only.

Explicitly, writing $\varphi(t,\bx)={(2m)^{-1/2}}\left[\psi(t,\bx)e^{-imt}+c.c\right]$
and assuming $|\partial_t\psi|\ll |m\psi|$, $|\partial_t\Psi|\ll |m\Psi|$, and  $H/m$, $|aH/\nabla|\ll 1$, we obtain the Schrodinger-Poisson system
\begin{align}
\label{eq:SPscalar}
i\left[\partial_t+\frac{3H}{2}\right]\psi=-\frac{\nabla^2\psi}{2ma^2}+m\Psi \psi\,,\quad \frac{\nabla^2\Psi}{a^2}=4\pi G \left[m\left(\psi^*\psi-\overline{\psi^*\psi}\right)+\overline{\rho}_\gamma \delta_\gamma+\hdots\right],
\end{align}
where $3H^2=8\pi G(\overline{|\psi|^2}+\overline{\rho}_\gamma+\hdots)$ and where $\hdots$ can include baryons, neutrinos, etc. These equations will have to be supplemented by equations for the perturbations in radiation, baryons, neutrinos, etc. The relevance of the matter content will depend on the era under consideration. Note that $\overline{\psi^2}\ne 0$ even though $\psi$ has no homogeneous mode. That is, there is a homogeneous mode for mass density even though there is no homogeneous field.
In upcoming work this is the approach we intend to take for including self-gravity of the field at sufficiently late times during radiation domination, all the way into and through matter domination.\footnote{A careful derivation, including all the leading order corrections (in the small quantities) is provided in \cite{Namjoo:2017nia,Salehian:2021khb}.} It is possible to include non-gravitational/non-minimal self-interactions in this picture as well (e.g.~\cite{Chavanis:2011zi, Amin:2019ums, Glennon:2020dxs,Sankharva:2021spi,Jain:2023tsr,Chen:2024pyr}). 

The initial conditions for this non-relativistic treatment can be taken as the final result of our present work. The results for $\varphi(t,\bx)$ and $\dot{\varphi}(t,\bx)$ obtained here (after $k_*/am\ll 1$), can be translated to $\psi$ (which is a complex field) as follows:
\begin{equation}
\psi(t,\bx)\approx \sqrt{\frac{m}{2}}\left[\varphi(t,\bx)+\frac{i}{m}\dot{\varphi}(t,\bx)\right].
\end{equation}
One can also start with the non-relativistic field and set the initial conditions using a modified version of the procedure outlined in Sec.~\ref{sec:generating_fields_density}, so that appropriate density and velocity perturbations on sufficiently large (but not necessarily superhorizon) length scales are included.  We will present details in an upcoming work.

\paragraph{Vector, Tensor fields and multiple scalar fields}
For multiple scalar fields $\varphi_i(t,\bx)$ ($i=1,2\hdots N$), the energy density in each field as well as the mass $m$ for each field can be different. This allows much richer possibilities for the isocurvature peak and the effect of free-streaming. There is a significant amount of literature motivating the existence of dark matter being made up of multiple fields, and an exploration of their phenomenology. For example, see \cite{Arvanitaki:2009fg,PhysRevD.86.055013,Dienes:2024wnu}. 

Similarly, instead of multiple scalars, we can also consider vector field ($s=1$) or tensor field ($s=2$) dark matter, which has $N=2s+1$ components. Production mechanisms and self-interactions can potentially favor an initial condition with asymmetry between different field polarizations or helicities, which would then likely be erased by gravitational clustering \cite{Amaral:2024tjg}. The effects of both the initial isocurvature peak as well as free-streaming suppression would again be worth investigating for $s>0$ fields. For existing work on small-scale structure in such dark matter, see  \cite{Adshead:2021kvl,Jain:2021pnk,Amin:2022pzv,Gorghetto:2022sue,Jain:2023ojg,Chen:2023bqy,Amaral:2024tjg,Chen:2024vgh}. Also see \cite{Zhang:2023fhs} for the cases with non-minimal coupling to gravity, and \cite{Zhang:2021xxa,Jain:2022kwq,Jain:2022agt,Zhang:2024bjo} for including non-gravitational self-interactions.

\paragraph{Detailed Models}
In this work, we have used some toy spectra of fields and their derivatives at time $t_i$ for initial conditions.  Realistic models will give different spectra depending on the underlying dynamics. For example, the spectra can arise from parametric resonance transferring energy from a parent field to a daughter dark matter field (e.g.~\cite{Amin:2014eta,Lozanov:2019jxc,Agrawal,Dror:2018pdh,Co:2018lka,Adshead:2023qiw,Cyncynates:2023zwj}, from gravitational production \cite{Graham:2015rva,Kolb:2023ydq}, or from topological defects. Production of axions and dark photons from defects can have particularly rich phenomenology \cite{Long:2019lwl,Gorghetto:2020qws,Buschmann:2021sdq,East:2022rsi,Saikawa:2024bta,Gorghetto:2024vnp}. While these avenues have been pursued before for understanding the nonlinear dynamics on subhorizon scales, the impact of free-streaming on large length scale structure has rarely been addressed. It would be useful to revisit these topics, in particular by employing our algorithm for initializing the adiabatic density perturbations on large length scales.

\section{Summary}
\label{sec:summary}
Light dark matter fields produced after inflation can have observationally accessible consequences due to a free-streaming suppression and/or a white noise enhancement of the density fluctuations in the field. 

Using lattice simulations of dark matter fields, we numerically investigated the impact of free streaming of dark matter fields on the dark matter density power spectrum, as well as the general evolution of the adiabatic and isocurvature density perturbations (in the radiation dominated era).  To this end, we  developed a framework for generating appropriate initial conditions for the DM field in the early universe, and evolved the field numerically including the impact of expansion and gravitational potentials on the field.

Motivated by post-inflationary production mechanisms, we considered an initial field configuration without a spatially homogeneous mode. The field configuration is dominated by small-scale, subhorizon field variations. The small-scale field variations result in a white noise isocurvature spectrum in the density perturbations. However, we also need to make the field configurations consistent with large length scale, roughly scale-invariant adiabatic density perturbations.
\begin{tcolorbox}[colback=lightgray!25, colframe=white, width=\linewidth, arc=4mm, boxrule=0mm]
For a general initial field power spectrum which peaks on subhorizon scales, we provide a systematic algorithm to achieve consistency with adiabatic density perturbations on large length scales. We achieve this by initializing the field as a Gaussian random field albeit with {\it spatially-dependent variances}. Our scheme works for relativistic and non-relativistic fields.
\end{tcolorbox}
This initialized field includes both the small-scale isocurvature density perturbations, and large length scale adiabatic ones. To the best of our knowledge, a systematic numerical framework for initial conditions for such DM fields has not been discussed before in the literature. 

In Sec.~\ref{sec:evolution_with_gravity}, we used a lattice simulation to evolve the scalar field in an expanding universe with spacetime dependent metric perturbations. The lack of a homogeneous field mode and presence of gravitational potentials, complicates the analytic treatment of the evolution. Our lattice simulation results in Sec.~\ref{sec:evolution_with_gravity} are consistent with approximate analytic predictions in \cite{Amin:2022nlh}. Specifically, we observed the following:
\begin{tcolorbox}[colback=lightgray!25, colframe=white, width=\linewidth, arc=4mm, boxrule=0mm]
Adiabatic density perturbations see initial growth after horizon entry, followed by free-streaming suppression. This free-streaming suppression reveals more of the white noise isocurvature perturbations at progressively larger length scales.  We also confirmed the lack of evolution of the white noise isocurvature density spectrum during radiation domination. 
\end{tcolorbox}
One can derive mode-by-mode WKB solutions when the modes are deep within the horizon and the gravitational potential can be neglected. We use this approach to extrapolate our simulation results to significantly later times.

In Sec.~\ref{sec:monodromy_potential_evolution}, we also used lattice simulations to study the impact of strong self-interactions on free-streaming. We found oscillon formation and motion, and verified similar free-streaming effects in the density spectrum: suppression of the initial adiabatic spectrum and non-evolution of the white noise isocurvature spectrum.

Finally, we described how our results can be used as initial conditions for late time simulations when the self-gravity of the field is important and the fields have become sufficiently non-relativistic. We intend to pursue the growth of structure around and after matter-radiation equality using a Schr\"{o}dinger-Poisson system. The evolution of the isocurvature part, as well as the impact of free-streaming will be be the key novelty -- eventually providing more accurate modelling of such warm, wavelike dark matter in the late universe.

\section*{Acknowledgements}
We would like to thank Kimberly Boddy, Anthony Challinor, Daniel Chung, Sten Delos, Wayne Hu, Mudit Jain, Eiichiro Komatsu, Rayne Liu, Andrew Long, Mehrdad Mirbabayi, Ethan Nadler, Moira Venegas, David Wands, Wisha Wanichwecharungruang and Huangyu Xiao for insightful questions and discussions. We especially thank Hong-Yi Zhang for their work during the early stages of this project, as well as Wayne Hu, Andrew Long, Rayne Liu, Simon May and Mehrdad Mirbabayi for their valuable feedback. We also thank an anonymous referee for their insightful questions, which led us to think critically about the applicability of our framework. MA is supported by a DOE award DE-SC0021619. Siyang Ling is supported through the NASA ATP award 80NSSC22K0825.  Part of this research was supported by grant NSF PHY-2309135 to the Kavli Institute for Theoretical Physics (KITP). This research was also supported  in part by the Munich Institute for Astro-, Particle and BioPhysics (MIAPbP), which is funded by the Deutsche Forschungsgemeinschaft (DFG, German Research Foundation) under Germany's Excellence Strategy – EXC-2094 – 390783311. MA would like to acknowledge the hospitality of KICC (Cambridge), MiAPbP (Munich) and KITP (Santa Barbara) where part of this work was completed.
\bibliographystyle{JHEP}
\bibliography{main}

\providecommand{\href}[2]{#2}\begingroup\raggedright\begin{thebibliography}{100}

\bibitem{Planck:2018vyg}
{\scshape Planck} collaboration, \emph{{Planck 2018 results. VI. Cosmological parameters}}, \href{https://doi.org/10.1051/0004-6361/201833910}{\emph{Astron. Astrophys.} {\bfseries 641} (2020) A6} [\href{https://arxiv.org/abs/1807.06209}{{\ttfamily 1807.06209}}].

\bibitem{Cirelli:2024ssz}
M.~Cirelli, A.~Strumia and J.~Zupan, \emph{{Dark Matter}},  \href{https://arxiv.org/abs/2406.01705}{{\ttfamily 2406.01705}}.

\bibitem{ParticleDataGroup:2020ssz}
{\scshape Particle Data Group} collaboration, \emph{{Review of Particle Physics}}, \href{https://doi.org/10.1093/ptep/ptaa104}{\emph{PTEP} {\bfseries 2020} (2020) 083C01}.

\bibitem{Dalal:2022rmp}
N.~Dalal and A.~Kravtsov, \emph{{Excluding fuzzy dark matter with sizes and stellar kinematics of ultrafaint dwarf galaxies}}, \href{https://doi.org/10.1103/PhysRevD.106.063517}{\emph{Phys. Rev. D} {\bfseries 106} (2022) 063517} [\href{https://arxiv.org/abs/2203.05750}{{\ttfamily 2203.05750}}].

\bibitem{Amin:2022nlh}
M.A.~Amin and M.~Mirbabayi, \emph{{A Lower Bound on Dark Matter Mass}}, \href{https://doi.org/10.1103/PhysRevLett.132.221004}{\emph{Phys. Rev. Lett.} {\bfseries 132} (2024) 221004} [\href{https://arxiv.org/abs/2211.09775}{{\ttfamily 2211.09775}}].

\bibitem{Tremaine}
S.~{Tremaine} and J.E.~{Gunn}, \emph{{Dynamical role of light neutral leptons in cosmology}}, \href{https://doi.org/10.1103/PhysRevLett.42.407}{\emph{Physical Review Letters} {\bfseries 42} (1979) 407}.

\bibitem{DiPaolo:2017geq}
C.~Di~Paolo, F.~Nesti and F.L.~Villante, \emph{{Phase space mass bound for fermionic dark matter from dwarf spheroidal galaxies}}, \href{https://doi.org/10.1093/mnras/sty091}{\emph{Mon. Not. Roy. Astron. Soc.} {\bfseries 475} (2018) 5385} [\href{https://arxiv.org/abs/1704.06644}{{\ttfamily 1704.06644}}].

\bibitem{PhysRevD.103.055014}
H.~Davoudiasl, P.B.~Denton and D.A.~McGady, \emph{Ultralight fermionic dark matter}, \href{https://doi.org/10.1103/PhysRevD.103.055014}{\emph{Phys. Rev. D} {\bfseries 103} (2021) 055014}.

\bibitem{Hui:2021tkt}
L.~Hui, \emph{{Wave Dark Matter}}, \href{https://doi.org/10.1146/annurev-astro-120920-010024}{\emph{Ann. Rev. Astron. Astrophys.} {\bfseries 59} (2021) 247} [\href{https://arxiv.org/abs/2101.11735}{{\ttfamily 2101.11735}}].

\bibitem{Ferreira:2020fam}
E.G.M.~Ferreira, \emph{{Ultra-light dark matter}}, \href{https://doi.org/10.1007/s00159-021-00135-6}{\emph{Astron. Astrophys. Rev.} {\bfseries 29} (2021) 7} [\href{https://arxiv.org/abs/2005.03254}{{\ttfamily 2005.03254}}].

\bibitem{OHare:2024nmr}
C.A.J.~O'Hare, \emph{{Cosmology of axion dark matter}}, \href{https://doi.org/10.22323/1.454.0040}{\emph{PoS} {\bfseries COSMICWISPers} (2024) 040} [\href{https://arxiv.org/abs/2403.17697}{{\ttfamily 2403.17697}}].

\bibitem{Antypas:2022asj}
D.~Antypas et~al., \emph{{New Horizons: Scalar and Vector Ultralight Dark Matter}},  \href{https://arxiv.org/abs/2203.14915}{{\ttfamily 2203.14915}}.

\bibitem{Dine:1982ah}
M.~Dine and W.~Fischler, \emph{{The Not So Harmless Axion}}, \href{https://doi.org/10.1016/0370-2693(83)90639-1}{\emph{Phys. Lett. B} {\bfseries 120} (1983) 137}.

\bibitem{Sikivie:1982qv}
P.~Sikivie, \emph{{Of Axions, Domain Walls and the Early Universe}}, \href{https://doi.org/10.1103/PhysRevLett.48.1156}{\emph{Phys. Rev. Lett.} {\bfseries 48} (1982) 1156}.

\bibitem{Hogan:1988mp}
C.~Hogan and M.~Rees, \emph{{AXION MINICLUSTERS}}, \href{https://doi.org/10.1016/0370-2693(88)91655-3}{\emph{Phys. Lett. B} {\bfseries 205} (1988) 228}.

\bibitem{Gorghetto:2020qws}
M.~Gorghetto, E.~Hardy and G.~Villadoro, \emph{{More axions from strings}}, \href{https://doi.org/10.21468/SciPostPhys.10.2.050}{\emph{SciPost Phys.} {\bfseries 10} (2021) 050} [\href{https://arxiv.org/abs/2007.04990}{{\ttfamily 2007.04990}}].

\bibitem{OHare:2021zrq}
C.A.J.~O'Hare, G.~Pierobon, J.~Redondo and Y.Y.Y.~Wong, \emph{{Simulations of axionlike particles in the postinflationary scenario}}, \href{https://doi.org/10.1103/PhysRevD.105.055025}{\emph{Phys. Rev. D} {\bfseries 105} (2022) 055025} [\href{https://arxiv.org/abs/2112.05117}{{\ttfamily 2112.05117}}].

\bibitem{Buschmann:2021sdq}
M.~Buschmann, J.W.~Foster, A.~Hook, A.~Peterson, D.E.~Willcox, W.~Zhang et~al., \emph{{Dark matter from axion strings with adaptive mesh refinement}}, \href{https://doi.org/10.1038/s41467-022-28669-y}{\emph{Nature Commun.} {\bfseries 13} (2022) 1049} [\href{https://arxiv.org/abs/2108.05368}{{\ttfamily 2108.05368}}].

\bibitem{Mondino:2020rkn}
C.~Mondino, A.-M.~Taki, K.~Van~Tilburg and N.~Weiner, \emph{{First Results on Dark Matter Substructure from Astrometric Weak Lensing}}, \href{https://doi.org/10.1103/PhysRevLett.125.111101}{\emph{Phys. Rev. Lett.} {\bfseries 125} (2020) 111101} [\href{https://arxiv.org/abs/2002.01938}{{\ttfamily 2002.01938}}].

\bibitem{Drlica-Wagner:2022lbd}
A.~Drlica-Wagner et~al., \emph{{Report of the Topical Group on Cosmic Probes of Dark Matter for Snowmass 2021}},  \href{https://arxiv.org/abs/2209.08215}{{\ttfamily 2209.08215}}.

\bibitem{Chung:2023syw}
D.J.H.~Chung, M.~M\"unchmeyer and S.C.~Tadepalli, \emph{{Search for isocurvature with large-scale structure: A forecast for Euclid and MegaMapper using EFTofLSS}}, \href{https://doi.org/10.1103/PhysRevD.108.103542}{\emph{Phys. Rev. D} {\bfseries 108} (2023) 103542} [\href{https://arxiv.org/abs/2306.09456}{{\ttfamily 2306.09456}}].

\bibitem{Irsic:2023equ}
V.~Ir\v{s}i\v{c} et~al., \emph{{Unveiling dark matter free streaming at the smallest scales with the high redshift Lyman-alpha forest}}, \href{https://doi.org/10.1103/PhysRevD.109.043511}{\emph{Phys. Rev. D} {\bfseries 109} (2024) 043511} [\href{https://arxiv.org/abs/2309.04533}{{\ttfamily 2309.04533}}].

\bibitem{Delos:2023dwq}
M.S.~Delos, \emph{{An analytical description of substructure-induced gravitational perturbations in stellar systems}}, \href{https://doi.org/10.1093/mnras/stae715}{\emph{Mon. Not. Roy. Astron. Soc.} {\bfseries 529} (2024) 2349} [\href{https://arxiv.org/abs/2312.13338}{{\ttfamily 2312.13338}}].

\bibitem{Nadler:2024ims}
E.O.~Nadler, V.~Gluscevic, T.~Driskell, R.H.~Wechsler, L.A.~Moustakas, A.~Benson et~al., \emph{{Forecasts for Galaxy Formation and Dark Matter Constraints from Dwarf Galaxy Surveys}}, \href{https://doi.org/10.3847/1538-4357/ad3bb1}{\emph{Astrophys. J.} {\bfseries 967} (2024) 61} [\href{https://arxiv.org/abs/2401.10318}{{\ttfamily 2401.10318}}].

\bibitem{Xiao:2024qay}
H.~Xiao, L.~Dai and M.~McQuinn, \emph{{Detecting dark matter substructures on small scales with fast radio bursts}}, \href{https://doi.org/10.1103/PhysRevD.110.023516}{\emph{Phys. Rev. D} {\bfseries 110} (2024) 023516} [\href{https://arxiv.org/abs/2401.08862}{{\ttfamily 2401.08862}}].

\bibitem{Ji:2024ott}
L.~Ji and L.~Dai, \emph{{Effects of Subhalos on Interpreting Highly Magnified Sources Near Lensing Caustics}},  \href{https://arxiv.org/abs/2407.09594}{{\ttfamily 2407.09594}}.

\bibitem{deKruijf:2024voc}
J.~de~Kruijf, E.~Vanzan, K.K.~Boddy, A.~Raccanelli and N.~Bartolo, \emph{{Searching for blue in the dark}},  \href{https://arxiv.org/abs/2408.04991}{{\ttfamily 2408.04991}}.

\bibitem{Liu:2024pjg}
R.~Liu, W.~Hu and H.~Xiao, \emph{{Warm and Fuzzy Dark Matter: Free Streaming of Wave Dark Matter}},  \href{https://arxiv.org/abs/2406.12970}{{\ttfamily 2406.12970}}.

\bibitem{Irsic:2019iff}
V.~Ir\v{s}i\v{c}, H.~Xiao and M.~McQuinn, \emph{{Early structure formation constraints on the ultralight axion in the postinflation scenario}}, \href{https://doi.org/10.1103/PhysRevD.101.123518}{\emph{Phys. Rev. D} {\bfseries 101} (2020) 123518} [\href{https://arxiv.org/abs/1911.11150}{{\ttfamily 1911.11150}}].

\bibitem{Hwang:2009js}
J.-c.~Hwang and H.~Noh, \emph{{Axion as a Cold Dark Matter candidate}}, \href{https://doi.org/10.1016/j.physletb.2009.08.031}{\emph{Phys. Lett.} {\bfseries B680} (2009) 1} [\href{https://arxiv.org/abs/0902.4738}{{\ttfamily 0902.4738}}].

\bibitem{Marsh:2010wq}
D.J.E.~Marsh and P.G.~Ferreira, \emph{{Ultra-Light Scalar Fields and the Growth of Structure in the Universe}}, \href{https://doi.org/10.1103/PhysRevD.82.103528}{\emph{Phys. Rev. D} {\bfseries 82} (2010) 103528} [\href{https://arxiv.org/abs/1009.3501}{{\ttfamily 1009.3501}}].

\bibitem{Nelson:2011sf}
A.E.~Nelson and J.~Scholtz, \emph{{Dark Light, Dark Matter and the Misalignment Mechanism}}, \href{https://doi.org/10.1103/PhysRevD.84.103501}{\emph{Phys. Rev. D} {\bfseries 84} (2011) 103501} [\href{https://arxiv.org/abs/1105.2812}{{\ttfamily 1105.2812}}].

\bibitem{Arias:2012az}
P.~Arias, D.~Cadamuro, M.~Goodsell, J.~Jaeckel, J.~Redondo and A.~Ringwald, \emph{{WISPy Cold Dark Matter}}, \href{https://doi.org/10.1088/1475-7516/2012/06/013}{\emph{JCAP} {\bfseries 06} (2012) 013} [\href{https://arxiv.org/abs/1201.5902}{{\ttfamily 1201.5902}}].

\bibitem{Noh:2017sdj}
H.~Noh, J.-c.~Hwang and C.-G.~Park, \emph{{Axion as a cold dark matter candidate: Proof to fully nonlinear order}}, \href{https://doi.org/10.3847/1538-4357/aa8366}{\emph{Astrophys. J.} {\bfseries 846} (2017) 1} [\href{https://arxiv.org/abs/1707.08568}{{\ttfamily 1707.08568}}].

\bibitem{Cembranos:2015oya}
J.A.R.~Cembranos, A.L.~Maroto and S.J.~N\'u\~nez Jare\~no, \emph{{Cosmological perturbations in coherent oscillating scalar field models}}, \href{https://doi.org/10.1007/JHEP03(2016)013}{\emph{JHEP} {\bfseries 03} (2016) 013} [\href{https://arxiv.org/abs/1509.08819}{{\ttfamily 1509.08819}}].

\bibitem{Hu:1998kj}
W.~Hu, \emph{{Structure formation with generalized dark matter}}, \href{https://doi.org/10.1086/306274}{\emph{Astrophys. J.} {\bfseries 506} (1998) 485} [\href{https://arxiv.org/abs/astro-ph/9801234}{{\ttfamily astro-ph/9801234}}].

\bibitem{Poulin:2018dzj}
V.~Poulin, T.L.~Smith, D.~Grin, T.~Karwal and M.~Kamionkowski, \emph{{Cosmological implications of ultralight axionlike fields}}, \href{https://doi.org/10.1103/PhysRevD.98.083525}{\emph{Phys. Rev. D} {\bfseries 98} (2018) 083525} [\href{https://arxiv.org/abs/1806.10608}{{\ttfamily 1806.10608}}].

\bibitem{Blas:2011rf}
D.~Blas, J.~Lesgourgues and T.~Tram, \emph{{The Cosmic Linear Anisotropy Solving System (CLASS) II: Approximation schemes}}, \href{https://doi.org/10.1088/1475-7516/2011/07/034}{\emph{JCAP} {\bfseries 07} (2011) 034} [\href{https://arxiv.org/abs/1104.2933}{{\ttfamily 1104.2933}}].

\bibitem{2022ascl.soft03026G}
D.~{Grin}, D.J.E.~{Marsh} and R.~{Hlozek}, \emph{{axionCAMB: Modification of the CAMB Boltzmann code}}, .

\bibitem{Hlozek:2014lca}
R.~Hlozek, D.~Grin, D.J.E.~Marsh and P.G.~Ferreira, \emph{{A search for ultralight axions using precision cosmological data}}, \href{https://doi.org/10.1103/PhysRevD.91.103512}{\emph{Phys. Rev. D} {\bfseries 91} (2015) 103512} [\href{https://arxiv.org/abs/1410.2896}{{\ttfamily 1410.2896}}].

\bibitem{Marsh:2015xka}
D.J.E.~Marsh, \emph{{Axion Cosmology}}, \href{https://doi.org/10.1016/j.physrep.2016.06.005}{\emph{Phys. Rept.} {\bfseries 643} (2016) 1} [\href{https://arxiv.org/abs/1510.07633}{{\ttfamily 1510.07633}}].

\bibitem{Hu:2000ke}
W.~Hu, R.~Barkana and A.~Gruzinov, \emph{{Cold and fuzzy dark matter}}, \href{https://doi.org/10.1103/PhysRevLett.85.1158}{\emph{Phys. Rev. Lett.} {\bfseries 85} (2000) 1158} [\href{https://arxiv.org/abs/astro-ph/0003365}{{\ttfamily astro-ph/0003365}}].

\bibitem{Irsic:2017yje}
V.~Ir\v{s}i\v{c}, M.~Viel, M.G.~Haehnelt, J.S.~Bolton and G.D.~Becker, \emph{{First constraints on fuzzy dark matter from Lyman-$\alpha$ forest data and hydrodynamical simulations}}, \href{https://doi.org/10.1103/PhysRevLett.119.031302}{\emph{Phys. Rev. Lett.} {\bfseries 119} (2017) 031302} [\href{https://arxiv.org/abs/1703.04683}{{\ttfamily 1703.04683}}].

\bibitem{PhysRevD.101.123026}
K.~Schutz, \emph{Subhalo mass function and ultralight bosonic dark matter}, \href{https://doi.org/10.1103/PhysRevD.101.123026}{\emph{Phys. Rev. D} {\bfseries 101} (2020) 123026}.

\bibitem{PhysRevLett.126.071302}
K.K.~Rogers and H.V.~Peiris, \emph{Strong bound on canonical ultralight axion dark matter from the lyman-alpha forest}, \href{https://doi.org/10.1103/PhysRevLett.126.071302}{\emph{Phys. Rev. Lett.} {\bfseries 126} (2021) 071302}.

\bibitem{Passaglia:2022bcr}
S.~Passaglia and W.~Hu, \emph{{Accurate effective fluid approximation for ultralight axions}}, \href{https://doi.org/10.1103/PhysRevD.105.123529}{\emph{Phys. Rev. D} {\bfseries 105} (2022) 123529} [\href{https://arxiv.org/abs/2201.10238}{{\ttfamily 2201.10238}}].

\bibitem{Schive:2014dra}
H.-Y.~Schive, T.~Chiueh and T.~Broadhurst, \emph{{Cosmic Structure as the Quantum Interference of a Coherent Dark Wave}}, \href{https://doi.org/10.1038/nphys2996}{\emph{Nature Phys.} {\bfseries 10} (2014) 496} [\href{https://arxiv.org/abs/1406.6586}{{\ttfamily 1406.6586}}].

\bibitem{Mocz:2019pyf}
P.~Mocz et~al., \emph{{First star-forming structures in fuzzy cosmic filaments}}, \href{https://doi.org/10.1103/PhysRevLett.123.141301}{\emph{Phys. Rev. Lett.} {\bfseries 123} (2019) 141301} [\href{https://arxiv.org/abs/1910.01653}{{\ttfamily 1910.01653}}].

\bibitem{Chan:2021bja}
H.Y.J.~Chan, E.G.M.~Ferreira, S.~May, K.~Hayashi and M.~Chiba, \emph{{The diversity of core\textendash{}halo structure in the fuzzy dark matter model}}, \href{https://doi.org/10.1093/mnras/stac063}{\emph{Mon. Not. Roy. Astron. Soc.} {\bfseries 511} (2022) 943} [\href{https://arxiv.org/abs/2110.11882}{{\ttfamily 2110.11882}}].

\bibitem{May:2021wwp}
S.~May and V.~Springel, \emph{{Structure formation in large-volume cosmological simulations of fuzzy dark matter: impact of the non-linear dynamics}}, \href{https://doi.org/10.1093/mnras/stab1764}{\emph{Mon. Not. Roy. Astron. Soc.} {\bfseries 506} (2021) 2603} [\href{https://arxiv.org/abs/2101.01828}{{\ttfamily 2101.01828}}].

\bibitem{May:2022gus}
S.~May and V.~Springel, \emph{{The halo mass function and filaments in full cosmological simulations with fuzzy dark matter}}, \href{https://doi.org/10.1093/mnras/stad2031}{\emph{Mon. Not. Roy. Astron. Soc.} {\bfseries 524} (2023) 4256} [\href{https://arxiv.org/abs/2209.14886}{{\ttfamily 2209.14886}}].

\bibitem{Kolb:1993hw}
E.W.~Kolb and I.I.~Tkachev, \emph{{Nonlinear axion dynamics and formation of cosmological pseudosolitons}}, \href{https://doi.org/10.1103/PhysRevD.49.5040}{\emph{Phys. Rev.} {\bfseries D49} (1994) 5040} [\href{https://arxiv.org/abs/astro-ph/9311037}{{\ttfamily astro-ph/9311037}}].

\bibitem{Amin:2019ums}
M.A.~Amin and P.~Mocz, \emph{{Formation, gravitational clustering, and interactions of nonrelativistic solitons in an expanding universe}}, \href{https://doi.org/10.1103/PhysRevD.100.063507}{\emph{Phys. Rev. D} {\bfseries 100} (2019) 063507} [\href{https://arxiv.org/abs/1902.07261}{{\ttfamily 1902.07261}}].

\bibitem{Lozanov:2017hjm}
K.D.~Lozanov and M.A.~Amin, \emph{{Self-resonance after inflation: oscillons, transients and radiation domination}}, \href{https://doi.org/10.1103/PhysRevD.97.023533}{\emph{Phys. Rev.} {\bfseries D97} (2018) 023533} [\href{https://arxiv.org/abs/1710.06851}{{\ttfamily 1710.06851}}].

\bibitem{Garcia:2022vwm}
M.A.G.~Garcia, M.~Pierre and S.~Verner, \emph{{Scalar dark matter production from preheating and structure formation constraints}}, \href{https://doi.org/10.1103/PhysRevD.107.043530}{\emph{Phys. Rev. D} {\bfseries 107} (2023) 043530} [\href{https://arxiv.org/abs/2206.08940}{{\ttfamily 2206.08940}}].

\bibitem{Agrawal:2018vin}
P.~Agrawal, N.~Kitajima, M.~Reece, T.~Sekiguchi and F.~Takahashi, \emph{{Relic Abundance of Dark Photon Dark Matter}}, \href{https://doi.org/10.1016/j.physletb.2019.135136}{\emph{Phys. Lett. B} {\bfseries 801} (2020) 135136} [\href{https://arxiv.org/abs/1810.07188}{{\ttfamily 1810.07188}}].

\bibitem{Narayanan:2000tp}
V.K.~Narayanan, D.N.~Spergel, R.~Dave and C.-P.~Ma, \emph{{Constraints on the mass of warm dark matter particles and the shape of the linear power spectrum from the Ly$\alpha$ forest}}, \href{https://doi.org/10.1086/317269}{\emph{Astrophys. J. Lett.} {\bfseries 543} (2000) L103} [\href{https://arxiv.org/abs/astro-ph/0005095}{{\ttfamily astro-ph/0005095}}].

\bibitem{Hansen:2001zv}
S.H.~Hansen, J.~Lesgourgues, S.~Pastor and J.~Silk, \emph{{Constraining the window on sterile neutrinos as warm dark matter}}, \href{https://doi.org/10.1046/j.1365-8711.2002.05410.x}{\emph{Mon. Not. Roy. Astron. Soc.} {\bfseries 333} (2002) 544} [\href{https://arxiv.org/abs/astro-ph/0106108}{{\ttfamily astro-ph/0106108}}].

\bibitem{Lewis:2002nc}
A.~Lewis and A.~Challinor, \emph{{Evolution of cosmological dark matter perturbations}}, \href{https://doi.org/10.1103/PhysRevD.66.023531}{\emph{Phys. Rev. D} {\bfseries 66} (2002) 023531} [\href{https://arxiv.org/abs/astro-ph/0203507}{{\ttfamily astro-ph/0203507}}].

\bibitem{Green:2003un}
A.M.~Green, S.~Hofmann and D.J.~Schwarz, \emph{{The power spectrum of SUSY - CDM on sub-galactic scales}}, \href{https://doi.org/10.1111/j.1365-2966.2004.08232.x}{\emph{Mon. Not. Roy. Astron. Soc.} {\bfseries 353} (2004) L23} [\href{https://arxiv.org/abs/astro-ph/0309621}{{\ttfamily astro-ph/0309621}}].

\bibitem{Viel:2005qj}
M.~Viel, J.~Lesgourgues, M.G.~Haehnelt, S.~Matarrese and A.~Riotto, \emph{{Constraining warm dark matter candidates including sterile neutrinos and light gravitinos with WMAP and the Lyman-alpha forest}}, \href{https://doi.org/10.1103/PhysRevD.71.063534}{\emph{Phys. Rev. D} {\bfseries 71} (2005) 063534} [\href{https://arxiv.org/abs/astro-ph/0501562}{{\ttfamily astro-ph/0501562}}].

\bibitem{Lesgourgues:2006nd}
J.~Lesgourgues and S.~Pastor, \emph{{Massive neutrinos and cosmology}}, \href{https://doi.org/10.1016/j.physrep.2006.04.001}{\emph{Phys. Rept.} {\bfseries 429} (2006) 307} [\href{https://arxiv.org/abs/astro-ph/0603494}{{\ttfamily astro-ph/0603494}}].

\bibitem{Viel:2007mv}
M.~Viel, G.D.~Becker, J.S.~Bolton, M.G.~Haehnelt, M.~Rauch and W.L.W.~Sargent, \emph{{How cold is cold dark matter? Small scales constraints from the flux power spectrum of the high-redshift Lyman-alpha forest}}, \href{https://doi.org/10.1103/PhysRevLett.100.041304}{\emph{Phys. Rev. Lett.} {\bfseries 100} (2008) 041304} [\href{https://arxiv.org/abs/0709.0131}{{\ttfamily 0709.0131}}].

\bibitem{Boyarsky:2008xj}
A.~Boyarsky, J.~Lesgourgues, O.~Ruchayskiy and M.~Viel, \emph{{Lyman-alpha constraints on warm and on warm-plus-cold dark matter models}}, \href{https://doi.org/10.1088/1475-7516/2009/05/012}{\emph{JCAP} {\bfseries 05} (2009) 012} [\href{https://arxiv.org/abs/0812.0010}{{\ttfamily 0812.0010}}].

\bibitem{Erickcek:2011us}
A.L.~Erickcek and K.~Sigurdson, \emph{{Reheating Effects in the Matter Power Spectrum and Implications for Substructure}}, \href{https://doi.org/10.1103/PhysRevD.84.083503}{\emph{Phys. Rev. D} {\bfseries 84} (2011) 083503} [\href{https://arxiv.org/abs/1106.0536}{{\ttfamily 1106.0536}}].

\bibitem{Lancaster:2017ksf}
L.~Lancaster, F.-Y.~Cyr-Racine, L.~Knox and Z.~Pan, \emph{{A tale of two modes: Neutrino free-streaming in the early universe}}, \href{https://doi.org/10.1088/1475-7516/2017/07/033}{\emph{JCAP} {\bfseries 07} (2017) 033} [\href{https://arxiv.org/abs/1704.06657}{{\ttfamily 1704.06657}}].

\bibitem{Irsic:2017ixq}
V.~Ir\v{s}i\v{c} et~al., \emph{{New Constraints on the free-streaming of warm dark matter from intermediate and small scale Lyman-$\alpha$ forest data}}, \href{https://doi.org/10.1103/PhysRevD.96.023522}{\emph{Phys. Rev. D} {\bfseries 96} (2017) 023522} [\href{https://arxiv.org/abs/1702.01764}{{\ttfamily 1702.01764}}].

\bibitem{wdm}
V.~Ir\v{s}i\v{c} et~al., \emph{{New Constraints on the free-streaming of warm dark matter from intermediate and small scale Lyman-$\alpha$ forest data}}, \href{https://doi.org/10.1103/PhysRevD.96.023522}{\emph{Phys. Rev. D} {\bfseries 96} (2017) 023522} [\href{https://arxiv.org/abs/1702.01764}{{\ttfamily 1702.01764}}].

\bibitem{Erickcek:2021fsu}
A.L.~Erickcek, P.~Ralegankar and J.~Shelton, \emph{{Cannibalism's lingering imprint on the matter power spectrum}}, \href{https://doi.org/10.1088/1475-7516/2022/01/017}{\emph{JCAP} {\bfseries 01} (2022) 017} [\href{https://arxiv.org/abs/2106.09041}{{\ttfamily 2106.09041}}].

\bibitem{Ballesteros:2020adh}
G.~Ballesteros, M.A.G.~Garcia and M.~Pierre, \emph{{How warm are non-thermal relics? Lyman-$\alpha$ bounds on out-of-equilibrium dark matter}}, \href{https://doi.org/10.1088/1475-7516/2021/03/101}{\emph{JCAP} {\bfseries 03} (2021) 101} [\href{https://arxiv.org/abs/2011.13458}{{\ttfamily 2011.13458}}].

\bibitem{Sarkar:2021pqh}
A.K.~Sarkar, K.L.~Pandey and S.K.~Sethi, \emph{{Using the redshift evolution of the Lyman-\ensuremath{\alpha} effective opacity as a probe of dark matter models}}, \href{https://doi.org/10.1088/1475-7516/2021/10/077}{\emph{JCAP} {\bfseries 10} (2021) 077} [\href{https://arxiv.org/abs/2101.09917}{{\ttfamily 2101.09917}}].

\bibitem{Garcia:2023qab}
M.A.G.~Garcia, M.~Pierre and S.~Verner, \emph{{New window into gravitationally produced scalar dark matter}}, \href{https://doi.org/10.1103/PhysRevD.108.115024}{\emph{Phys. Rev. D} {\bfseries 108} (2023) 115024} [\href{https://arxiv.org/abs/2305.14446}{{\ttfamily 2305.14446}}].

\bibitem{Ma:1995ey}
C.-P.~Ma and E.~Bertschinger, \emph{{Cosmological perturbation theory in the synchronous and conformal Newtonian gauges}}, \href{https://doi.org/10.1086/176550}{\emph{Astrophys. J.} {\bfseries 455} (1995) 7} [\href{https://arxiv.org/abs/astro-ph/9506072}{{\ttfamily astro-ph/9506072}}].

\bibitem{Lozanov:2019jxc}
K.D.~Lozanov, \emph{{Lectures on Reheating after Inflation}},  \href{https://arxiv.org/abs/1907.04402}{{\ttfamily 1907.04402}}.

\bibitem{Amin:2014eta}
M.A.~Amin, M.P.~Hertzberg, D.I.~Kaiser and J.~Karouby, \emph{{Nonperturbative Dynamics Of Reheating After Inflation: A Review}}, \href{https://doi.org/10.1142/S0218271815300037}{\emph{Int. J. Mod. Phys.} {\bfseries D24} (2014) 1530003} [\href{https://arxiv.org/abs/1410.3808}{{\ttfamily 1410.3808}}].

\bibitem{Planck:2018jri}
{\scshape Planck} collaboration, \emph{{Planck 2018 results. X. Constraints on inflation}}, \href{https://doi.org/10.1051/0004-6361/201833887}{\emph{Astron. Astrophys.} {\bfseries 641} (2020) A10} [\href{https://arxiv.org/abs/1807.06211}{{\ttfamily 1807.06211}}].

\bibitem{Ganjoo:2023fgg}
H.~Ganjoo and M.S.~Delos, \emph{{Simulations of gravitational heating due to early matter domination}}, \href{https://doi.org/10.1088/1475-7516/2024/04/015}{\emph{JCAP} {\bfseries 04} (2024) 015} [\href{https://arxiv.org/abs/2306.14961}{{\ttfamily 2306.14961}}].

\bibitem{Hu:1995en}
W.~Hu and N.~Sugiyama, \emph{{Small scale cosmological perturbations: An Analytic approach}}, \href{https://doi.org/10.1086/177989}{\emph{Astrophys. J.} {\bfseries 471} (1996) 542} [\href{https://arxiv.org/abs/astro-ph/9510117}{{\ttfamily astro-ph/9510117}}].

\bibitem{Baumann:2022mni}
D.~Baumann, \emph{{Cosmology}}, Cambridge University Press (7, 2022), \href{https://doi.org/10.1017/9781108937092}{10.1017/9781108937092}.

\bibitem{Gleiser:1993pt}
M.~Gleiser, \emph{{Pseudostable bubbles}}, \href{https://doi.org/10.1103/PhysRevD.49.2978}{\emph{Phys. Rev.} {\bfseries D49} (1994) 2978} [\href{https://arxiv.org/abs/hep-ph/9308279}{{\ttfamily hep-ph/9308279}}].

\bibitem{Copeland:1995fq}
E.J.~Copeland, M.~Gleiser and H.R.~Muller, \emph{{Oscillons: Resonant configurations during bubble collapse}}, \href{https://doi.org/10.1103/PhysRevD.52.1920}{\emph{Phys. Rev.} {\bfseries D52} (1995) 1920} [\href{https://arxiv.org/abs/hep-ph/9503217}{{\ttfamily hep-ph/9503217}}].

\bibitem{Kasuya:2002zs}
S.~Kasuya, M.~Kawasaki and F.~Takahashi, \emph{{I-balls}}, \href{https://doi.org/10.1016/S0370-2693(03)00344-7}{\emph{Phys. Lett.} {\bfseries B559} (2003) 99} [\href{https://arxiv.org/abs/hep-ph/0209358}{{\ttfamily hep-ph/0209358}}].

\bibitem{Amin:2010jq}
M.A.~Amin and D.~Shirokoff, \emph{{Flat-top oscillons in an expanding universe}}, \href{https://doi.org/10.1103/PhysRevD.81.085045}{\emph{Phys. Rev.} {\bfseries D81} (2010) 085045} [\href{https://arxiv.org/abs/1002.3380}{{\ttfamily 1002.3380}}].

\bibitem{Zhang:2020bec}
H.-Y.~Zhang, M.A.~Amin, E.J.~Copeland, P.M.~Saffin and K.D.~Lozanov, \emph{{Classical Decay Rates of Oscillons}}, \href{https://doi.org/10.1088/1475-7516/2020/07/055}{\emph{JCAP} {\bfseries 07} (2020) 055} [\href{https://arxiv.org/abs/2004.01202}{{\ttfamily 2004.01202}}].

\bibitem{Cyncynates:2021rtf}
D.~Cyncynates and T.~Giurgica-Tiron, \emph{{Structure of the oscillon: The dynamics of attractive self-interaction}}, \href{https://doi.org/10.1103/PhysRevD.103.116011}{\emph{Phys. Rev. D} {\bfseries 103} (2021) 116011} [\href{https://arxiv.org/abs/2104.02069}{{\ttfamily 2104.02069}}].

\bibitem{Levkov:2023ncb}
D.G.~Levkov and V.E.~Maslov, \emph{{Analytic description of monodromy oscillons}}, \href{https://doi.org/10.1103/PhysRevD.108.063514}{\emph{Phys. Rev. D} {\bfseries 108} (2023) 063514} [\href{https://arxiv.org/abs/2306.06171}{{\ttfamily 2306.06171}}].

\bibitem{Amin:2011hj}
M.A.~Amin, R.~Easther, H.~Finkel, R.~Flauger and M.P.~Hertzberg, \emph{{Oscillons After Inflation}}, \href{https://doi.org/10.1103/PhysRevLett.108.241302}{\emph{Phys. Rev. Lett.} {\bfseries 108} (2012) 241302} [\href{https://arxiv.org/abs/1106.3335}{{\ttfamily 1106.3335}}].

\bibitem{Lozanov:2019ylm}
K.D.~Lozanov and M.A.~Amin, \emph{{Gravitational perturbations from oscillons and transients after inflation}}, \href{https://doi.org/10.1103/PhysRevD.99.123504}{\emph{Phys. Rev.} {\bfseries D99} (2019) 123504} [\href{https://arxiv.org/abs/1902.06736}{{\ttfamily 1902.06736}}].

\bibitem{Gleiser:2014ipa}
M.~Gleiser and N.~Graham, \emph{{Transition To Order After Hilltop Inflation}}, \href{https://doi.org/10.1103/PhysRevD.89.083502}{\emph{Phys. Rev. D} {\bfseries 89} (2014) 083502} [\href{https://arxiv.org/abs/1401.6225}{{\ttfamily 1401.6225}}].

\bibitem{Shafi:2024jig}
M.~Shafi, E.J.~Copeland, R.~Mahbub, S.S.~Mishra and S.~Basak, \emph{{Formation and decay of oscillons after inflation in the presence of an external coupling, Part-I: Lattice simulations}},  \href{https://arxiv.org/abs/2406.00108}{{\ttfamily 2406.00108}}.

\bibitem{Ballesteros:2024hhq}
G.~Ballesteros, J.~Iguaz~Juan, P.D.~Serpico and M.~Taoso, \emph{{Primordial black hole formation from self-resonant preheating?}},  \href{https://arxiv.org/abs/2406.09122}{{\ttfamily 2406.09122}}.

\bibitem{Arvanitaki:2019rax}
A.~Arvanitaki, S.~Dimopoulos, M.~Galanis, L.~Lehner, J.O.~Thompson and K.~Van~Tilburg, \emph{{Large-misalignment mechanism for the formation of compact axion structures: Signatures from the QCD axion to fuzzy dark matter}}, \href{https://doi.org/10.1103/PhysRevD.101.083014}{\emph{Phys. Rev. D} {\bfseries 101} (2020) 083014} [\href{https://arxiv.org/abs/1909.11665}{{\ttfamily 1909.11665}}].

\bibitem{Sakharov:1994id}
A.S.~Sakharov and M.Y.~Khlopov, \emph{{The Nonhomogeneity problem for the primordial axion field}}, {\emph{Phys. Atom. Nucl.} {\bfseries 57} (1994) 485}.

\bibitem{Sakharov:1996xg}
A.S.~Sakharov, D.D.~Sokoloff and M.Y.~Khlopov, \emph{{Large scale modulation of the distribution of coherent oscillations of a primordial axion field in the universe}}, {\emph{Phys. Atom. Nucl.} {\bfseries 59} (1996) 1005}.

\bibitem{Khlopov:1998uj}
M.Y.~Khlopov, A.S.~Sakharov and D.D.~Sokoloff, \emph{{The large scale modulation of the density distribution in standard axionic CDM and its cosmological and physical impact}},  in \emph{{2nd International Workshop on Birth of the Universe and Fundamental Physics}}, 12, 1998 [\href{https://arxiv.org/abs/hep-ph/9812286}{{\ttfamily hep-ph/9812286}}].

\bibitem{Imagawa:2021sxt}
K.~Imagawa, M.~Kawasaki, K.~Murai, H.~Nakatsuka and E.~Sonomoto, \emph{{Free streaming length of axion-like particle after oscillon/I-ball decays}}, \href{https://doi.org/10.1088/1475-7516/2023/02/024}{\emph{JCAP} {\bfseries 02} (2023) 024} [\href{https://arxiv.org/abs/2110.05790}{{\ttfamily 2110.05790}}].

\bibitem{Vaquero:2018tib}
A.~Vaquero, J.~Redondo and J.~Stadler, \emph{{Early seeds of axion miniclusters}}, \href{https://doi.org/10.1088/1475-7516/2019/04/012}{\emph{JCAP} {\bfseries 04} (2019) 012} [\href{https://arxiv.org/abs/1809.09241}{{\ttfamily 1809.09241}}].

\bibitem{Visinelli:2018wza}
L.~Visinelli and J.~Redondo, \emph{{Axion Miniclusters in Modified Cosmological Histories}}, \href{https://doi.org/10.1103/PhysRevD.101.023008}{\emph{Phys. Rev. D} {\bfseries 101} (2020) 023008} [\href{https://arxiv.org/abs/1808.01879}{{\ttfamily 1808.01879}}].

\bibitem{Eggemeier:2019khm}
B.~Eggemeier, J.~Redondo, K.~Dolag, J.C.~Niemeyer and A.~Vaquero, \emph{{First Simulations of Axion Minicluster Halos}}, \href{https://doi.org/10.1103/PhysRevLett.125.041301}{\emph{Phys. Rev. Lett.} {\bfseries 125} (2020) 041301} [\href{https://arxiv.org/abs/1911.09417}{{\ttfamily 1911.09417}}].

\bibitem{Smith:2023fob}
T.L.~Smith, J.T.~Giblin, Jr., M.A.~Amin, M.~Gerhardinger, E.~Florio, M.~Cerep et~al., \emph{{Novel integrated Sachs-Wolfe effect from early dark energy}}, \href{https://doi.org/10.1103/PhysRevD.108.123534}{\emph{Phys. Rev. D} {\bfseries 108} (2023) 123534} [\href{https://arxiv.org/abs/2304.02028}{{\ttfamily 2304.02028}}].

\bibitem{Namjoo:2017nia}
M.H.~Namjoo, A.H.~Guth and D.I.~Kaiser, \emph{{Relativistic Corrections to Nonrelativistic Effective Field Theories}}, \href{https://doi.org/10.1103/PhysRevD.98.016011}{\emph{Phys. Rev.} {\bfseries D98} (2018) 016011} [\href{https://arxiv.org/abs/1712.00445}{{\ttfamily 1712.00445}}].

\bibitem{Salehian:2021khb}
B.~Salehian, H.-Y.~Zhang, M.A.~Amin, D.I.~Kaiser and M.H.~Namjoo, \emph{{Beyond Schr\"odinger-Poisson: nonrelativistic effective field theory for scalar dark matter}}, \href{https://doi.org/10.1007/JHEP09(2021)050}{\emph{JHEP} {\bfseries 09} (2021) 050} [\href{https://arxiv.org/abs/2104.10128}{{\ttfamily 2104.10128}}].

\bibitem{Chavanis:2011zi}
P.-H.~Chavanis, \emph{{Mass-radius relation of Newtonian self-gravitating Bose-Einstein condensates with short-range interactions: I. Analytical results}}, \href{https://doi.org/10.1103/PhysRevD.84.043531}{\emph{Phys. Rev. D} {\bfseries 84} (2011) 043531} [\href{https://arxiv.org/abs/1103.2050}{{\ttfamily 1103.2050}}].

\bibitem{Glennon:2020dxs}
N.~Glennon and C.~Prescod-Weinstein, \emph{{Modifying PyUltraLight to model scalar dark matter with self-interactions}}, \href{https://doi.org/10.1103/PhysRevD.104.083532}{\emph{Phys. Rev. D} {\bfseries 104} (2021) 083532} [\href{https://arxiv.org/abs/2011.09510}{{\ttfamily 2011.09510}}].

\bibitem{Sankharva:2021spi}
K.~Sankharva and S.~Sethi, \emph{{Nonminimally coupled ultralight axions as cold dark matter}}, \href{https://doi.org/10.1103/PhysRevD.105.103517}{\emph{Phys. Rev. D} {\bfseries 105} (2022) 103517} [\href{https://arxiv.org/abs/2110.04322}{{\ttfamily 2110.04322}}].

\bibitem{Jain:2023tsr}
M.~Jain, W.~Wanichwecharungruang and J.~Thomas, \emph{{Kinetic relaxation and nucleation of Bose stars in self-interacting wave dark matter}}, \href{https://doi.org/10.1103/PhysRevD.109.016002}{\emph{Phys. Rev. D} {\bfseries 109} (2024) 016002} [\href{https://arxiv.org/abs/2310.00058}{{\ttfamily 2310.00058}}].

\bibitem{Chen:2024pyr}
J.~Chen and H.-Y.~Zhang, \emph{{Novel structures and collapse of solitons in nonminimally gravitating dark matter halos}},  \href{https://arxiv.org/abs/2407.09265}{{\ttfamily 2407.09265}}.

\bibitem{Arvanitaki:2009fg}
A.~Arvanitaki, S.~Dimopoulos, S.~Dubovsky, N.~Kaloper and J.~March-Russell, \emph{{String Axiverse}}, \href{https://doi.org/10.1103/PhysRevD.81.123530}{\emph{Phys. Rev. D} {\bfseries 81} (2010) 123530} [\href{https://arxiv.org/abs/0905.4720}{{\ttfamily 0905.4720}}].

\bibitem{PhysRevD.86.055013}
K.R.~Dienes and B.~Thomas, \emph{Phenomenological constraints on axion models of dynamical dark matter}, \href{https://doi.org/10.1103/PhysRevD.86.055013}{\emph{Phys. Rev. D} {\bfseries 86} (2012) 055013}.

\bibitem{Dienes:2024wnu}
K.R.~Dienes, L.~Heurtier, F.~Huang, T.M.P.~Tait and B.~Thomas, \emph{{Cosmological Stasis from Dynamical Scalars: Tracking Solutions and the Possibility of a Stasis-Induced Inflation}},  \href{https://arxiv.org/abs/2406.06830}{{\ttfamily 2406.06830}}.

\bibitem{Amaral:2024tjg}
D.W.P.~Amaral, M.~Jain, M.A.~Amin and C.~Tunnell, \emph{{Vector wave dark matter and terrestrial quantum sensors}}, \href{https://doi.org/10.1088/1475-7516/2024/06/050}{\emph{JCAP} {\bfseries 06} (2024) 050} [\href{https://arxiv.org/abs/2403.02381}{{\ttfamily 2403.02381}}].

\bibitem{Adshead:2021kvl}
P.~Adshead and K.D.~Lozanov, \emph{{Self-gravitating Vector Dark Matter}}, \href{https://doi.org/10.1103/PhysRevD.103.103501}{\emph{Phys. Rev. D} {\bfseries 103} (2021) 103501} [\href{https://arxiv.org/abs/2101.07265}{{\ttfamily 2101.07265}}].

\bibitem{Jain:2021pnk}
M.~Jain and M.A.~Amin, \emph{{Polarized solitons in higher-spin wave dark matter}}, \href{https://doi.org/10.1103/PhysRevD.105.056019}{\emph{Phys. Rev. D} {\bfseries 105} (2022) 056019} [\href{https://arxiv.org/abs/2109.04892}{{\ttfamily 2109.04892}}].

\bibitem{Amin:2022pzv}
M.A.~Amin, M.~Jain, R.~Karur and P.~Mocz, \emph{{Small-scale structure in vector dark matter}}, \href{https://doi.org/10.1088/1475-7516/2022/08/014}{\emph{JCAP} {\bfseries 08} (2022) 014} [\href{https://arxiv.org/abs/2203.11935}{{\ttfamily 2203.11935}}].

\bibitem{Gorghetto:2022sue}
M.~Gorghetto, E.~Hardy, J.~March-Russell, N.~Song and S.M.~West, \emph{{Dark Photon Stars: Formation and Role as Dark Matter Substructure}},  \href{https://arxiv.org/abs/2203.10100}{{\ttfamily 2203.10100}}.

\bibitem{Jain:2023ojg}
M.~Jain, M.A.~Amin, J.~Thomas and W.~Wanichwecharungruang, \emph{{Kinetic relaxation and Bose-star formation in multicomponent dark matter}}, \href{https://doi.org/10.1103/PhysRevD.108.043535}{\emph{Phys. Rev. D} {\bfseries 108} (2023) 043535} [\href{https://arxiv.org/abs/2304.01985}{{\ttfamily 2304.01985}}].

\bibitem{Chen:2023bqy}
J.~Chen, X.~Du, M.~Zhou, A.~Benson and D.J.E.~Marsh, \emph{{Gravitational Bose-Einstein condensation of vector or hidden photon dark matter}}, \href{https://doi.org/10.1103/PhysRevD.108.083021}{\emph{Phys. Rev. D} {\bfseries 108} (2023) 083021} [\href{https://arxiv.org/abs/2304.01965}{{\ttfamily 2304.01965}}].

\bibitem{Chen:2024vgh}
J.~Chen, L.H.~Nguyen, X.~Du and D.J.E.~Marsh, \emph{{Vector Dark Matter Halo: From Polarization Dynamics to Direct Detection}},  \href{https://arxiv.org/abs/2407.17315}{{\ttfamily 2407.17315}}.

\bibitem{Zhang:2023fhs}
H.-Y.~Zhang and S.~Ling, \emph{{Phenomenology of wavelike vector dark matter nonminimally coupled to gravity}}, \href{https://doi.org/10.1088/1475-7516/2023/07/055}{\emph{JCAP} {\bfseries 07} (2023) 055} [\href{https://arxiv.org/abs/2305.03841}{{\ttfamily 2305.03841}}].

\bibitem{Zhang:2021xxa}
H.-Y.~Zhang, M.~Jain and M.A.~Amin, \emph{{Polarized vector oscillons}}, \href{https://doi.org/10.1103/PhysRevD.105.096037}{\emph{Phys. Rev. D} {\bfseries 105} (2022) 096037} [\href{https://arxiv.org/abs/2111.08700}{{\ttfamily 2111.08700}}].

\bibitem{Jain:2022kwq}
M.~Jain, \emph{{Soliton stars in Yang-Mills-Higgs theories}}, \href{https://doi.org/10.1103/PhysRevD.106.085011}{\emph{Phys. Rev. D} {\bfseries 106} (2022) 085011} [\href{https://arxiv.org/abs/2205.03418}{{\ttfamily 2205.03418}}].

\bibitem{Jain:2022agt}
M.~Jain and M.A.~Amin, \emph{{i-SPin: an integrator for multicomponent Schr\"odinger-Poisson systems with self-interactions}}, \href{https://doi.org/10.1088/1475-7516/2023/04/053}{\emph{JCAP} {\bfseries 04} (2023) 053} [\href{https://arxiv.org/abs/2211.08433}{{\ttfamily 2211.08433}}].

\bibitem{Zhang:2024bjo}
H.-Y.~Zhang, \emph{{Unified view of scalar and vector dark matter solitons}},  \href{https://arxiv.org/abs/2406.05031}{{\ttfamily 2406.05031}}.

\bibitem{Agrawal}
P.~Agrawal, N.~Kitajima, M.~Reece, T.~Sekiguchi and F.~Takahashi, \emph{{Relic Abundance of Dark Photon Dark Matter}}, \href{https://doi.org/10.1016/j.physletb.2019.135136}{\emph{Phys. Lett. B} {\bfseries 801} (2020) 135136} [\href{https://arxiv.org/abs/1810.07188}{{\ttfamily 1810.07188}}].

\bibitem{Dror:2018pdh}
J.A.~Dror, K.~Harigaya and V.~Narayan, \emph{{Parametric Resonance Production of Ultralight Vector Dark Matter}}, \href{https://doi.org/10.1103/PhysRevD.99.035036}{\emph{Phys. Rev. D} {\bfseries 99} (2019) 035036} [\href{https://arxiv.org/abs/1810.07195}{{\ttfamily 1810.07195}}].

\bibitem{Co:2018lka}
R.T.~Co, A.~Pierce, Z.~Zhang and Y.~Zhao, \emph{{Dark Photon Dark Matter Produced by Axion Oscillations}}, \href{https://doi.org/10.1103/PhysRevD.99.075002}{\emph{Phys. Rev. D} {\bfseries 99} (2019) 075002} [\href{https://arxiv.org/abs/1810.07196}{{\ttfamily 1810.07196}}].

\bibitem{Adshead:2023qiw}
P.~Adshead, K.D.~Lozanov and Z.J.~Weiner, \emph{{Dark photon dark matter from an oscillating dilaton}}, \href{https://doi.org/10.1103/PhysRevD.107.083519}{\emph{Phys. Rev. D} {\bfseries 107} (2023) 083519} [\href{https://arxiv.org/abs/2301.07718}{{\ttfamily 2301.07718}}].

\bibitem{Cyncynates:2023zwj}
D.~Cyncynates and Z.J.~Weiner, \emph{{Detectable, defect-free dark photon dark matter}},  \href{https://arxiv.org/abs/2310.18397}{{\ttfamily 2310.18397}}.

\bibitem{Graham:2015rva}
P.W.~Graham, J.~Mardon and S.~Rajendran, \emph{{Vector Dark Matter from Inflationary Fluctuations}}, \href{https://doi.org/10.1103/PhysRevD.93.103520}{\emph{Phys. Rev. D} {\bfseries 93} (2016) 103520} [\href{https://arxiv.org/abs/1504.02102}{{\ttfamily 1504.02102}}].

\bibitem{Kolb:2023ydq}
E.W.~Kolb and A.J.~Long, \emph{{Cosmological gravitational particle production and its implications for cosmological relics}},  \href{https://arxiv.org/abs/2312.09042}{{\ttfamily 2312.09042}}.

\bibitem{Long:2019lwl}
A.J.~Long and L.-T.~Wang, \emph{{Dark Photon Dark Matter from a Network of Cosmic Strings}}, \href{https://doi.org/10.1103/PhysRevD.99.063529}{\emph{Phys. Rev. D} {\bfseries 99} (2019) 063529} [\href{https://arxiv.org/abs/1901.03312}{{\ttfamily 1901.03312}}].

\bibitem{East:2022rsi}
W.E.~East and J.~Huang, \emph{{Dark photon vortex formation and dynamics}}, \href{https://doi.org/10.1007/JHEP12(2022)089}{\emph{JHEP} {\bfseries 12} (2022) 089} [\href{https://arxiv.org/abs/2206.12432}{{\ttfamily 2206.12432}}].

\bibitem{Saikawa:2024bta}
K.~Saikawa, J.~Redondo, A.~Vaquero and M.~Kaltschmidt, \emph{{Spectrum of global string networks and the axion dark matter mass}},  \href{https://arxiv.org/abs/2401.17253}{{\ttfamily 2401.17253}}.

\bibitem{Gorghetto:2024vnp}
M.~Gorghetto, E.~Hardy and G.~Villadoro, \emph{{More Axion Stars from Strings}},  \href{https://arxiv.org/abs/2405.19389}{{\ttfamily 2405.19389}}.

\bibitem{WidrowKaiser:1993}
L.M.~{Widrow} and N.~{Kaiser}, \emph{{Using the Schroedinger Equation to Simulate Collisionless Matter}}, \href{https://doi.org/10.1086/187073}{\emph{Astrophyical Journal Letters} {\bfseries 416} (1993) L71}.

\bibitem{Mocz:2018ium}
P.~Mocz, L.~Lancaster, A.~Fialkov, F.~Becerra and P.-H.~Chavanis, \emph{{Schr\"odinger-Poisson\textendash{}Vlasov-Poisson correspondence}}, \href{https://doi.org/10.1103/PhysRevD.97.083519}{\emph{Phys. Rev. D} {\bfseries 97} (2018) 083519} [\href{https://arxiv.org/abs/1801.03507}{{\ttfamily 1801.03507}}].

\end{thebibliography}\endgroup

\appendix
\newpage
\section{Energy density, pressure and equation of state}
\label{sec:EOS}
Let us assume that the density and pressure are dominated by subhorizon scalar field modes. Then using $p$ and $\rho$ defined in Eq.~\eqref{eq:varphi_stress_energy}, the spatial averages (ignoring metric potentials), are given by
\begin{align}
\overline{p}(t)&=\int_{\bm{q}}\left[\frac{1}{2}\underline{P}_{\dot{\varphi}}(t,q)-\frac{1}{2}\left(\frac{q^2}{3a^2}+m^2\right)\underline{P}_\varphi(t,q)\right]\,,\\
\overline{\rho}(t)&=\int_{\bm{q}}\left[\frac{1}{2}\underline{P}_{\dot{\varphi}}(t,q)+\frac{1}{2}\left(\frac{q^2}{a^2}+m^2\right)\underline{P}_\varphi(t,q)\right]\,.
\end{align}
If $\underline{P}_{\dot{\varphi}}= \omega_q^2\underline{P}_\varphi$ where $\omega_q^2=q^2/a^2+m^2$, then 
\begin{align}
\overline{p}(t)= \int_{\bm{q}}\frac{q^2}{3a^2}\underline{P}_\varphi(t,q)\,,\qquad
\overline{\rho}(t)=\int_{\bm{q}}\left(\frac{q^2}{a^2}+m^2\right)\underline{P}_\varphi(t,q)\,.
\end{align}
If the density is dominated by relativistic modes so that $m^2$ can be ignored, we get $\overline{p}/\overline{\rho}\rightarrow 1/3$, whereas if it is dominated by  non-relativistic modes, we get $\overline{p}/\overline{\rho}\rightarrow 0$, as expected. It is also worth noting that
\beq
\overline{\rho}(t)+\overline{p}(t)= \int_\bq\left(\frac{4}{3}\frac{q^2}{a^2}+m^2\right)\underline{P}_\varphi(t,q).
\eeq
Using the form of the initial field spectrum $\underline{\Delta}^2_\varphi(t_i,q)=q^3/(2\pi^2)\underline{P}_\varphi(t_i,q)$, in \eqref{eq:P_varphi_candidates}, we have
\beq
\overline{\rho}(t_i)&=m^2\underline{\Delta}^2_\varphi(k_*) \left(\frac{1}{\alpha}+\frac{1}{\nu }\right)\left[1+ \frac{({k_*}/{a_im})^2}{ (1 -2\alpha^{-1}) (1 +2\nu^{-1})}\right],\\
\overline{p}(t_i)&=\frac{k_*^2}{3a_i^2}\underline{\Delta}^2_\varphi(k_*)\left(\frac{1}{\alpha}+\frac{1}{\nu }\right)\frac{ 1}{(1 -2\alpha^{-1}) (1 +2\nu^{-1})},\\
w(t_i)&=\frac{\overline{p}(t_i)}{\overline{\rho}(t_i)} =\frac{1}{3}\frac{(k_*/a_i m)^2} {(k_*/a_i m)^2+(1-2\alpha^{-1})(1+{2}{\nu}^{-1})},\\
  \dot{w}(t_i) &= -\frac{2}{3} \frac{(k_*/a_i m)^2 (1-2\alpha^{-1})(1+{2}{\nu}^{-1}) H_i} {\left((k_*/a_i m)^2+(1-2\alpha^{-1})(1+{2}{\nu}^{-1})\right)^2} .
\eeq

\subsection*{The Kernel}
\label{sec:kernel}
It is possible to evaluate the Kernel function in Sec.~\ref{sec:generating_fields_density} analytically for the initial field spectrum given by Eq.~\eqref{eq:P_varphi_candidates}:
\beq
K(r)&=
\frac{k_\ast^3 \sqrt{\underline{P}_\varphi(k_\ast)}}{2\pi^2}
\left[\frac{2}{(\nu+3)}\,{}_1 F_2\left(\frac{\nu+3}{4},\frac{3}{2},\frac{\nu+7}{4},-\frac{(k_*r)^2}{4}\right)-(\nu\rightarrow -\alpha)\right.\,\\
&\qquad\qquad+\left.\cos\left(\frac{1}{4} \pi  (\alpha +1)\right) \Gamma \left(\frac{1-\alpha}{2}\right) (k_* r)^{\frac{\alpha -3}{2}}\right].\\
\eeq
For $\alpha<3$, the kernel function diverges as $r\rightarrow 0$. Even for $\alpha>3$, the function falls rapidly beyond $k_*r=1$. 

\section{WKB Solutions, with long wavelength metric perturbations}
\label{sec:WKB}
At linear order in the metric perturbations, the evolution of the field is governed by ($\Psi=\Phi)$
\beq\label{eom1}
\frac{1}{a^3} \partial_t[a^3(1-4\Psi) \dot\varphi]- \frac{1}{a^2} \nabla^2 \varphi
+m^2 (1- 2\Psi) \varphi = 0.
\eeq
First, we will assume that the gradients in the metric potentials are negligible. We will also assume that the metric potentials are varying slowly in time. In ``Fourier" space, $\varphi(t,\bx)=\int_\bq \varphi_{\bq}(t;\Psi)e^{i\bq\cdot\bx}$, the equations of motion become
\beq\label{eom2}
\frac{1}{A} \partial_t[A \dot\varphi_\bq]+\left[\frac{q^2}{a^2}(1+4\Psi)+m^2 (1+2\Psi)\right]
 \varphi_\bq = 0\,\qquad A\equiv a^3(1-4\Psi)\,,
\eeq
We can eliminate the first time-derivative in this equation by defining $\chi_\bq=\sqrt{A}\varphi_q$. Then, finding the leading order WKB solution for the $\chi_\bq$ equation and then converting back to $\varphi_\bq$, at leading order in the WKB approximation, the solutions are
\beq
    \varphi_\bq(t;\Psi)= c_\bq^+(\Psi)\frac{e^{- i\int^t\omega(t')}}{\sqrt{2A\omega}}+c_\bq^-(\Psi)\frac{e^{i\int^t dt'\omega(t')}}{\sqrt{2A\omega}},
\eeq
where
\beq
    \qquad \omega^2=\frac{q^2}{a^2}(1+4\Psi)+m^2 (1+2\Psi)-\frac{1}{2}\frac{d^2}{dt^2}\ln A-\frac{1}{4}\left(\frac{d}{dt}\ln A\right)^{\!2}.
\eeq
The factor of $\sqrt{2}$ in the denominator of the WKB solutions is an arbitrary choice at this point. The reality of $\varphi$ implies that $\varphi_\bq^*=\varphi_{-\bq}$, which in turn implies that $(c_\bq^+)^*=c_{-\bq}^-$ and $(c_\bq^-)^*=c_{-\bq}^+$. In what follows, we drop the log derivative terms which are of order $H^2$. This means that either we are operating in the subhorizon {\it or} in $m\gg H$ regime. 

The solution can be written in terms of sines and cosines as follows:\footnote{Note that the $g_q$ here is different from the one in the main text, it differs by a factor of $\omega$.}
\beq
    \varphi_\bq(t;\Psi)= F_\bq(\Psi) f_q(t;\Psi)+G_\bq(\Psi) g_q(t;\Psi)\,,
\eeq
where
\beq
 &F_\bq(\Psi)=c_\bq^+(\Psi)+c_\bq^-(\Psi), \quad G_\bq(\Psi)=i(c_\bq^-(\Psi)-c_\bq^+(\Psi))\,,\\
&f_q(t;\Psi)=\frac{1}{\sqrt{2A\omega}} \cos \int^t\omega(t';\Psi),\,\, g_q(t;\Psi)=\frac{1}{\sqrt{2A\omega}} \sin \int^t\omega(t';\Psi).
\eeq
We will assume the following for the random variables $F_\bq$ and $G_\bq$:
\begin{align}
&\langle F_\bq(\Psi) F^*_{\bq'}(\Psi)\rangle=(2\pi)^3\delta_D(\bq-\bq')\underline{P}_F(q;\Psi)\,,\quad \langle G_\bq(\Psi) G^*_{\bq'}(\Psi)\rangle=(2\pi)^3\delta_D(\bq-\bq')\underline{P}_G(q;\Psi),\nonumber\\
&\langle G_\bq F^*_{\bq'}\rangle=0\,.
\end{align}
Note that this assumption needs to be revisited because we are assuming spatial homogeneity here apart from the $\Psi$ dependence. In what follows we will assume that $\underline{P}_F=\underline{P}_G=2\underline{P}$.\footnote{Note that these conditions imply: $\langle(c_\bq^++c_\bq^-)(c_\bp^++c_\bp^-)^*\rangle=\langle(c_\bq^--c_\bq^+)(c_\bp^--c_\bp^+)^*\rangle=(2\pi)^3 \underline{P}(q)\delta_D(\bq-\bp)$ and $\langle(c_\bq^--c_\bq^+)(c_\bp^++c_\bp^-)^* \rangle=0$. These can be satisfied if $\langle c_\bq^+(c_\bp^+)^*\rangle =(2\pi)^3 \underline{P}(q)\delta_D(\bq-\bp)$ and $\langle c_\bq^+c_{-\bp}^+\rangle=0.$}
\\ \\
The correlation function for the time-dependent field is then given by
\begin{align}
\label{eq:phiqphiq'}
\langle \varphi_\bq(t;\Psi)\varphi^*_{\bq'}(t;\Psi)\rangle
= (2\pi)^3\delta_D(\bq-\bq')\underline{P}_\varphi(t,q;\Psi)\quad\textrm{where}\quad \underline{P}_\varphi(t,q;\Psi)=\frac{1}{A\omega}\underline{P}(q;\Psi)\,.
\end{align}
Similarly
\begin{align}
\langle \dot\varphi_\bq(t;\Psi)\dot\varphi^*_{\bq'}(t;\Psi)\rangle
&=(2\pi)^3\delta_D(\bq-\bq') \underline{P}_{\dot\varphi}(t,q;\Psi)\quad\textrm{where}\quad
\underline{P}_{\dot\varphi}(t,q;\Psi)\approx \frac{\omega}{A} \underline{P}(q;\Psi).
\end{align}
In the last line we assumed that $\omega^{-1}$ is the fastest time-scale, so that $\dot{f}_q\approx-\omega g_q$ and $\dot{g}_q\approx \omega f_q$.
We can now use this in the energy density expression and take the expectation value of the field to get
\beq
\langle\rho\rangle
    =\frac{1}{2}\left[(1-2\Psi)\langle\dot{\varphi}^2\rangle+(1+2\Psi)\langle\dfrac{(\nabla\varphi)^2}{a^2}\rangle+m^2\langle\varphi^2\rangle\right]\,.
\eeq
The expectation values
\beq
\label{eq:expval3}
\langle \dot{\varphi}^2\rangle\approx \int_\bq \frac{\omega^2}{A\omega} \underline{P}(q;\Psi),\quad
\langle \frac{(\nabla{\varphi})^2}{a^2}\rangle= \int_\bq \frac{1}{A\omega} \frac{q^2}{a^2} \underline{P}(q;\Psi),\quad
m^2\langle\varphi^2\rangle= \int_\bq \frac{1}{A\omega} m^2 \underline{P}(q;\Psi),
\eeq
which yields (to leading order in metric potentials)
\beq
\label{eq:exprhoNewt}
\langle\rho\rangle
    &\approx\frac{1}{2}\int_\bq\frac{\underline{P}(q;\Psi)}{A\omega}\left[(1-2\Psi)\omega^2+(1+2\Psi)\frac{q^2}{a^2}+m^2\right],\\
    &= \frac{1}{a^3}\int_\bq \sqrt{q^2/a^2+m^2}\underline{P}(q;\Psi)\left[1+\frac{4q^2/a^2+3m^2}{q^2/a^2+m^2}\Psi\right].
\eeq
Note that we have not made any assumptions about radiation or matter domination here. We only assumed that gradients in the metric potentials can be neglected and that they are slowly varying in time (appropriate for superhorizon scales).

\section{Numerical details}
\label{sec:numerical_details}

In Sec.~\ref{sec:evol}, we provided the results for three lattice simulations, addressing the free streaming for a Klein Gordon field, a scalar field in external gravitational potential, and a scalar field that forms oscillons due to self-interactions.
We describe the details of our numerical method in this appendix. 

We first describe our scheme for spatial discretization.
The simulations are all carried out on a 3D lattice with $N^3$ grid points, where $N = 384$.
For the free field simulation and the external gravity simulation (Sec.~\ref{sec:W/GravPertEvol} and \ref{sec:evolution_with_gravity}), we set the box volume to be $L^3$, where $mL = 307.2$, and the resolution is $m\Delta{x} = mL / N = 0.8$.
To validate our numerical scheme, we also ran two extra external gravity simulations, one with $N = 256$, $m\Delta{x}=0.8$, and another with $N = 256$, $m\Delta{x}=1.2$, and found no qualitative difference in the power spectra compared to the $N = 384$, $m\Delta{x}=0.8$ one.
For the oscillon simulation, we take $N=384$, $mL = 768$, or $m\Delta{x} = 2$.  Given that the typical size of the oscillons in our simulations is $mL_{\mathrm{oscillon}} = 15$, each oscillon is resolved by around $8$ lattice points. 
Laplacians are evaluated using finite difference on the standard 7-point stencil.

For time evolution, we use Runge-Kutta 4 for all simulations.
For the free field simulation and the external gravity simulation, we use time step $\Delta{t} = 0.1 m^{-1}$ to evolve from $mt=10$ to $mt=36000$.
For the oscillon simulation, we use $\Delta{t} = 0.05 m^{-1}$ to evolve from $mt=0$ to $mt=12500$.
We found that the total energy is well-preserved for our given time steps.
For the free field simulation and the external gravity simulation, the total energy $a^3 \overline{\rho}$ fell by around $25\%$ from $mt=10$ to $mt=1510$, and fell less than $1\%$ from $mt=1510$ to $mt=36000$.
This behavior is expected, since the energy density should fall as $a^{-3(1+w)}$ with $w>0$ when the field is still relativistic, and fall as $a^{-3}$ after the field becomes nonrelativistic.
Note that for the simulation with gravity, the time-varying gravitational potential $\Psi$ should also affect energy conservation, though the effect is not discernible.
Finally, for the self-interaction simulation, the total energy dropped by around $0.5\%$ throughout the simulation.

In order to speed up the simulation with external gravity, we apply two approximations on Eq.~\eqref{eq:varphi_eom}.
The first approximation is linearization of the exponentials, e.g. $e^{\pm 2\Psi} \approx 1 \pm 2 \Psi$.
The other approximation is to use a coarse grid to store the gravitational potential $\Psi$.
This is a valid approximation since $\Psi$ has a spectrum cut off at around $k=aH \ll k_{\mathrm{UV}}$ (see Eq.~\eqref{eq:gravitational_potentials_in_radiation_domination}), and contains negligible small length scale modes throughout the simulation.
More specifically, we use a coarse grid with grid size $M^3$ and $M=128=N/3$.
Lattice site $(a,b,c)$ on the coarse grid corresponds to site $(3a,3b,3c)$ on the full grid.
When evaluating $\Psi$ on the full grid, we use nearest neighbor interpolation, such that the values of $\Psi$ at site $(3a\pm1,3b\pm1,3c\pm1)$ on the full grid are taken to be the value at site $(a,b,c)$ on the coarse grid.
The difference in norm between the interpolated $\Psi$ and the actual $\Psi$ is $\norm{\Psi_{\textrm{interpolated}} - \Psi}_2 / \norm{\Psi}_2 = 0.05$ at $t=t_i$.
This approximation significantly reduces the time spent on the computation $\mathcal{R}_\bk \mapsto \Psi_\bk \mapsto \Psi(\bx)$, which is done at each RK4 stage.
We checked the validity of the above approximations by comparing the simulation against a shorter one without approximations.
We found that the evolution of field and density spectra in the two simulations are almost identical, except for a high-$k$ tail in the field spectrum at around $k/a_im = 2$. See Fig.~\ref{fig:varphi_spectrum_evolution_2}.
This tail only appears when we use the coarse grid approximation on $\Psi$, and is likely an artifact of the interpolation scheme.

We use binning to produce the power spectrum plots (e.g. Fig.~\ref{fig:delta_spectrum_without_gravity} and \ref{fig:delta_spectrum_with_gravity}) from simulation outputs.
More specifically, we bin spectral power into fixed logarithmic intervals of size $\Delta{\ln{k}} = 0.12$, with the bins given by $\ln(k/k_{\mathrm{IR}}) \in [n \Delta{\ln{k}}, (n+1)\Delta{\ln{k}})$ for $n \geq 0$.
For each bin, we sum the Fourier powers $\abs{f_\bk}^2$ and divide by the mode counting factor to get the average power $ \sum_{\bk \in \textrm{bin}} \abs{f_\bk}^2 / \textrm{(number of $\bk$'s in bin)} $.
We also associate with each bin a wavenumber $k_{\mathrm{bin}}$, which is simply the lowest-$k$ in the bin.
The points on the spectra are then given by $P_f(k_{\mathrm{bin}}) = \textrm{(average power of bin)}$.
The spectra are usually jagged at low-$k$ and smooth at high-$k$, since bins at low-$k$ contain only a small number of modes and thus have large ensemble variances. Some low-$k$ bins are in fact empty; those bins are dropped.

Finally, we comment on our numerical implementation.
We wrote CUDA code and ran our simulations on a RTX 3060 Ti graphics card.
The use of GPU led to more than $10$ times speed-up compared to our CPU implementations.
Specifically, the free field simulation took $6$ hours, averaging to $0.06\mathrm{s}$ per RK4 time step.
The external gravity simulation took $8.7$ hours, averaging to $0.087\mathrm{s}$ per RK4 time step.
The self interaction simulation took $6.8$ hours, averaging to $0.1\mathrm{s}$ per RK4 time step.
The link to our code is \href{https://github.com/hypermania/Cosmic-Fields-Lite}{https://github.com/hypermania/Cosmic-Fields-Lite}.

\section{Full expression for the density 2-point function}
\label{sec:full_expr_for_2_point_function}
The ``short'' contribution to the density 2-point function $\expval{\rho(\bx)\rho(\by)}$ is:
\begin{align}
  \xi_{\rho\rho}^{\mathrm{(short)}}(\bx, {\bm y}) &= e^{-2\Psi(\bx)-2\Psi({\bm y})} \frac12 \expval{\dot{\varphi}(\bx)\dot{\varphi}({\bm y})}^2 \nonumber \\
                                                  &\quad + e^{2\Psi(\bx)+2\Psi({\bm y})} \frac{1}{2 a^4} \expval{\partial_i\varphi(\bx) \partial_j\varphi({\bm y})} \expval{\partial_i\varphi(\bx) \partial_j\varphi({\bm y})} \nonumber \\
                                                  &\quad + e^{2\Psi(\bx)} \frac{1}{2 a^2} m^2 \expval{\partial_i\varphi(\bx) \varphi({\bm y})} \expval{\partial_i\varphi(\bx) \varphi({\bm y})} \nonumber \\
                                                  &\quad + e^{2\Psi({\bm y})} \frac{1}{2 a^2} m^2 \expval{\varphi(\bx) \partial_i\varphi({\bm y})} \expval{\varphi(\bx) \partial_i\varphi({\bm y})} \nonumber \\
                                                  &\quad + \frac12 m^4 \expval{\varphi(\bx)\varphi({\bm y})}^2
                                                    \;.
\end{align}
In Fourier space, the above terms are given (to first order in $\Psi$) by
\begin{align}
  & \int_{\bx,\by} e^{i \bk \cdot \bx - i \bk' \cdot \by} e^{-2\Psi(\bx)-2\Psi({\bm y})} \frac12 \expval{\dot{\varphi}(\bx)\dot{\varphi}({\bm y})}^2 \nonumber \\
  =& \frac12 \int_{{\bm q}} \Big[ (2\pi)^3 \delta_D(\bk-\bk') \underline{P}_{\dot\varphi}(q) \underline{P}_{\dot{\varphi}}(\abs{{\bm q} + \bk'}) \nonumber \\
                                               & -2 \Psi_{\bk'-\bk} \underline{P}_{\dot\varphi}(q) \underline{P}_{\dot{\varphi}}(\abs{{\bm q} + \bk'})
                                                 -2 \Psi_{\bk'-\bk} \underline{P}_{\dot\varphi}(q) \underline{P}_{\dot{\varphi}}(\abs{{\bm q} + \bk}) \nonumber \\
                                               & + 2 (f_2)_{\bk'-\bk} \sqrt{\underline{P}_{\dot\varphi}(\abs{{\bm q}+\bk}) \underline{P}_{\dot\varphi}(\abs{{\bm q}+\bk'})}  \underline{P}_{\dot{\varphi}}(q)
                                                 \Big]
                                                 \;,\\\nonumber \\
  &\int_{\bx,\by} e^{i \bk \cdot \bx - i \bk' \cdot \by} e^{2\Psi(\bx)+2\Psi({\bm y})} \frac{1}{2 a^4} \expval{\partial_i\varphi(\bx) \partial_j\varphi({\bm y})} \expval{\partial_i\varphi(\bx) \partial_j\varphi({\bm y})} \nonumber \\
  =& \frac{1}{2 a^4} \int_{\bm q} \Big[(2\pi)^3 \delta_D(\bk-\bk') \left({\bm q} \cdot (\bk+{\bm q})\right)^2 \underline{P}_\varphi(q) \underline{P}_\varphi(\abs{\bk+{\bm q}}) \nonumber \\
  &+ 2 \Psi_{\bk'-\bk} \left({\bm q} \cdot (\bk'+{\bm q})\right)^2 \underline{P}_\varphi(\abs{\bk'+{\bm q}}) \underline{P}_\varphi(q)
    + 2 \Psi_{\bk'-\bk} \left({\bm q} \cdot (\bk+{\bm q})\right)^2 \underline{P}_\varphi(\abs{\bk+{\bm q}}) \underline{P}_\varphi(q) \nonumber \\
  &+ 2 (f_1)_{\bk'-\bk} \left({\bm q} \cdot (\bk+{\bm q})\right) \left({\bm q} \cdot (\bk'+{\bm q})\right) \sqrt{\underline{P}_\varphi(\abs{\bk+{\bm q}}) \underline{P}_\varphi(\abs{\bk'+{\bm q}})} \underline{P}_\varphi(q) \Big],
    \;\\\nonumber\\
  & \int_{\bx,\by} e^{i \bk \cdot \bx - i \bk' \cdot \by} e^{2\Psi(\bx)} \frac{1}{2 a^2} m^2 \expval{\partial_i\varphi(\bx) \varphi({\bm y})} \expval{\partial_i\varphi(\bx) \varphi({\bm y})} \nonumber \\
  =& \frac{m^2}{2a^2} \int_{\bm q} \Big[
     (2\pi)^3\delta_D(\bk-\bk') {\bm q} \cdot (\bk+{\bm q}) \underline{P}_\varphi(\abs{\bk+{\bm q}}) \underline{P}_\varphi(q) \nonumber \\
  &+ 2 \Psi_{\bk'-\bk} {\bm q} \cdot (\bk'+{\bm q}) \underline{P}_\varphi(\abs{\bk'+{\bm q}}) \underline{P}_\varphi(q) \nonumber \\
  &+ 2 (f_1)_{\bk'-\bk} {\bm q} \cdot (\bk+{\bm q}) \sqrt{\underline{P}_\varphi(\abs{\bk+{\bm q}}) \underline{P}_\varphi(\abs{\bk'+{\bm q}})} \underline{P}_\varphi(q) \Big]
    \;,
 \end{align}
 \begin{align}
  & \int_{\bx,\by} e^{i \bk \cdot \bx - i \bk' \cdot \by} e^{2\Psi({\bm y})} \frac{1}{2 a^2} m^2 \expval{\varphi(\bx) \partial_i\varphi({\bm y})} \expval{\varphi(\bx) \partial_i\varphi({\bm y})} \nonumber \\
  =& \frac{m^2}{2a^2} \int_{\bm q} \Big[
     (2\pi)^3\delta_D(\bk-\bk') {\bm q} \cdot (\bk'+{\bm q}) \underline{P}_\varphi(\abs{\bk'+{\bm q}}) \underline{P}_\varphi(q) \nonumber \\
  &+ 2 \Psi_{\bk'-\bk} {\bm q} \cdot (\bk+{\bm q}) \underline{P}_\varphi(\abs{\bk+{\bm q}}) \underline{P}_\varphi(q) \nonumber \\
  &+ 2 (f_1)_{\bk'-\bk} {\bm q} \cdot (\bk'+{\bm q}) \sqrt{\underline{P}_\varphi(\abs{\bk+{\bm q}}) \underline{P}_\varphi(\abs{\bk'+{\bm q}})} \underline{P}_\varphi(q) \Big]
    \;,\\\nonumber\\
  & \int_{\bx,\by} e^{i \bk \cdot \bx - i \bk' \cdot \by} \frac12 m^4 \expval{\varphi(\bx)\varphi({\bm y})}^2 \nonumber \\
  =& \frac12 m^4 \int_{{\bm q}} \Big[ (2\pi)^3 \delta_D(\bk-\bk') \underline{P}_{\varphi}(q) \underline{P}_{\varphi}(\abs{\bk+{\bm q}}) \nonumber \\
                                               & + 2 (f_1)_{\bk'-\bk} \sqrt{\underline{P}_{\varphi}(\abs{{\bm q}+\bk}) \underline{P}_{\varphi}(\abs{{\bm q}+\bk'})}  \underline{P}_{\varphi}(q)
                                                 \Big]
                                                 \;.
\end{align}
For $\bk=\bk'$, the first order terms in $\Psi$ are all zero since $\overline{f_i} = 0$ and $\overline{\Psi} = 0$.
In this case the sum can be collected as
\begin{align}
  & \xi_{\rho\rho}^{\mathrm{(short)}}(\bk, \bk') \nonumber \\
  =& \int_{\bx,\by} e^{i \bk \cdot \bx - i \bk' \cdot \by} \xi_{\rho\rho}^{\mathrm{(short)}}(\bx, {\bm y}) \nonumber \\
  =& (2\pi)^3 \delta_D(\bk-\bk') \frac12 \int_{\bm q} \left[ \underline{P}_{\dot\varphi}(q) \underline{P}_{\dot{\varphi}}(\abs{\bk+{\bm q}})
     + \left(m^2 + \frac{{\bm q}\cdot(\bk+{\bm q})}{a^2} \right)^2 \underline{P}_{\varphi}(q) \underline{P}_{\varphi}(\abs{\bk+{\bm q}}) \right]\nonumber\\
     &\qquad\qquad \times\left( 1 + \order{\Psi^2} \right)
     \;.
\end{align}
Note that the higher order $\Psi$ correction is given as a $1 + \order{\Psi^2}$ factor instead of a $\order{\Psi^2}$ term.  This factor arises since the $\Psi$ dependence in $\xi_{\rho\rho}^{\mathrm{(short)}}(\bx, {\bm y})$ comes in the form $(1+\alpha_1\Psi)(1+\alpha_2\Psi)\hdots$, and we know that the first order terms in $\Psi$ vanish here.

\section{Free-Streaming Transfer Function}
\label{sec:smoothing}
To get a more concrete understanding of free-streaming transfer function, let us consider a simple scenario.

Let us begin with a phase space distribution function $f(t_i,\bx,\bq)$. In absence of any interactions, for non-relativistic particles in Minkowski Space, we have:
\beq
\frac{df}{dt}=\partial_tf+\frac{\bq}{m}\cdot \nabla f=0\Longrightarrow f(t,\bx,\bq)=f(t_i,\bx-\frac{\bq}{m}(t-t_i),\bq)\,,
\eeq
where $\bm{\lambda}_{q,\rm fs}(t)=(\bq/m)(t-t_i)$ and we used $\dot{\bx}=\bq/m$. If free-streaming is the only dynamics, then $f(t,\bx,\bq)=f(0,\bx-\bm{\lambda}_{q,\rm fs}(t),\bq)$. A second assumption is that $f(t_i,\bx,\bq)=n(t_i,\bx)\mathcal{F}(t_i,\bq)/\overline{n}(t_i)$. Note that $\overline{n}(t_i)=\int_\bq \mathcal{F}(t_i,\bq)$. The connection to the field picture is simply via the recognition that for the nonrelativistic particles, the field power spectrum $P_\varphi(t_i,\bq)\propto\mathcal{F}(t_i,\bq)$.\footnote{In a different context, for a detailed exploration of connection between phase space evolution in the particle picture and non-relativistic field picture, see \cite{WidrowKaiser:1993,Mocz:2018ium}.} In what follows, we  evaluate the time-evolution of the spatial number density perturbations in presence of free-streaming. Recall that:
\beq
n(t,\bx)&=\int_\bq f(t,\bx,\bq).
\eeq
In Fourier space:
\beq
n(t,\bk)&=\int_{\bq,\bx} e^{-i\bk\cdot\bx} f(t,\bx,\bq),\\
&=\int_{\bq,\bx} e^{-i\bk\cdot\bx} f(0,\bx-\bm{\lambda}_{q,\rm fs}(t),\bq)\quad\textrm{where}\quad\bm{\lambda}_{q,\rm fs}(t)=(\bq/m)(t-t_i),\\
&=\int_{\bq,\by} e^{-i\bk\cdot\by}e^{-i\bk\cdot\bm{\lambda}_{q,\rm fs}(t)} f(t_i,\by,\bq)\quad\textrm{where}\quad\by=\bx-\bm{\lambda}_{q,\rm fs}(t),\\
&=\int_{\bq,\by} e^{-i\bk\cdot\by}e^{-i\bq\cdot\bm{\lambda}_{k,\rm fs}(t)} f(t_i,\by,\bq)\quad\textrm{because}\quad\bk\cdot\bm{\lambda}_{q,\rm fs}(t)=\bq\cdot\bm{\lambda}_{k,\rm fs}(t),\\
&=\int_{\bq} e^{-i\bq\cdot\bm{\lambda}_{k,\rm fs}(t)} \tilde{f}(t_i,\bk,\bq),\\
&=\frac{n(t_i,\bk)}{\overline{n}(t_i)}\int_{\bq} e^{-i\bq\cdot\bm{\lambda}_{k,\rm fs}(t)} \mathcal{F}(t_i,\bq),\\
&=n(t_i,\bk) T_{\rm sf}(t,\bk)\quad\textrm{where}\quad T_{\rm fs}(t,\bk)\equiv {\overline{n}^{-1}(t_i)}\int_{\bq} e^{-i\bq\cdot\bm{\lambda}_{k,\rm fs}(t)} \mathcal{F}(t_i,\bq). \\
\eeq
Thus, we have a simple to understand result:
\beq
n(t,\bk)=n(t_i,\bk)T_{\rm fs}(t,\bk),
\eeq
where the transfer function $T_{\rm fs}(t,\bk)$ is simply the Fourier transform of the initial phase space distribution function provided in \eqref{eq:Tfs2}, with the free streaming displacement (for $\bq=\bk$) being the conjugate variable to $\bq$. The angular integration can be carried out if $\mathcal{F}(t_i,\bq)=\mathcal{F}(t_i,q)$. This would yield the $\sin(k/q_{\rm fs})/(k/q_{\rm fs})$ factor in the transfer function, where $q_{\rm fs}=\lambda_{q,\rm fs}^{-1}(t)$. Note that we assumed that the particles were already all non-relativistic at $t=t_i$. 

Returning back to position space
\beq
n(t,\bx)=\int_\bk e^{i\bk\cdot\bx}n(t_i,\bk)T_{\rm fs}(t,k)= \int_{\bm{r}} n(t,\bm{r})K_{\rm fs}(t,\bx-\bm{r}),
\eeq
where in position space, free-streaming is captured by the smoothing kernel:
\beq
K_{\rm fs}(t,\bx)=\int_\bk e^{i\bk\cdot\bx}T_{\rm fs}(t,\bk).
\eeq
For the simple case that $q^3/(2\pi^2)\mathcal{F}(t_i,q)=\mathcal{A}(q/k_*)^3\Theta(k_*-q)$, we can evaluate this explicitly to gain some intuition:
\beq
K_{\rm fs}(t,|\bx-\bm{r}|)=\frac{[k_{\rm fs}^*(t)]^3}{4\pi}\Theta\left(1-k_{\rm fs}^*(t)|\bx-\bm{r}|\right).
\eeq
This smoothing function is a top hat function with a width $[k_{\rm fs}^*(t)]^{-1}$, where $[k_{\rm fs}^*(t)]^{-1}=k_*/m(t-t_i)$.

\paragraph{Including Expansion}
Now, for an expanding universe $\bq/m\rightarrow \bq/a(t)m$, and $\bx$ is comoving, and the comoving free-streaming length $\bm{\lambda}_{q,\rm fs}(t)=(\bq/m)\int_{t_i}^t dt'/a^2(t').$ With the interpretation that number densities are comoving as well, everything else carries through. Again, we have assumed that all particles are non-relativistic at $t_i$.

More generally, by using the Boltzmann equation, one can include the effects of bulk velocities, as well as growth of density perturbations due to gravitational clustering into account. 

\end{document}